\documentclass[12pt]{JHEP3}

\usepackage{amsmath,epsfig}
\usepackage{amssymb,amsfonts}
\usepackage{latexsym}
\usepackage{graphicx}
\usepackage{longtable}
\relax
\renewcommand{\theequation}{\arabic{section}.\arabic{equation}}
\def\be{\begin{equation}}
\def\ee{\end{equation}}
\def\bea{\begin{eqnarray}}
\def\eea{\end{eqnarray}}
\def\qslash{{q\hspace{-6.3pt}/}}
\def\STr{{\rm STr}}
\def\Tr{{\rm Tr}}

\newcommand\fverb{\setbox\pippobox=\hbox\bgroup\verb}
\newcommand\fverbdo{\egroup\medskip\noindent%
                        \fbox{\unhbox\pippobox}\ }
\newcommand\fverbit{\egroup\item[\fbox{\unhbox\pippobox}]}

\newcommand{\rc}{\nonumber\\}

\newcommand{\bear}{\begin{eqnarray}}

\newcommand{\eear}{\end{eqnarray}}

\newbox\pippobox

\def\6{\partial}

\def\a{\alpha}

\def\half{\frac12}

\def\m{\mu}
\def\n{\nu}

\def\sp{\;\;\;,\;\;\;}

\def\sq
\def\a{\alpha}

\def\tr{{\rm Tr}}

\def\hri#1#2{\href{http://arxiv.org/abs/#1}{[ArXiv:#1]#2}}
\def\hre#1#2{\href{http://arxiv.org/abs/#1/#2}{[ArXiv:#1/#2]}}

\title{An AdS/QCD model from tachyon condensation: II}
\author{Ioannis Iatrakis$^a$, Elias Kiritsis$^{a,b}$,
\'Angel Paredes$^c$\\
~\\
$^a$ \href{http://hep.physics.uoc.gr}{Crete Center for Theoretical Physics},
Department of Physics, University of Crete, 71003 Heraklion, Greece\\
~\\
$^b$ \href{http://www.apc.univ-paris7.fr}{APC, Universit\'e Paris 7}, \\ B\^atiment Condorcet, F-75205, Paris Cedex 13, France (UMR du CNRS 7164).\\
~\\
$^c$  Departament de F\'\i sica Fonamental and ICCUB Institut de
Ci\`encies del Cosmos, Universitat de Barcelona, Mart\'\i\ i Franqu\`es, 1,
E-08028, Barcelona, Spain.}

\preprint{CCTP-2010-13~~~~~~~~~~~~~~\\}

\abstract{ A simple holographic model is presented and analyzed that describes
chiral symmetry breaking and the physics of the meson sector in QCD.
This is a bottom-up model that incorporates string theory ingredients
like tachyon condensation which is expected to be the main
manifestation of chiral symmetry breaking
in the holographic context. As a model for glue the
Kuperstein-Sonnenschein background is used.
The structure of the flavor vacuum is analyzed in the quenched
approximation. Chiral symmetry breaking is shown at
zero temperature. Above the deconfinement transition chiral symmetry
  is restored.  A complete holographic renormalization is performed and
the chiral condensate is calculated for different quark masses both at
zero and non-zero temperatures.
   The $0^{++},0^{-+},1^{++},1^{--}$ meson trajectories are analyzed and
their masses and decay constants are computed.
  The asymptotic trajectories are linear.
  The model has one phenomenological parameter beyond those of QCD that
affects the $1^{++},0^{-+}$ sectors. Fitting this parameter we obtain
very good agreement with data.
  The model improves in several ways the popular hard-wall and soft wall
bottom-up models. }


{
\begin{document}

\section{Introduction}

The AdS/CFT correspondence \cite{Maldacena:1997re} has been one of the most fruitful arenas for research
in fundamental physics in the last decade. Having the possibility of mapping strongly-coupled field
theories to weakly coupled gravity  has set the stage for a large amount of effort devoted to use
this idea in order to obtain  new results on the physics of the strong interactions (see \cite{Mateos:2007ay}, for an introduction).
It is clear that
a precise and controllable string dual of QCD is far from our present understanding. However, it is possible to build
models that describe interesting strong coupling phenomena which share many similarities with real world
physics.
For the physics of pure glue, models descending from string theory, \cite{D4,KS,MN} have provided important clues towards the confinement
of color.
Phenomenological models for glue inspired and motivated from the AdS/CFT correspondence ranged from very simple like AdS/QCD \cite{ps}
to more sophisticated versions, namely ``Improved Holographic QCD" that capture rather accurately the dynamics of glue at both zero \cite{ihqcd} and finite temperature \cite{ihqcd2}.

In this context, the holographic description of chiral symmetry breaking ($\chi$SB) has been thoroughly studied. Early
examples were studied in the Maldacena-Nunez \cite{MN} and Klebanov-Strassler backgrounds \cite{KS}.
 Later, more QCD-like $\chi$SB, in the sense that the operator
condensing was bilinear of fundamental fields, was described in \cite{topdown,Kruczenski:2003uq}.
In the beginning chiral symmetry breaking involved abelian chiral symmetry,
\cite{MN,KS,topdown,Kruczenski:2003uq}.
A major breakthrough was the Sakai-Sugimoto model \cite{Sakai:2004cn} where the broken symmetry is non-abelian
$U(N_f)_L\times U(N_f)_R \to U(N_f)_V$, as opposed to just a $U(1)$.
However, the model of \cite{Sakai:2004cn} and generalizations of it have its own limitations.
Just as an example, one can mention the absence of a tower of excited pions.

All the aforementioned models come from well controlled approximations
 to string theory and, accordingly,  are top-down approaches.
A different possibility is to use string theory just as inspiration and
define {\it ad hoc} holographic models using some QCD features as inputs. This is the bottom-up approach.
The obvious question is whether the output extracted from such  models is larger than their input, and
we believe that experience has shown that the answer is yes.
The benchmark bottom-up models for chiral symmetry breaking and meson physics are the hard wall model
\cite{Erlich:2005qh,Da Rold:2005zs} based on the Polchinski-Strassler background
\cite{ps} and the soft wall model \cite{Karch:2006pv}.

In the present work, we analyze in detail  several aspects of a bottom-up model presented first in \cite{Iatrakis:2010zf}.
It gives a explicit realization of the framework for chiral symmetry breaking first advocated in \cite{ckp}: the quarks
and antiquarks are introduced
by a brane-antibrane pair. A key point of the dynamics is the condensation of
the lowest lying bifundamental scalar which
comes from the strings connecting the brane to the antibrane. Around flat space  this scalar  has negative mass squared and this is called the
tachyon. This corresponds precisely to a QCD-like chiral symmetry breaking.
We need an effective action to describe this dynamics and we will resort to the tachyon-DBI action proposed
by Sen in \cite{Sen:2003tm}. Once we have decided that we will use this action, we still have to choose
an expression for the tachyon potential and a holographic geometry in order to use as curved background.
For these choices, we will constrain ourselves to the (arguably) simplest possibilities.
Another interesting property of considering brane-antibrane tachyonic actions is that there is a natural
Wess-Zumino term, which correctly incorporates to the model features like parity and charge
conjugation symmetries and anomalies; see the discussions in
\cite{ckp}.

So far our ingredients are those of a top-down approach.
However, it is not possible to stay in a limit in which the approximation
to string theory is controlled if one wants to reproduce some QCD features in this context. For instance,
we will need a background with curvature comparable to the string scale. We do not regard this point
as a negative feature, but just as a sign that our approach should be considered of the bottom-up type.
This has its own advantages, since for instance it seems impossible to get Regge trajectories for
excited mesons $m_n^2 \sim n$ from any top-down approach, because in those cases the meson mass scale
is parametrically smaller than the QCD string tension scale, see for instance \cite{Kruczenski:2003be}.
Therefore, the model discussed in this work should be regarded as a phenomenological model, partly
inspired by top-down consideration and in particular by Sen's action \cite{Sen:2003tm}. These top-down
inputs will generate some dynamics (compared to other bottom-up approaches) which will be
crucial in the successful modeling of several QCD features,
see the discussions in \cite{ckp} and also in section \ref{sec:summary}.

We will constrain ourselves to the abelian case of a single quark flavor but, unlike \cite{topdown,Kruczenski:2003uq},
this is not an essential limitation, since we can make the model non-abelian by piling up branes and
antibranes, in the spirit of \cite{Sakai:2004cn}. This elaboration, however, is left for future work.

In section \ref{sec: backgr}, we will discuss the backgrounds (both for confined and deconfined
phases) and the gravity action of which they are solutions. In section \ref{section3}, we will study
in detail the equation for the tachyon modulus $\tau$ and its bulk vacuum expectation value.
In other words, we find the open string vacuum and show how it dynamically breaks the chiral
symmetry. In section \ref{sec:mesons}, we discuss in detail linearized
open string excitations around the vacuum,
namely the meson physics. A good review of this kind of analysis
in different holographic frameworks is \cite{Erdmenger:2007cm}. Apart from remarking several
general qualitative properties, we end the section with a quantitative phenomenological analysis.
Section \ref{sec:melting} provides a brief analysis of the same kind of excitations, but in the
deconfined phase. We conclude in section \ref{sec:discussion} with several  discussions;
we will convey the pros and cons of the present model and give some ideas for future directions.
We have relegated various technical comments to eight appendices. In particular,
in order to facilitate
the reading of the text,
 we review the meaning of the different constants and parameters that will
appear throughout the paper and for which physical reasons some of them are fixed in appendix
\ref{app:bookkeeing} .

\section{The gravitational background}
\label{sec: backgr}

As acknowledged in the introduction, the model we will discuss does not come from any controlled
approximation to string theory. Notwithstanding, we will follow general insights coming from string
theory and effective actions developed in that framework, especially in the non-critical setting, \cite{ks1}-\cite{Casero:2005se}. In this sense, the meson physics
(in the quenched approximation) is described
by the dynamics of a D4-anti D4 system in a fixed closed string background, \cite{BCCKP}.

We take the following gravitational two-derivative action \cite{ks1} for the background
fields:
\be
S= \int d^6x \sqrt{g_{(6)}}\,\left[ e^{-2\phi}
\left( {\cal R} + 4 (\partial \phi)^2 + \frac{c}{\alpha'}\right)
- \frac12 \frac{1}{6!} F_{(6)}^2
\right]\,\,,
\ee
with a constant $c$.
We consider the solution discussed in \cite{ks2}
whose metric is given by:
\be
ds_6^2 \equiv-g_{tt}dt^2+g_{zz}dz^2+g_{xx}dx_{3}^2+g_{\eta\eta}d\eta^2=\frac{R^2}{z^2} \left[ dx_{1,3}^2 +  f_\Lambda^{-1} dz^2 + f_\Lambda\, d\eta^2 \right]
\label{adsbh6}
\ee
with:
\be
f_\Lambda=1-\frac{z^5}{z_\Lambda^5}
\ee
This is the AdS soliton, a double Wick rotation of an AdS$_6$ Schwartszchild black hole.
The only active RR-form we consider is:
\be
F_{(6)} = \frac{Q_c}{\sqrt{\alpha'}} \sqrt{-g_{(6)}}\,d^6x
\label{F6val}
\ee
for some constant $Q_c$ which is proportional to the number of colors and that will not
be important in the following.
The dilaton is constant:
\be
e^\phi = \frac{1}{Q_c} \sqrt{\frac{2c}{3}}
\label{dilatonval}
\ee
The coordinate $\eta$ is compactified and regularity of the metric at $z=z_\Lambda$ requires
the following periodicity condition:
\be
\eta \sim \eta + \delta \eta\,\,,\qquad\qquad
\delta\eta= \frac{4\pi}{5} z_\Lambda = \frac{2\pi}{M_{KK}}\,\,.
\label{etaperiod}
\ee
The AdS radius is given by:
\be
R^2 = \frac{30}{c} \alpha'
\label{Rads}
\ee
The application of this geometry for a phenomenological non-critical strings/gauge
duality  was first discussed  in \cite{ks1,ks2}.
The solution is dual to 1+4 dimensional gauge theory compactified in a circle with
(susy-breaking) antiperiodic
boundary conditions for the fermions. Thus, the low energy theory is 1+3 dimensional confining gauge theory
coupled to a set of massive Kaluza-Klein fields.

One can consider the theory at non-zero temperature by compactifying to Euclidean time $t_E$.
When both circles $t_E$ and $\eta$ are compactified, there is a second solution competing
with (\ref{adsbh6}):
\be
ds_6^2 = \frac{R^2}{z^2} \left[ -f_T dt^2 + dx_{3}^2 +  f_T^{-1} dz^2 + \, d\eta^2 \right]
\label{adsbh6b}
\ee
while (\ref{F6val}), (\ref{dilatonval}), (\ref{Rads}) still hold.
We have introduced:
\be
f_T=1-\frac{z^5}{z_T^5}
\ee
and $z_T$ is related to the period of the euclidean time and therefore to the temperature as:
\be
t_E \sim t_E + \delta t_E\,\,,\qquad\qquad
\delta t_E= \frac{4\pi}{5} z_T = \frac{1}{T}\,\,.
\label{tEperiod}
\ee
Since when we Euclideanize (\ref{adsbh6}), (\ref{adsbh6b}) both solutions are related
by the interchange $t_E \leftrightarrow \eta$, $z_T \leftrightarrow z_\Lambda$, the symmetry
makes obvious that there is a deconfining first order phase transition at
\be
T_c = \frac{M_{KK}}{2\pi} = \frac{5}{4\pi \,z_\Lambda}
\label{tdeconf}
\ee
For $T<T_c$, the confining solution (\ref{adsbh6}) is preferred and, conversely,
(\ref{adsbh6b}) dominates for $T>T_c$.
Of course, this discussion is just a straightforward generalization of
\cite{Witten:1998zw}.

\section{The tachyon vacuum expectation value}
\label{section3}

Our main interest will be to study a ``tachyon-DBI" action for
a single brane-antibrane pair
of the form advocated in \cite{Sen:2003tm}. In section \ref{othertachyons} we will
comment about the literature related to effective actions including open string tachyon
fields and the possible impact of different choices of actions in a holographic model of this kind.

We take the brane-antibrane pair to be at a fixed value of $\eta$ and we will not
consider oscillations of the transverse scalar, which has no
QCD counterpart\footnote{A different construction involving D4-anti D4 in this
background was considered in \cite{Casero:2005se}. The present scenario is more successful
in describing different features of QCD.}. The brane and antibrane are at zero distance and are therefore overlapping.
We have therefore a 5D model for the quarks embedded
in a 6D model for the glue. The Sen action reads:
\be
S= - \int d^4x dz  V(|T|)
\left(\sqrt{-\det {\bf A}_L}+\sqrt{-\det {\bf A}_R}\right)
\label{generalact}
\ee
 The quantities inside the
square roots are defined as:
\be
{\bf A}_{(i)MN}=g_{MN} + {2\pi \alpha' \over g_{V}^{2}} F^{(i)}_{MN}
+ \pi \alpha' \lambda \left((D_M T)^* (D_N T)+
(D_N T)^* (D_M T)\right)
\label{Senaction}
\ee
where $(i)=L,R$ and
 the complex tachyon will be denoted $T=\tau e^{i \theta}$.
Indices $M,N$ run over the 5 world-volume dimensions while we will use
$\mu,\nu$ for the Minkowski directions (indices to be contracted
using $\eta_{\m\n}$).
With respect to \cite{Sen:2003tm},
we have included two constants $g_V$, $\lambda$ in (\ref{Senaction}), which are related
to the normalization of the fields to be discussed later.

The covariant derivative
of the tachyon field is defined as:
\be
D_M T = (\partial_M + i A_M^L- i A_M^R)T
\ee

For the tachyon potential we take:
\be
V= {\cal K}\, e^{-\frac12 \m^2 \tau^2}
\label{tachyonpot}
\ee
where ${\cal K}$ is a constant\footnote{We have included the constant dilaton
in ${\cal K}$, in order to avoid unnecessary cluttering of formulae.}
 which in principle should be related to the tension of the
D4-branes. The gaussian is a simple choice that has been discussed in different situations
for instance in \cite{Kutasov:2000aq,Takayanagi:2000rz,Garousi:2007fn},
but we warn the reader that it is not at all top-down
derived for the present situation and thus should be considered as an ingredient of the bottom-up
approach. We will comment further in section \ref{othertachyons}.
For book-keeping, let us enumerate here the constants that have been introduced up to now:
$R,\alpha',\lambda, g_V, {\cal K},\mu , z_\Lambda$. In the following, we will impose, on physical
grounds, some relations among these constants and in appendix \ref{app:bookkeeing} we will summarize
these arguments.

We must first  find the vacuum of the theory.
We should set $\theta, A_L, A_R$ to zero because of Lorentz invariance, but $\tau$
must have non-trivial dynamics, at least in the confined phase, as will
be argued below.
We thus discuss here the function $\tau(z)$ that defines the vacuum.
The corresponding reduced action reads:
\be
S=-2 {\cal K}\int d^4x dz e^{-\frac12 \m^2 \tau^2}
g_{tt}^\frac12 g_{xx}^\frac32 \sqrt{g_{zz} + 2\pi \alpha' \lambda (\partial_z \tau)^2}
\label{actitau}
\ee
and the corresponding equation of motion:
\be
\tau'' +\frac{\pi \alpha' \lambda}{g_{zz}} \tau'^3\left(\frac{g_{tt}'}{g_{tt}}
+3\frac{g_{xx}'}{g_{xx}}\right)
+\frac{\tau'}{2} \left(\frac{g_{tt}'}{g_{tt}}
+3\frac{g_{xx}'}{g_{xx}}-\frac{g_{zz}'}{g_{zz}}\right)
+ \left(\frac{g_{zz}}{2\pi \alpha' \lambda} + \tau'^2 \right) \mu^2 \tau =0
\label{taueqgen}
\ee

We want to study this equation in both the confined and deconfined backgrounds of section
\ref{sec: backgr}. For this, we need to explicitly substitute the components of the metric
of each background, as given in section \ref{sec: backgr}.
We will make these studies separately in the following subsections.
Before that, since the UV of both solutions is identical (up to ${\cal O}(z^5)$), the
analysis of the UV asymptotics of (\ref{taueqgen}) is the same. We find that the near-boundary limit
$z\to 0$ limit is given in terms of the two integration constants as:
\be
\tau = c_1 z + \frac{\mu^2}{6}c_1^3 z^3 \log z + c_3 z^3 + {\cal O}(z^5)
\label{tauUVexpan}
\ee
In order to find this expansion, we have imposed that:
\be
\frac{R^2 \mu^2}{2\pi \alpha'\lambda} = 3 \,\,.
\label{dimcon}
\ee
This enforces that the scalar bifundamental operator dual to the scalar field  (which has mas $m_\tau^2= -\mu^2/(2\pi \alpha'\lambda)$)
has UV dimension  3 matching the dimension of $\bar q q$ in QCD. This is in agreement with the usual AdS/CFT rule
$\Delta (\Delta - 4) = m_\tau^2 R^2$.
It is worth stressing that (\ref{dimcon}) should be understood as a
bottom-up condition on the parameters
determining the open string data $\mu,\alpha',\lambda$ and not on $R$, since in the quenched
approximation one should not think of the flavor branes affecting the closed string background.

The asymptotic expansions for $\tau$ in the confined and deconfined backgrounds start
differing at order ${\cal O}(z^6)$. On the other hand, the IR behaviour
for both cases is very different, as will be discussed below.

\subsection{The confined phase}
\label{sec:confvev}

Inserting the metric for the confining background (\ref{adsbh6}) into
(\ref{taueqgen}), we obtain the following equation of motion for the order parameter:
\be
\tau'' - \frac{4 \mu^2 z f_\Lambda}{3} \tau'^3
+ ( - \frac{3}{z}  + \frac{f_\Lambda'}{2f_\Lambda})\tau'
+ \left(\frac{3}{  z^2 f_\Lambda} +\mu^2 \tau'^2 \right) \tau =0
\label{taueq}
\ee
Before going on, notice that equation (\ref{taueq}) depends on two constants $z_\Lambda$ and
$\mu$. However, such dependence can be easily reabsorbed by redefining the field and radial
coordinate as $z\to \tilde z = z/z_\Lambda$, $\tau \to \tilde \tau = \mu \tau$.  The plots in this section will
be performed by taking $z_\Lambda=1$, $\mu^2 = \pi$, but it is automatic to find
the solution for different values of the constants by rescaling as mentioned above.

According to the discussion of \cite{ckp}, since the background is confining,
we must require the tachyon to blow up somewhere.
Heuristically, one can think of the diverging tachyon as a brane-antibrane recombination;
if the tachyon were finite until the bottom of the space one would have an open brane (and
antibrane). In \cite{ckp}, it was argued that this would lead to  bulk flavor
anomalies that do not match those of QCD\footnote{Anomalies in the hard wall model have been
discussed in \cite{hwanomalies}. In that case, appropriate IR boundary conditions have to
be imposed on the gauge
fields in order to get rid of the IR contribution to the gauge variation of the
Chern-Simons term. In our case, that contribution is killed due to the divergent tachyon.}.
The fact that confinement requires brane recombination (and therefore, chiral
symmetry breaking) is a Coleman-Witten-like theorem \cite{Coleman:1980mx}
  for the present set-up, and it is analogous
to a similar discussion of
\cite{Aharony:2006da} for the Sakai-Sugimoto model \cite{Sakai:2004cn}. The difference is that the realization of
chiral symmetry breaking in \cite{Aharony:2006da} is geometrical while here it is driven by the
field $\tau$.

Equation (\ref{taueq}) only allows the tachyon to diverge at exactly the end of space (the tip of the cigar)
$z=z_\Lambda$, see appendix \ref{app:singul} for details.

In the IR, generically the two linearly independent solutions behave as a constant and $\sqrt{z-z_{\Lambda}}$ and they are regular at the tip.
There is however a one parameter ``boundary" family of solutions that (1) depends on a single parameter (2) diverges at the tip.
This is the solution we should allow in the IR.
If we  call the single parameter  $C$ then the acceptable IR solution is:
\be
\tau=\sum_{n=0}^{\infty}(z_\Lambda-z)^{3(2n-1)\over 20}C_n~g_n(z)
\label{tauIRexpan}\ee
where
\be
g_{n}(z)=1+\sum_{m=1}^{\infty}D_{n,m}\left(1-\frac{z}{z_\Lambda}\right)^m
\ee
For the first few constants we have
\be
C_0=C\sp C_1=-{13\over 6\mu^2 C}\sp C_2={247\over 72\mu^4 C^3}\sp C_3=-{26975\over 1296\mu^6 C^5}
\ee
$$
C_4={6381505\over 31104\mu^8 C^7}\sp C_5=-{276207997\over 103680\mu^{10}C^9}\sp
C_6={1402840243831\over 33592320\mu^{12} C^{11}}
$$
and the first few
functions
\be
g_0(z)=1-{9\over 20}\left(1-\frac{z}{z_\Lambda}\right)+{\cal O}\left(\left(1-\frac{z}{z_\Lambda}\right)^2\right)
\ee
\be
g_1(z)=1 - {1479\over 3380}\left((1-\frac{z}{z_\Lambda}\right)+{\cal O}\left(\left(1-\frac{z}{z_\Lambda}\right)^2\right)
\ee
\be
g_2(z)=1-{8481\over 4940}\left((1-\frac{z}{z_\Lambda}\right)+{\cal O}\left(\left(1-\frac{z}{z_\Lambda}\right)^2\right)
\ee
\be
g_3(z)=1-{396189\over 82004}\left((1-\frac{z}{z_\Lambda}\right)+{\cal O}\left(\left(1-\frac{z}{z_\Lambda}\right)^2\right)
\ee
As $C$ increases, the radius of convergence of this series increases.

The condition that the solution should end up in the one parameter family described above is our ```regularity condition".
It relates the two UV initial conditions, the source (mass) $c_1$ and the vev (chiral condensate) $c_3$.
This is a dynamical determination of the condensate as a function of the mass by the condition $\tau (z=z_\Lambda) = \infty$.
This relation will be found numerically.

\begin{figure}[ht]
\centering
	\includegraphics[width=.4\textwidth]{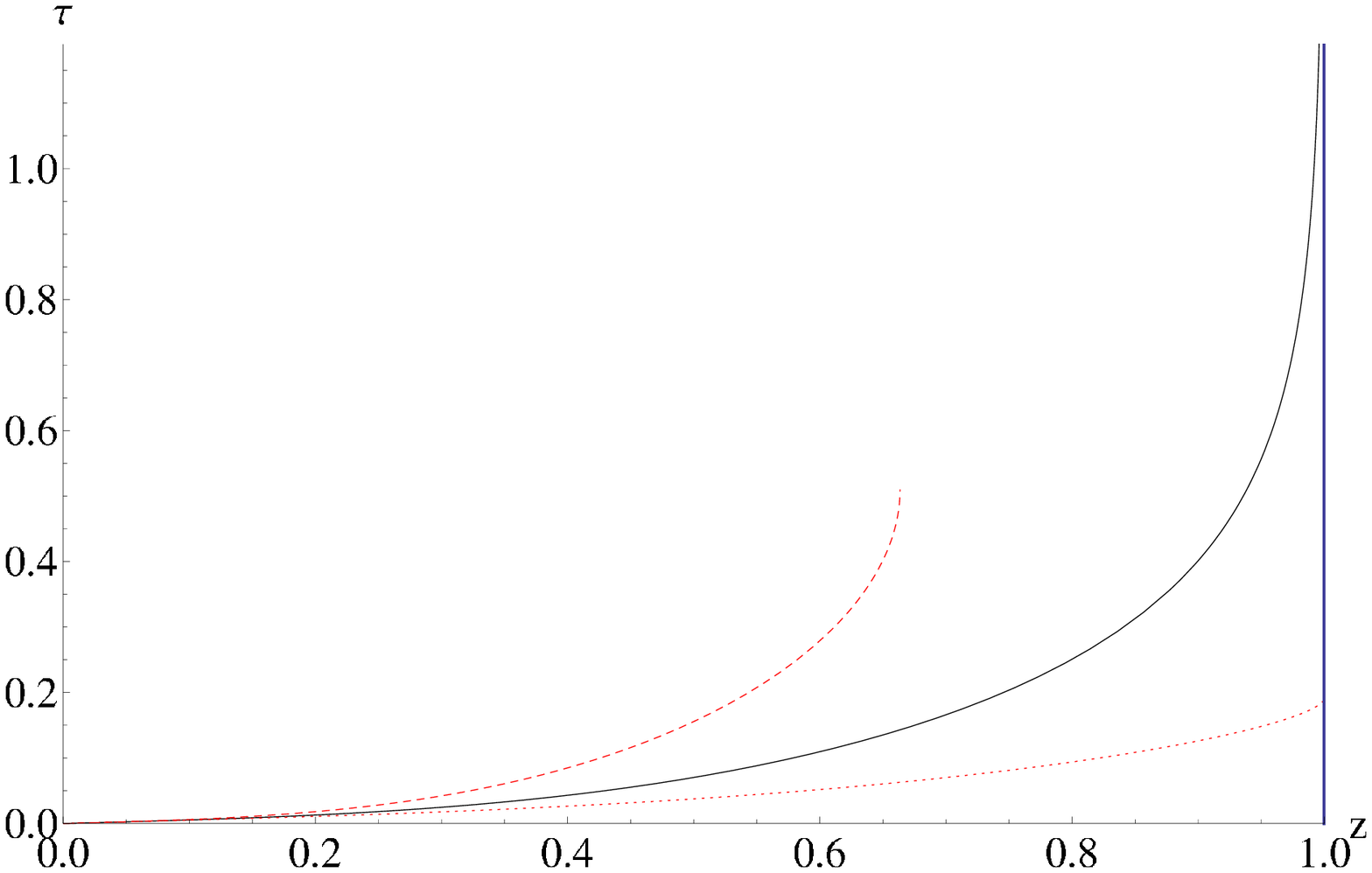}
	\includegraphics[width=.4\textwidth]{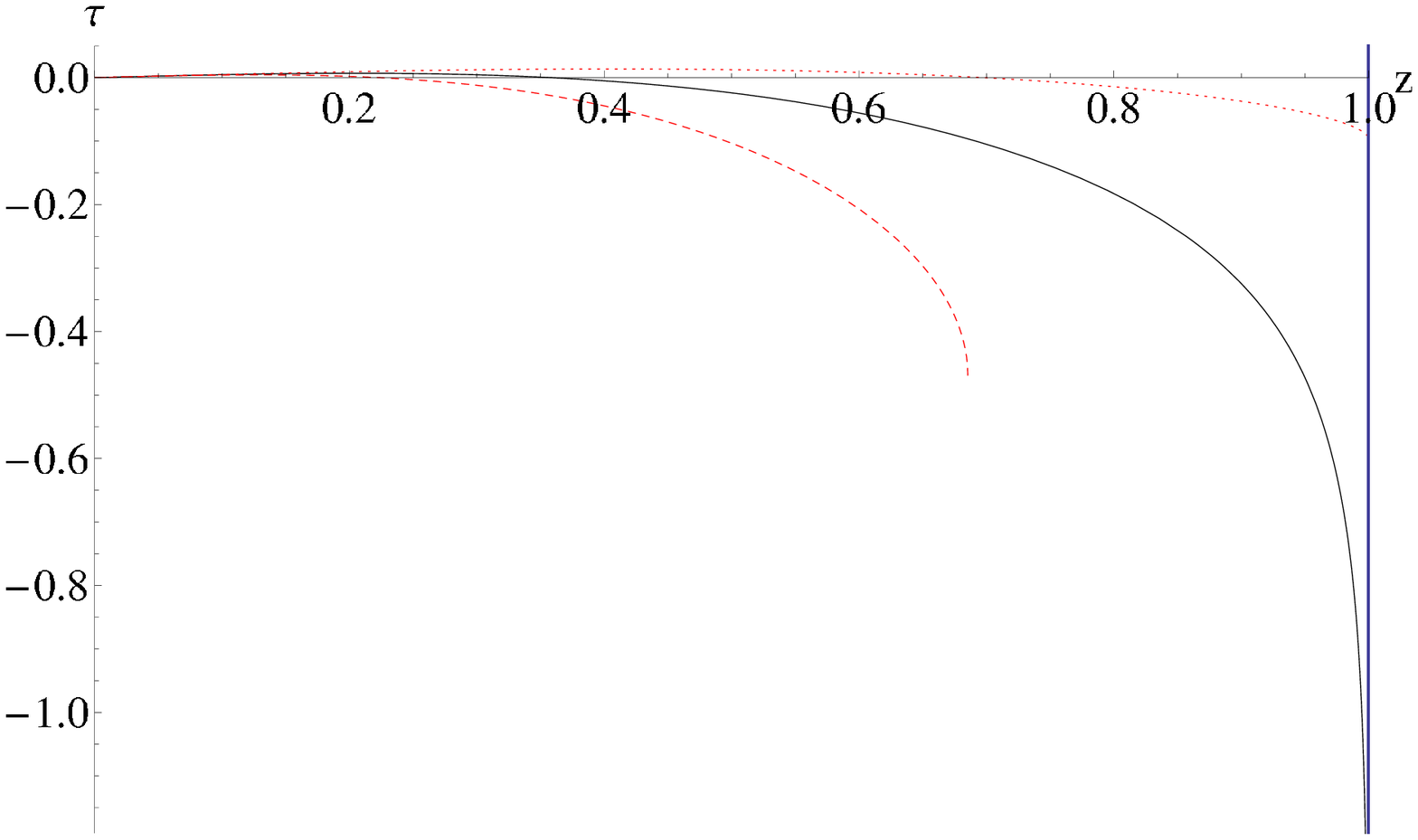}
	\caption{All the graphs are plotted using $z_\Lambda = 1$, $\mu^2 = \pi$ and
	$c_1=0.05$. The vertical line at $z=z_\Lambda=1$ represents the IR end of space (tip of the cigar).
	On the left, the solid black line represents a solution with
	$c_3\approx 0.3579$ for which $\tau$ diverges
	at $z_\Lambda$. The red dashed line has a too large $c_3$
	- in particular, it corresponds to $c_3=1$ -
	such that there is a singularity at
	 $z = z_s $ where $\partial_z \tau$ diverges while $\tau$ stays finite (a behaviour of the
	type $\tau = k_1 - k_2 \sqrt{z_s - z}$, that is unacceptable since the solution stops
	at $z=z_s$ where the energy density of the flavor branes diverges). The red dotted line corresponds to $c_3=0.1$; this
	kind of solution
	ought to be discarded because the tachyon stays finite everywhere. The plot in the right
	is done with the same conventions but with negative values of $c_3=-0.1,-0.3893,-1$.
	For $c_3 \approx -0.3893$ there is a solution of the differential equation such that
	$\tau$ diverges to $-\infty$. As explained in the text, this solution is unstable.
Thus,	the physical solution for this particular value of $c_1$ is uniquely determined to be
	the solid line of the graph on the left.  }
	\label{div_tac_fig}
\end{figure}

 In practice, one has to solve numerically the equation of motion (\ref{taueq})  arranging
  the asymptotics to be (\ref{tauUVexpan})
  in the UV and (\ref{tauIRexpan}) in the IR.
  One can implement a standard shooting routine
  whose inputs are $c_{1}$ and some UV and IR cutoffs, where the numerical solution is required to match
  the mentioned asymptotics. The value of $c_3$ leading to (\ref{tauIRexpan}) is the limiting point
  between a behavior of diverging derivative of $\tau$ and a behaviour where $\tau$ remains finite
  everywhere, see figure \ref{div_tac_fig}.

In fact, for fixed $c_1$ there are two values of $c_3$ for which $\tau$ diverges at $z_\Lambda$,
 since $\tau$ can diverge
to $+\infty$, for a particular $c_3>0$; or to $-\infty$, for a particular $c_3<0$, (we are
assuming by convention that $c_1 > 0$).
However, the $c_3<0$ solution is unstable and should be
discarded. This can be understood by comparing
the free energy of both solutions or, alternatively, by realizing that there is a tachyonic mode
in the pseudoscalar sector.
In the massless quark  case $c_1=0$, both solutions are related by $\tau \to -\tau$ and are physically
equivalent. They are just related by a rotation in the direction of the Goldstone pion, which
is exactly massless.
This behaviour is completely analogous to the one described in \cite{Kruczenski:2003uq}.
For illustrative purposes, we plot in figure \ref{div_tac_fig} the result of numerically integrating
(\ref{taueq}) and the behaviour of $\tau(z)$ for different values of $c_3$.

\begin{figure}[ht]
\centering
	\includegraphics[width=.3\textwidth]{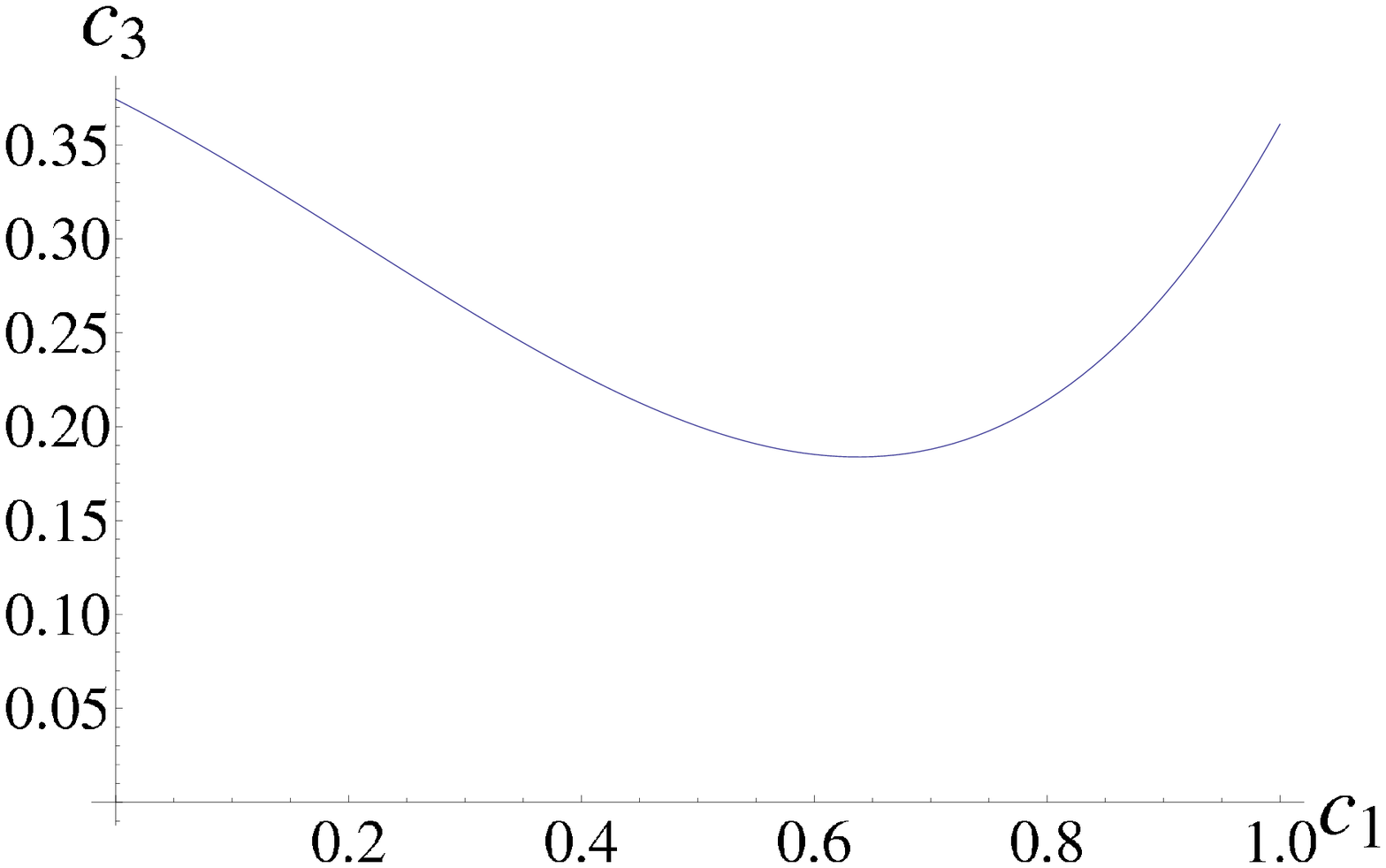} \quad
	\includegraphics[width=.3\textwidth]{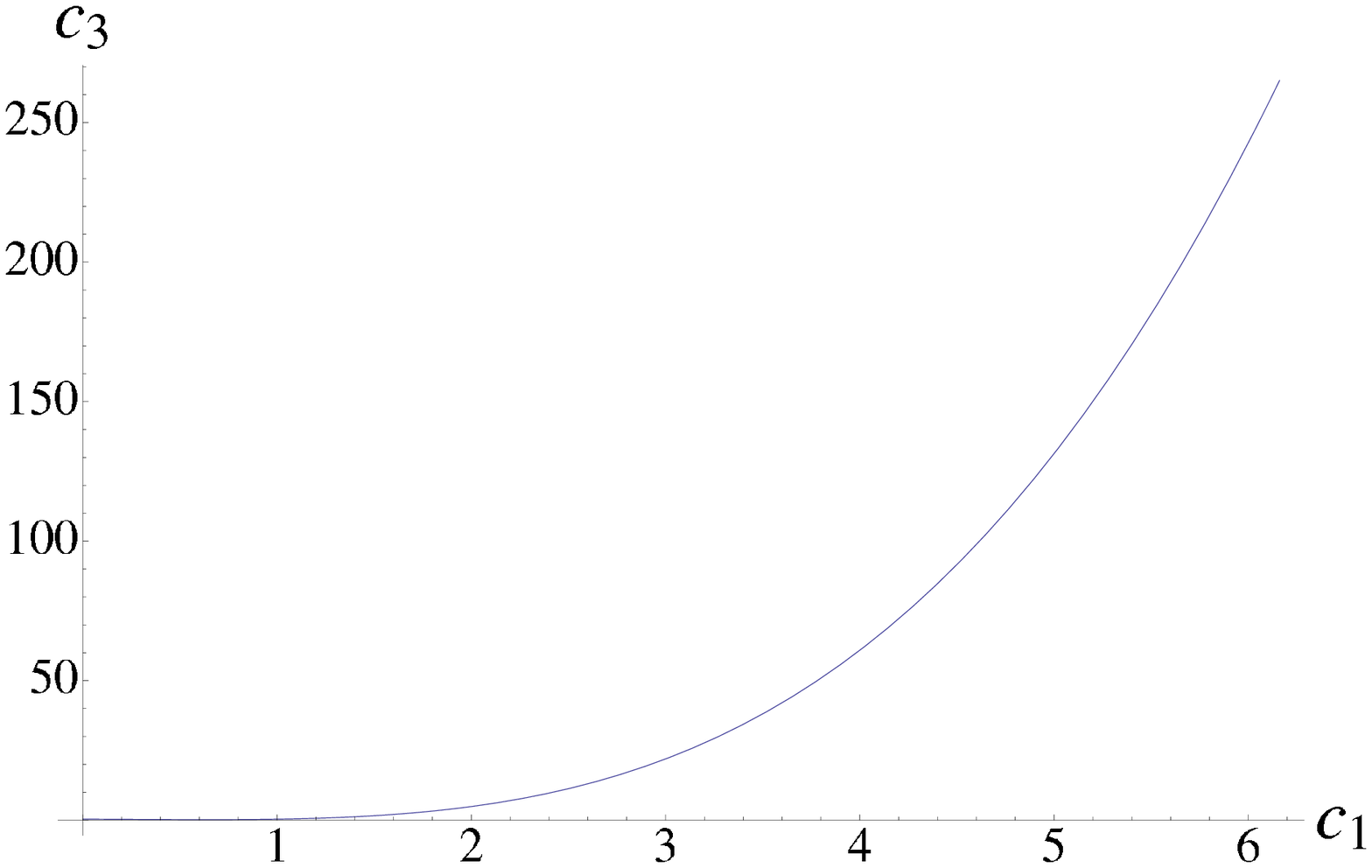}\quad
	\includegraphics[width=.3\textwidth]{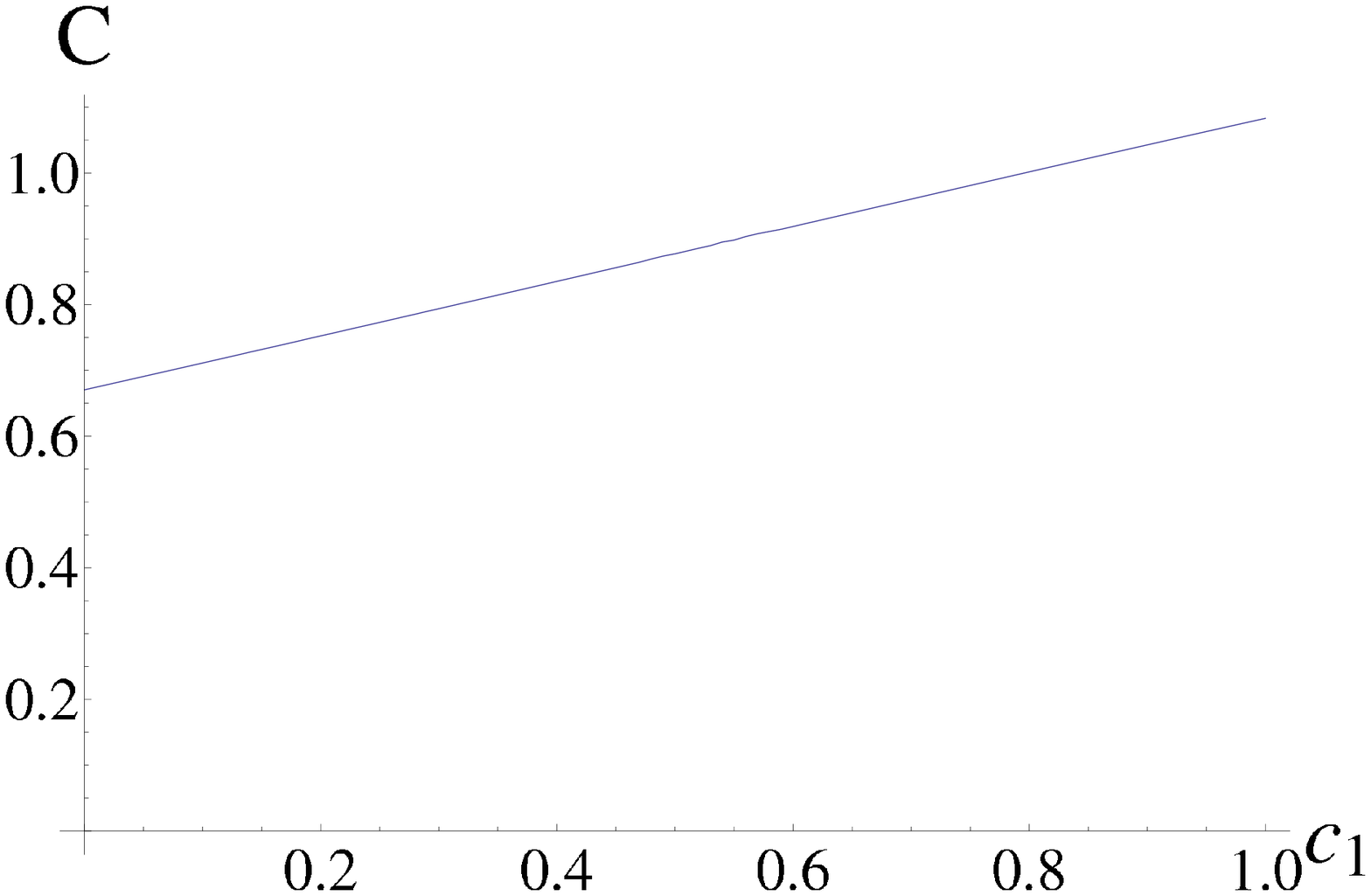}
	\caption{
	The values of $c_3$ and $C$ determined numerically as a
        function of $c_1$. In the first plot we portray $c_3$ in terms
        of $c_1$, for $c_1 \leq 1$. In the second plot we show
        $c_3$ for a larger range of $c_1$.
The third plot depicts the constant $C$ entering the IR expansion
as a function of $c_1$.
	Again, we have used $z_\Lambda = 1$, $\mu^2 = \pi$ for the plots.}
	\label{div_tac_fig_2}
\end{figure}

In figure \ref{div_tac_fig_2}, we plot the
values of $c_3$ and $C$ obtained dynamically, as a function of $c_1$.

\subsection{The deconfined phase}
\label{secdeconf}

Inserting the metric (\ref{adsbh6b}) in
(\ref{taueqgen}), we obtain the following equation for $\tau$ in the deconfined phase:
\be
\tau''  +  \frac{\mu^2 z^2 f_T}{3} \tau'^3\left(-\frac{4}{z} +
\frac{f_T'}{2 f_T} \right)
+ ( - \frac{3}{z}  + \frac{f_T'}{f_T})\tau'
+ \left(\frac{3}{  z^2 f_T} +\mu^2 \tau'^2 \right) \tau =0
\label{taueqbh}
\ee
The IR behaviour of this equation is quite different from the one of (\ref{taueq}).
First of all $\tau$ is not allowed to diverge at any point.
The difference with respect to the confining case is that since there is a horizon,
one can allow the
branes not to recombine as long as they end
on the horizon. Then, they will not generate any anomaly.

\begin{figure}[ht]
\centering
\includegraphics[width=.3\textwidth]{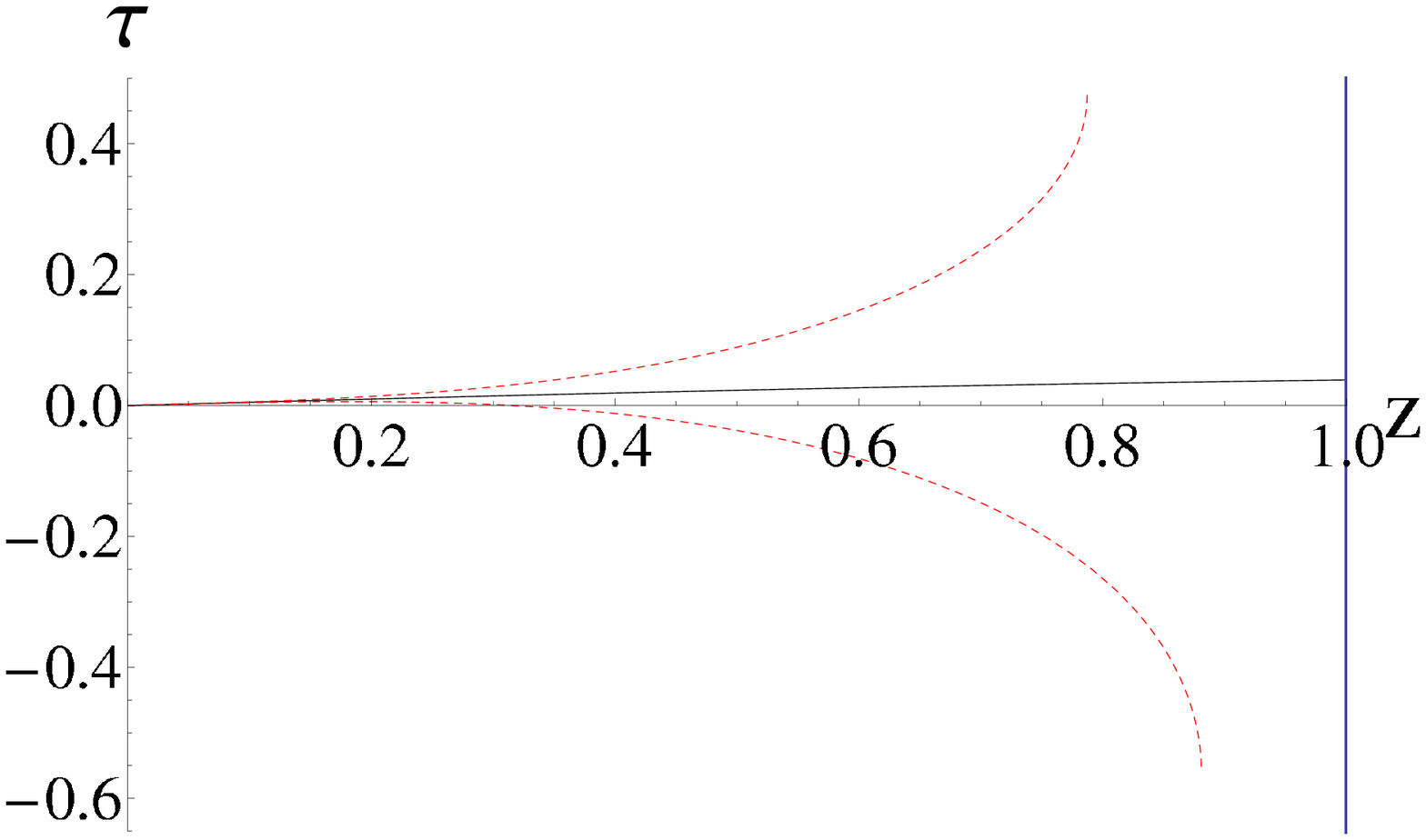}\quad
	\includegraphics[width=.3\textwidth]{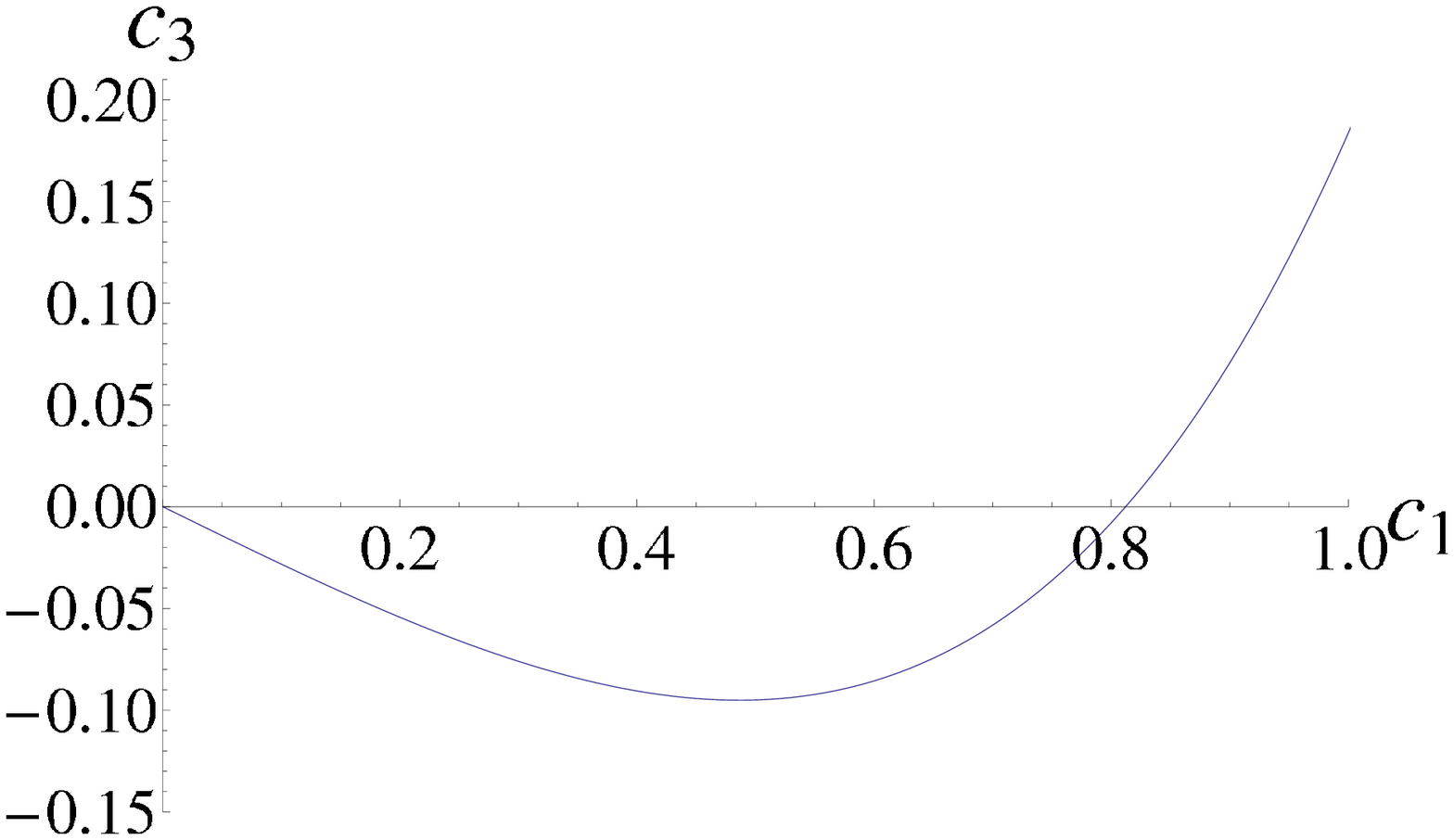}\quad
	\includegraphics[width=.3\textwidth]{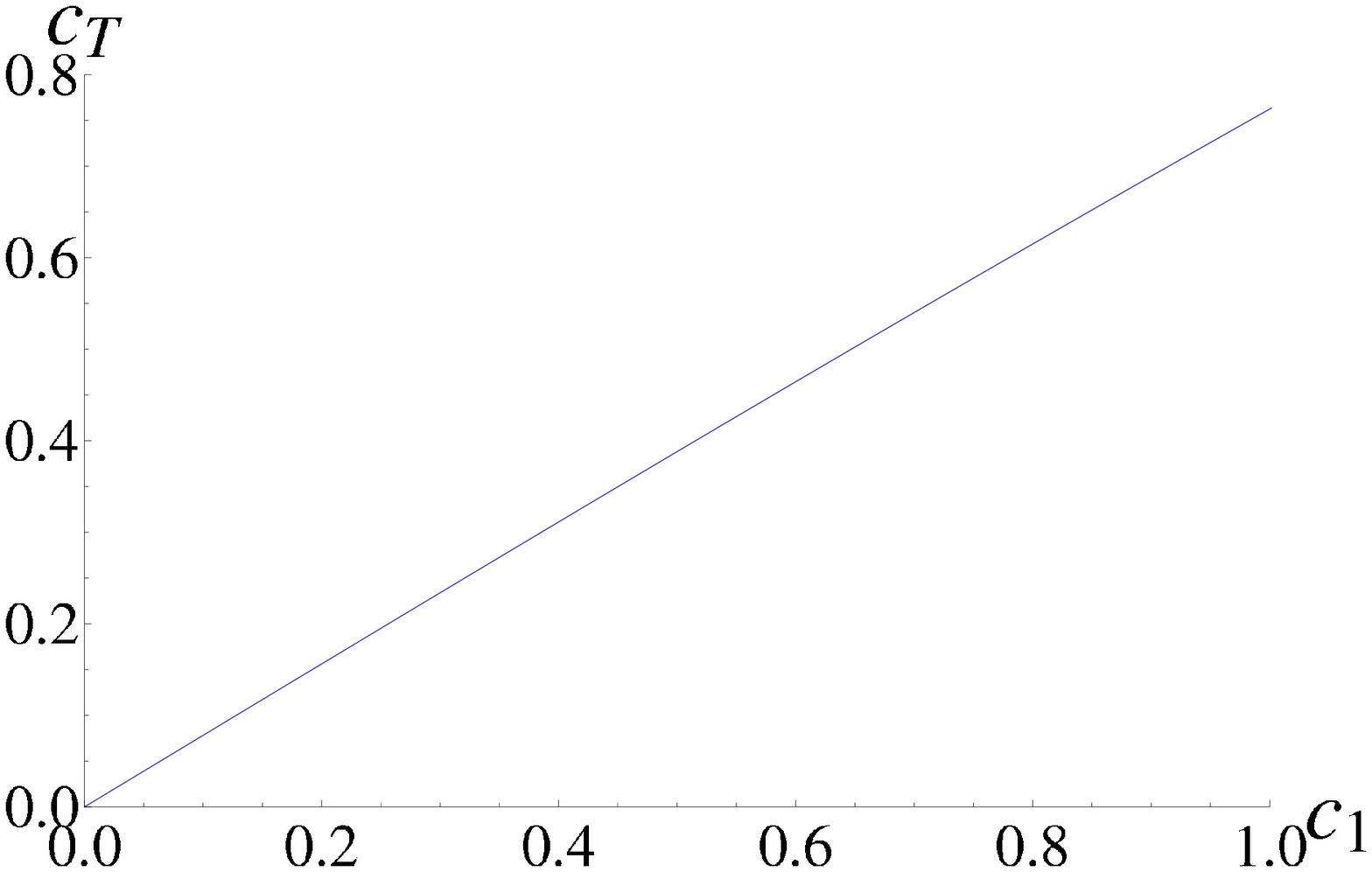}
	\caption{Plots corresponding to the
	deconfined phase. All the graphs are plotted using $z_T = 1$, $\mu^2 = \pi$.
For the first plot we have taken $c_1=0.05$. The solid line displays the physical solution
$c_3=-0.0143$ whereas the dashed lines ($c_3=-0.5$ and $c_3=0.5$) are unphysical and end
with a behaviour of the type $\tau = k_1 - k_2 \sqrt{z_s - z}$. The second and third
plots give the values of $c_3$ and $c_T$ determined numerically by demanding the correct IR
behaviour of the solution, as a function of $c_1$.}
\label{fig:taudeconf}
\end{figure}

Still, one has to discard solutions for which $\tau'$ diverges at some $z<z_T$
(with $\tau$ remaining finite). Those solutions yield infinite energy density and
are physically inconsistent, just as in the confining case.
It turns out that this condition uniquely selects a value
for $c_3$, for which $\tau$  reaches the horizon at $z=z_T$ taking there a finite
value, say $\tau|_{z=z_T}=c_T$, as shown in the first plot of figure \ref{fig:taudeconf}.

Thus, we have a
one parameter family of physical solutions (which again fix $c_3$ in terms of $c_1$ as
expected on physical grounds). In a similar fashion to the confined case,
by redefining the field and radial
coordinate as $z\to \tilde z = z/z_T$, $\tau \to \tilde \tau = \mu \tau$, the dependence
of the equation on these two parameters can be reabsorbed.
Near the IR, these solutions read, in terms of
the parameter $\tau(z_T) \equiv c_T$:
\be
\tau = c_T - \frac{3c_T}{5z_T} (z_T -z) - \frac{9c_T}{200z_T}(8 + \mu^2c_T^2) (z_T -z)^2
+\dots
\label{tauIRexpanbh}
\ee

Once $c_1$ is fixed, $c_3$ and $c_T$ are dynamically determined by this IR condition,
and their values can be found
numerically; using a standard shooting technique.
Notice that for $c_1=0$, the solution is simply $\tau=0$ and chiral symmetry is unbroken.
We display some plots with numerical results in figure \ref{fig:taudeconf}.

\subsection{Holographic renormalization, the quark mass and the quark condensate}

On general AdS/CFT grounds, we expect the integration constants $c_1$, $c_3$ of the
UV expansion (\ref{tauUVexpan}) to be related to the source and vacuum expectation value
of the boundary operator associated to the bulk field $\tau$, which is the scalar quark
bilinear $\bar q q$. Namely $c_1$ should be, essentially, the quark mass and $c_3$ the
quark condensate. In this section, we will make this connection precise.

As has been pointed out many times -see for instance \cite{DaRold:2005vr}, \cite{Cherman:2008eh}-,
in QCD the quark mass runs all the way to zero in the far UV, a fact that cannot be matched
in a holographic model with AdS asymptotics (such that the $m_q$ we will define is the UV value,
which does not run further).
 If we want to make a phenomenological analysis, the most natural
option is to identify the $m_q$ of the model with the QCD quark mass measured at a scale around 1 or
2 GeV. It is conceivable that this feature can be ameliorated
 by using the tachyon action in a holographic setup
which incorporates asymptotic freedom, as in  ``Improved holographic QCD"
\cite{ihqcd}.

The quark condensate is defined as:
\be
\langle \bar q\,q\rangle = -\frac{\delta S_{ren}}{\delta m_q}
\label{condensatedef}
\ee
where in order to find $S_{ren}$ we have to follow the procedure of holographic renormalization, see
for instance \cite{Skenderis:2002wp}. The first step is to regularize the action by placing a
UV cut-off at $z=\epsilon$, namely $S_{reg}=\int_\epsilon^{z_\Lambda} {\cal L}$, where we have
defined, from (\ref{actitau}):
\be
{\cal L}=-2 {\cal K} \, e^{-\frac12 \m^2 \tau^2}
g_{tt}^\frac12 g_{xx}^\frac32 \sqrt{g_{zz} + 2\pi \alpha' \lambda (\partial_z \tau)^2}
\label{actitau2}
\ee
Since we are just concerned with the variation of $S_{reg}$ with respect to $m_q$, we
 compute the functional derivative with respect to $\tau$:
\be
\delta S_{reg} = \int_\epsilon^{z_\Lambda} \left( \delta \tau \frac{\partial {\cal L}}{\partial
\tau} +\delta \tau' \frac{\partial {\cal L}}{\partial
\tau'}\right)dz = \int_\epsilon^{z_\Lambda} \frac{d}{dz}\left(\delta \tau \frac{\partial {\cal L}}
{\partial \tau'}\right)
\ee
and therefore
\be
\frac{\delta {S_{reg}}}{\delta \tau} = -\frac{\partial {\cal L}}
{\partial \tau'}{\Big|}_{z=\epsilon}
\ee
We are interested in  $\frac{\delta {S_{reg}}}{\delta c_1}=
\frac{\delta \tau}{\delta c_1} \frac{\delta {S_{reg}}}{\delta \tau}$. In order
to compute $\frac{\delta \tau}{\delta c_1}$, one should take into account
 that $c_3$ is a non-trivial function of $c_1$.
We find by explicit computation, using the UV expansion (\ref{tauUVexpan}):
\be
\frac{\delta {S_{reg}}}{\delta c_1} =
{\cal K} R^5 \mu^2 \left(  \frac{2c_1}{3\epsilon^2}
+ \frac23 c_1^3 \mu^2 \log \epsilon +2c_3 - \frac13 c_1^3 \mu^2 + \frac23 c_1 \partial_{c_1}c_3
 \right)
\label{sregoverc1}
\ee
where we have disregarded terms that vanish as $\epsilon \to 0$.
We now have to write the appropriate covariant counterterms that should be added to $S_{reg}$
in order to define the subtracted action $S_{sub}=S_{reg} + S_{ct}$:
\be
S_{ct}= - {\cal K}R\int d^4x \sqrt{-\gamma} \left(- \frac12 + \frac{\mu^2}{3} \tau^2
+\frac{\mu^4}{18} \tau^4 \log\epsilon + \frac{\mu^4}{12} \alpha\, \tau^4  \right)
\label{Sct}
\ee
where $\gamma$ corresponds to the induced metric at $z=\epsilon$, namely
$\sqrt{-\gamma}=R^4 \epsilon^{-4}$. We have introduced a constant $\alpha$
which captures the scheme dependence of the condensate and reflects an analogous scheme dependence in field theory.
It will be further discussed in appendix \ref{app: alpha}.
The renormalized action is just $S_{ren}=\lim_{\epsilon\to 0} S_{sub}$.
It is now straightforward to find:
\be
\frac{\delta S_{ren}}{{\delta c_1}} = -(2\pi \alpha' {\cal K } R^3 \lambda)  \left(
- 4 c_3 +  c_1^3 \mu^2 (1+\alpha) \right)
\ee
Notice that the term with $c_1 \partial_{c_1}c_3$ in (\ref{sregoverc1}) drops out because there
is one with the opposite sign in $\frac{\delta S_{ct}}{{\delta c_1}}$ that cancels it.
We now want to evaluate the quark condensate (\ref{condensatedef}).
The quark mass is proportional to $c_1$, and  we  take it to be
\be
 m_q = \beta \,c_1
 \label{mqc1}
 \ee
 where $\beta$ is a constant.

 The arbitrariness of this multiplicative constant
related to the normalization of the fields has been stressed (in analogous situations) in
\cite{Cherman:2008eh}, \cite{Jugeau:2009mn}.
We finally obtain
\be
\langle \bar q q \rangle = \frac{1}{\beta} (2\pi \alpha' {\cal K } R^3 \lambda)  \left(
- 4 c_3 +  \left(\frac{m_q}{\beta}\right)^3 \mu^2 (1+\alpha) \right)
\label{quarkcon}
\ee

\subsection{The jump of the condensate at the phase transition}

\begin{figure}[ht]
\centering
	{\includegraphics[width=.4\textwidth]{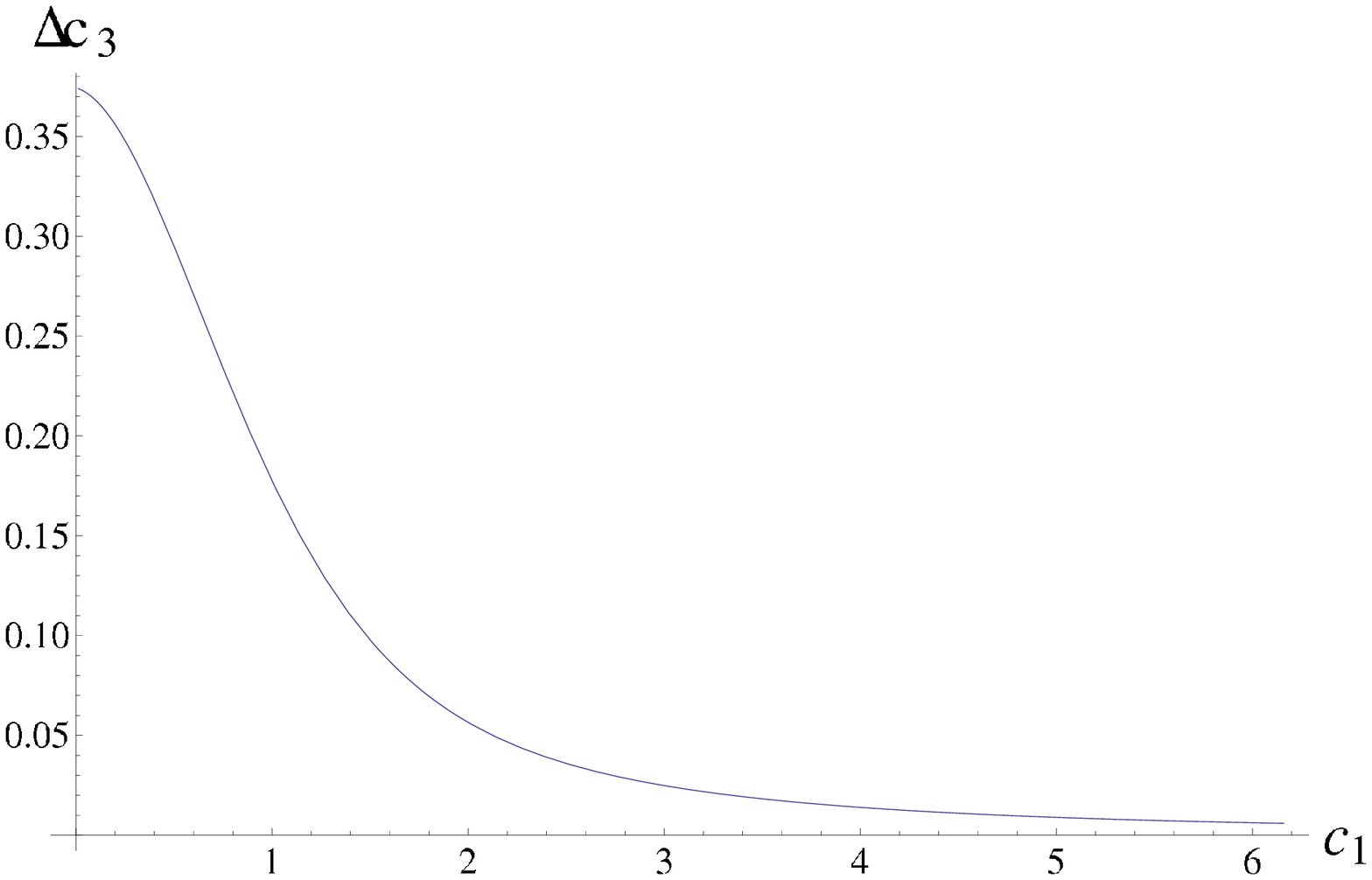},
         \includegraphics[width=.4\textwidth]{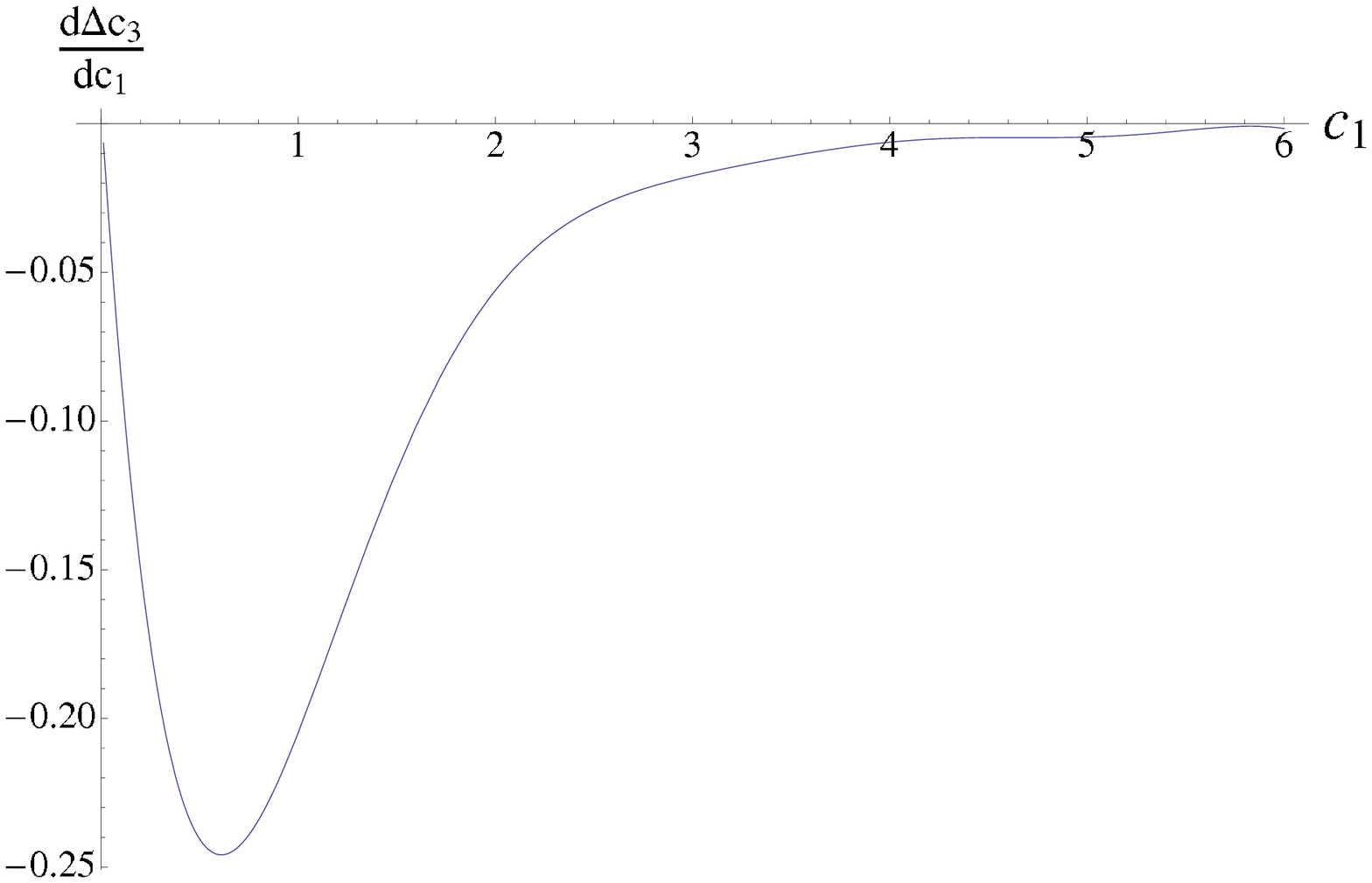}}
	\caption{The finite jump of the quark condensate and its
          derivative with respect to $c_1$ when the confinement-deconfinement
	transition takes place. The values $z_\Lambda=z_T=1$ and $\mu^2=\pi$ have been used in the plot.}
	\label{fig:jump}
\end{figure}

The first term of the expression for the quark condensate (\ref{quarkcon}) depends on the
quantity $c_3$ that is
determined dynamically via the numerical integration. The second term depends on the quark mass and
a scheme dependent constant $\alpha$. We now compute an observable which is independent of this second
term by finding the jump of the quark condensate when the theory is heated such that it undergoes the
deconfinement phase transition. Concretely, we take a fixed mass (fixed $c_1$) and compare $c_3$ for
a confined theory and deconfined theories, such that $z_\Lambda=z_T$, namely at the phase transition
point. We have that $\Delta \langle \bar q q \rangle \equiv \langle \bar q q \rangle_{conf} -
\langle \bar q q \rangle_{deconf} = -4  \frac{1}{\beta} (2\pi \alpha' {\cal K } R^3 \lambda)\Delta c_3$.
In figure \ref{fig:jump} we plot $\Delta c_3$, which in practice is nothing else that the difference
between the first plot in figure \ref{div_tac_fig_2} and the second plot of figure \ref{fig:taudeconf}.
It turns out to be a monotonously decreasing function, at least in the range of $c_1$ which we have
been able to study numerically. We plot the result in figure \ref{fig:jump}.

\begin{figure}[ht]
\centering
	{\includegraphics[width=.4\textwidth]{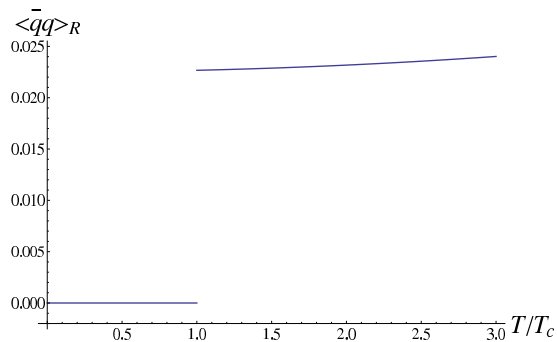}}
	\caption{Behaviour of $\langle \bar q q \rangle_R$ as a function of the temperature.
We have taken $N_c=3$, $\beta=1$, $m_q/T_c = 1/40$ for the plot.}
	\label{fig:qqR}
\end{figure}

Let us now discuss how the quark condensate changes when tuning the temperature, while keeping
fixed the quark mass and the QCD scale. We will plot the quantity:
\be
\langle \bar q q \rangle_R =  \frac{m_q }{T_c^4}(\langle \bar q q \rangle_T
-\langle \bar q q \rangle_0)\,,
\ee
where $\langle \bar q q \rangle_T$ is the condensate evaluated at temperature
$T$. We have included the power of $T_c$ in the denominator in order to make
the quotient dimensionless. Let us start by computing the explicit value of
$\langle \bar q q \rangle_0$ from (\ref{quarkcon}). We will consider small quark masses
(compared to the QCD scale or $M_{KK}$) so we can neglect the last term of (\ref{quarkcon})
and use the value $c_3 \approx 0.37 z_\Lambda^{-3}$ computed for small $c_1$ in the first
plot\footnote{We remind the reader that the plots were
done by fixing $z_\Lambda=1$ and $z_T=1$ respectively. The values for generic
$z_{\Lambda}$, $z_T$ are obtained just by rescaling
$c_1 = (c_1|_{z_\Lambda=1}) z_\Lambda^{-1}$ and $c_3 = (c_3|_{z_\Lambda=1}) z_\Lambda^{-3}$,
and similarly in the deconfined phase, substituting $z_\Lambda$ by $z_T$.} of figure \ref{div_tac_fig_2}.
Inserting the value of $z_\Lambda$ in terms of $T_c$ (\ref{tdeconf}) and advancing the value of the
normalization constant that will be found in (\ref{lamqcd}), we have
$\langle \bar qq\rangle_0 \approx - 0.3 N_c \beta\,T_c^3$.
Turning on the temperature but staying below the
phase transition, the functions of the metric do not depend on the temperature
and therefore $\langle \bar q q \rangle_R=0$ for $T<T_c$. This is just a consequence of
large-$N$ volume independence. In order to compute the result in the deconfined phase,
we would  like to use the values of $c_3$ as a function of $c_1$ plotted in
figure \ref{fig:taudeconf}. From the figure,
one can fit, for small $c_1 z_T$
the value of $c_3$ to be
$c_3 z_T^3 \approx  -0.286  c_1 z_T$.
Using this expression, together with (\ref{mqc1}), (\ref{tEperiod}), (\ref{lamqcd}),
we have $\langle \bar q q \rangle_T \approx 0.09 N_c m_q T^2$.
Finally, we reach the result:
\be
\langle \bar q q \rangle_R \approx N_c \frac{m_q }{T_c^4}( 0.3  \beta\,T_c^3
+0.09 m_q T^2
)\,\,,\qquad (T>T_c)
\ee
We illustrate the behaviour of $\langle \bar q q \rangle_R$ in figure  \ref{fig:qqR}.
Notice that, since we are considering light quarks,
the constant term is the largest until $T \gg T_c$.

This is in rough qualitative agreement with lattice results, see figure 4 of
\cite{Borsanyi:2010bp}.
In our case, the jump at the phase transition is sharp
due to the large $N$ limit.

\section{Meson excitations: the confined phase}
\label{sec:mesons}

Up to now, we have discussed the vacuum (saddle point) of the model.
We will now discuss in turn the different excitation modes,
by expanding the action (\ref{generalact})
up to quadratic order in all the fields.
In this section, we will only refer to the confined phase and therefore $\langle \tau \rangle (z)$
is computed as in section \ref{sec:confvev}.

We define the vector and axial vector fields as:
\be
V_M = \frac{A_M^L + A_M^R}{2}\,\,,\qquad
A_M = \frac{A_M^L - A_M^R}{2}
\label{VAdefs}
\ee
The notation for the associated field strengths will be
$V_{MN}$, $A_{MN}$. We use a gauge $A_z = V_z = 0$.
We split the relevant fields as:
\bear
V_\mu (x^\mu, z) &=&  \psi^V(z)\, {\cal V}_\mu (x^\mu)\,\,,\rc
A_\m (x^\mu, z)&=&A^\bot_\m (x^\mu, z)+ A^{\lVert}_\m(x^\mu, z)=  \psi^A(z)\, {\cal A}_\mu (x^\mu)
- \varphi(z)\, \partial_\m({\cal P}(x^\m))\,\,,\rc
\theta (x^\mu, z)&=& 2\,  \vartheta(z)\, {\cal P}(x^\mu)\,\,, \rc
\tau (x^\mu, z) &=&\langle\tau \rangle(z)\, +\,s(x^\mu,z)=
 \langle\tau \rangle(z)\, +\, \psi^S (z) \,{\cal S}(x^\mu)\,\,.
\label{somedefsnorm}
\eear
where ${\cal V}_\mu$ and ${\cal A}_\mu$ are transverse vectors
$\partial^\m {\cal V}_\mu = \partial^\m {\cal A}_\mu = 0$.
A few comments are in order: we have used the residual gauge freedom
to make $V_\m$ transverse. We have anticipated the behaviour of the
equations of motion in order to write down the terms containing
${\cal P} (x)$, associated to the pseudoscalars.
The symbol $\langle\tau \rangle(z)$ represents the tachyon vev
in the bulk, as discussed in section \ref{section3}.

The different bulk fields are dual to the field theory
quark bilinears due to the boundary couplings\footnote{The various discrete symmetries and their realization are detailed in \cite{ckp}.}:
$\int d^4x {\cal V}_\mu {\cal J}_V^\mu$,
$\int d^4x {\cal A}_\mu {\cal J}_A^\mu$,
$\int d^4x {\cal P} {\cal J}_P$,
$\int d^4x {\cal S} {\cal J}_S$, where
the ${\cal J}$'s are the different bilinear quark currents:
${\cal J}_V^\mu = \bar q \gamma^\mu q$,
${\cal J}_A^\mu = \bar q \gamma^\mu \gamma^5 q$,
${\cal J}_P = \bar q \gamma^5 q$,
${\cal J}_S = \bar q  q - \langle \bar q  q \rangle$.

We also define the useful quantity:
\be
\tilde g_{zz} = g_{zz} + 2\pi\alpha' \lambda (\partial_z \langle\tau\rangle)^2
\ee

In the rest of this section, we will discuss the explicit prescriptions to compute the
masses and decay constants for the different mesonic modes.
In particular, the decay constants will require computing two-point correlators
for which one has to holographically renormalize. We give here the complete set
of counterterms which make the on-shell action finite.
\bea
S_{ct}= - {\cal K}R\int d^4x \sqrt{-\gamma} \Bigg(-\frac12+ \frac{\mu^2}{3} \tau^2
+\frac{\mu^4}{18} \tau^4 \log\epsilon + \frac{\mu^4}{12} \alpha \tau^4  +\rc
+{(2\pi\alpha')^2 \over g_{V}^{4}} \frac12
  \gamma^{\mu\rho}\gamma^{\nu \delta}
 (  V_{\mu\nu}V_{\rho\delta}+
 A_{\mu\nu}A_{\rho\delta})
 (\log \epsilon + \frac12)
 + \frac{R^2 \mu^2}{3} \gamma^{\mu\nu} (D_\mu T)^* (D_\nu T)
 (\log \epsilon + \frac12)
\Bigg)
\label{allcounterterms}
\eea
This expression completes (\ref{Sct}) by including all the active fields we are
considering.
The terms of $1/2$ inside the brackets of the second line are finite contact terms that have been chosen
for convenience.

We now discuss in turn each of the modes.

\subsection{Vector mesons}
\label{vecmesoneqs}

The quadratic action corresponding to the vector mesons that comes from expanding
(\ref{generalact}) reads:
\be
S_V = - {(2\pi\alpha')^2 \over g_{V}^{4}}{\cal K}\int d^4x\, dz
e^{-\frac12 \mu^2 \tau^2}
\left[ \frac12 \tilde g_{zz}^\frac12 V_{\m\n}V^{\m\n} + g_{xx}
\tilde g_{zz}^{-\frac12} \partial_z V_\m \partial_z V^\m
\right]\,\,,
\label{vectoracti}
\ee
where we have constrained ourselves to the confining phase in which $g_{tt}=g_{xx}$.
Here and in the following, it should be understood that the $\mu,\nu$ indices are
contracted using the flat Minkowski metric, since we have explicitly written the factor
of $g_{xx}=R^2/z^2$.
The equation of motion can be easily derived:
\be
\frac{1}{e^{-\frac12 \mu^2 \tau^2} \tilde g_{zz}^{\frac12}  }
\partial_z \left( e^{-\frac12 \mu^2 \tau^2}
g_{xx}
\tilde g_{zz}^{-\frac12} \partial_z \psi^V (z)\right)
- q^2 \psi^V (z) = 0
\label{vectoreom}
\ee
where we have gone to Fourier space and defined the 4d-momentum such that
for the eigenmodes it corresponds to the mass eigenvalues $q^2 = -m_V^2$.
The above equation explicitly depends on only two parameters $z_\Lambda$ and $\mu^2$.
It is easy to check that $z_\Lambda$ just gives an overall scale to $m_V^2$
(and, in fact, to all dimensionful quantities that will appear later)
and
$\mu^2$ only enters through the combination $\tilde \tau^2 = \mu^2 \tau^2$.
This was the same combination in the tachyon equation (see the comment below
(\ref{taueq})), and in fact one can fix the value of $\mu^2$ without any loss of
generality. From now on, we will set $\mu^2=\pi$, $z_\Lambda=1$
in all the plots, although we will keep the parameters explicit
in the equations.

Finally, notice that (\ref{vectoreom}) depends implicitly on $c_1$ (the quark mass)
 through the bulk vacuum
expectation value of $\tau$.
In short, the vector spectrum given by the model depends just on a multiplicative constant
$z_\Lambda$ and the parameter $c_1$, namely the quark mass.
All the other parameters that we have defined drop out from this computation.

\subsubsection{Schr\"odinger formalism and the mass spectrum}
\label{subsecschrovec}

\begin{figure}[ht]
\centering
	\includegraphics[width=.42\textwidth]{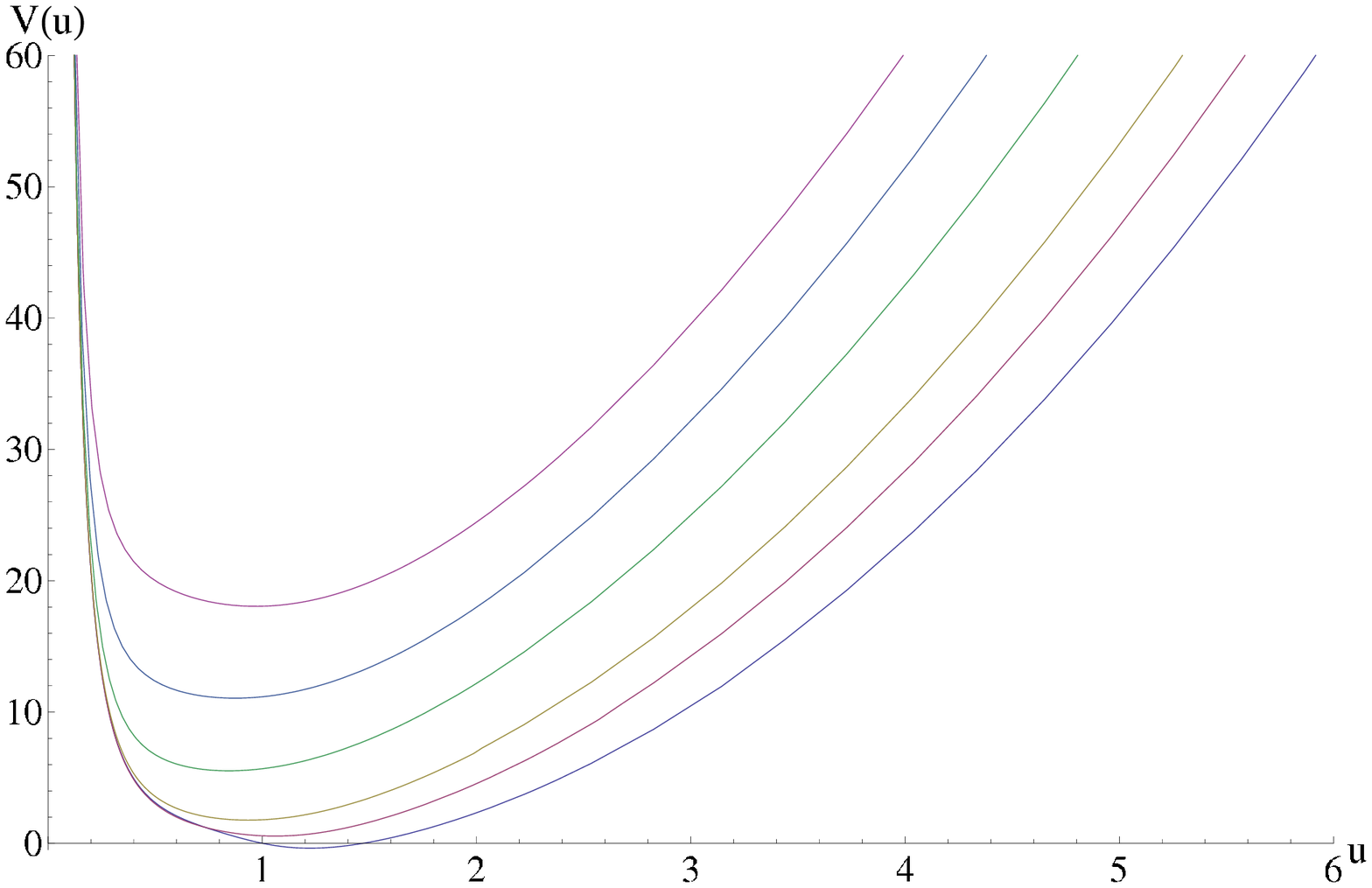}\
	\includegraphics[width=.42\textwidth]{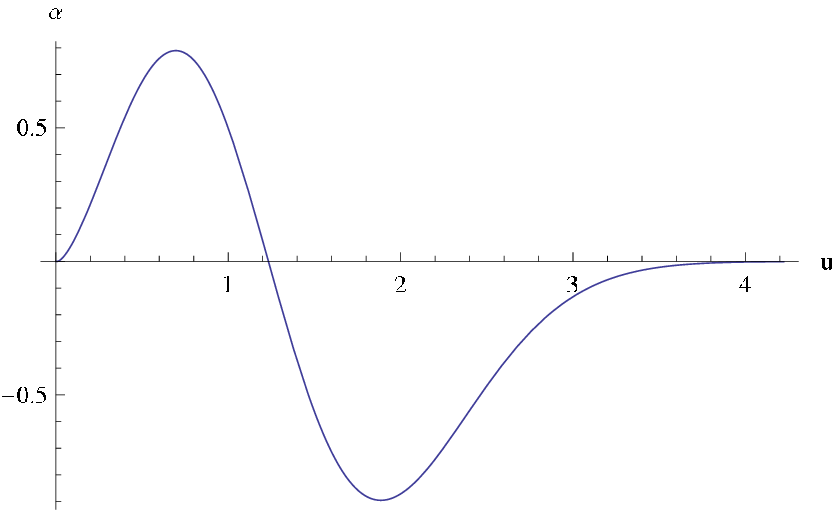}
	\caption{The Schr\"odinger potential associated to the vector excitation for different values
	of the quark mass. From bottom to top $c_1=0,0.5,1,2,3,4$. For illustrative purposes, on the
	right we plot the second normalizable ``wavefunction" for $c_1=1$. }
	\label{fig:schrovectors}
\end{figure}

In order to gain some insight in the problem, let us transform equation (\ref{vectoreom})
to a Schr\"odinger problem, following appendix \ref{app: schro}.
We immediately read $C(z)=M(z)=0$ and:
\begin{equation}
A(z)=e^{-\frac12 \mu^2 \tau^2}
g_{xx}
\tilde g_{zz}^{-\frac12}\,\,,\qquad
B(z)=e^{-\frac12 \mu^2 \tau^2} \tilde g_{zz}^{\frac12} \,\,,
\label{ABdefs}
\end{equation}
such that the Schr\"odinger radial variable is defined by:
\be
u=\int_0^z \sqrt{\frac{B(\tilde z)}{A(\tilde z)}} d\tilde z=
\int_0^z \sqrt{\frac{\tilde g_{zz}(\tilde z)}{g_{xx}(\tilde z)}} d\tilde z\,\,.
\label{uofz}
\ee
Notice that $u\in [0,\infty)$. It is now a straightforward exercise to obtain the Schr\"odinger-like
potential (\ref{schrlike}), for a given $c_1$.  One has to compute numerically $\tau(z)$
as in section \ref{sec:confvev}, then evaluate (\ref{schrlike}) and finally implement the variable change
(\ref{uofz}). Some examples are plotted in figure \ref{fig:schrovectors}.

We observe that the potentials move up as we increase $c_1$. This is of course expected on
general grounds, since meson masses should grow with increasing quark masses, but this feature
 is missing from the hard wall or soft wall models.
In \cite{Iatrakis:2010zf}, we made a phenomenological fit including the strange-strange
meson masses in the analysis, finding good agreement with experimental data.

One can  check that the leading contribution to $V(u)$ near $u=0$ is of the
form $V(u)=\frac34 u^{-2} + \dots$. This just comes from the UV AdS asymptotics.
Let us now find the leading IR contribution to $V(u)$. For large $u$ (namely near
$z=z_\Lambda$), we have that $g_{xx} \approx R^2/z_\Lambda^2$ and
$\tilde g_{zz}\approx 2\pi\alpha' \lambda (\partial_z \tau)^2$. Using the
expressions in appendix \ref{app: schro}, a little
algebra shows that for large $u$, we have $\frac{du}{dz}\approx \frac{\mu}{\sqrt3}z_\Lambda
\partial_z \tau$ and therefore $u\sim  \frac{\mu}{\sqrt3}z_\Lambda \tau$. The function
$\Xi$ behaves as $\Xi\approx  \left(\frac{R}{z_\Lambda}\right)^\frac12 e^{-\frac{3u^2}{4z_\Lambda^2}}$
what finally leads to $V(u)= \frac{9}{4 z_\Lambda^4} u^2 + {\cal O}(u)$.
Therefore, $V(u)$ grows quadratically at large $u$, a fact that leads
to standard Regge trajectories for large excitation number $n$ \cite{Karch:2006pv}.
Asymptotically, the slope of these trajectories is
$\lim_{n\to \infty} \frac{dm_n^2}{dn}= \frac{6}{z_\Lambda^2}$,
as can be found by evaluating (\ref{WKB}).

\begin{figure}[ht]
\centering
	\includegraphics[width=.45\textwidth]{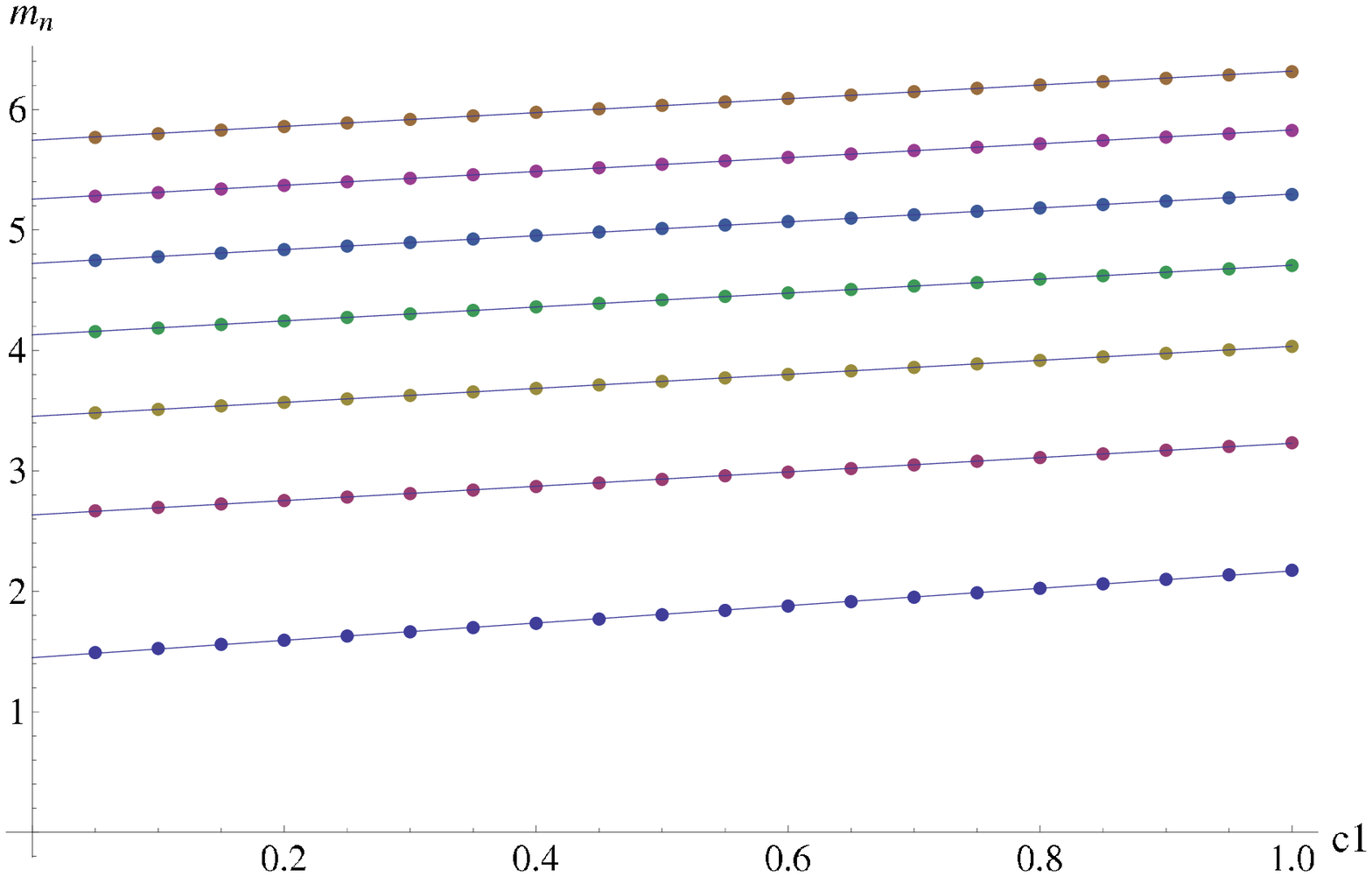}\
	\includegraphics[width=.45\textwidth]{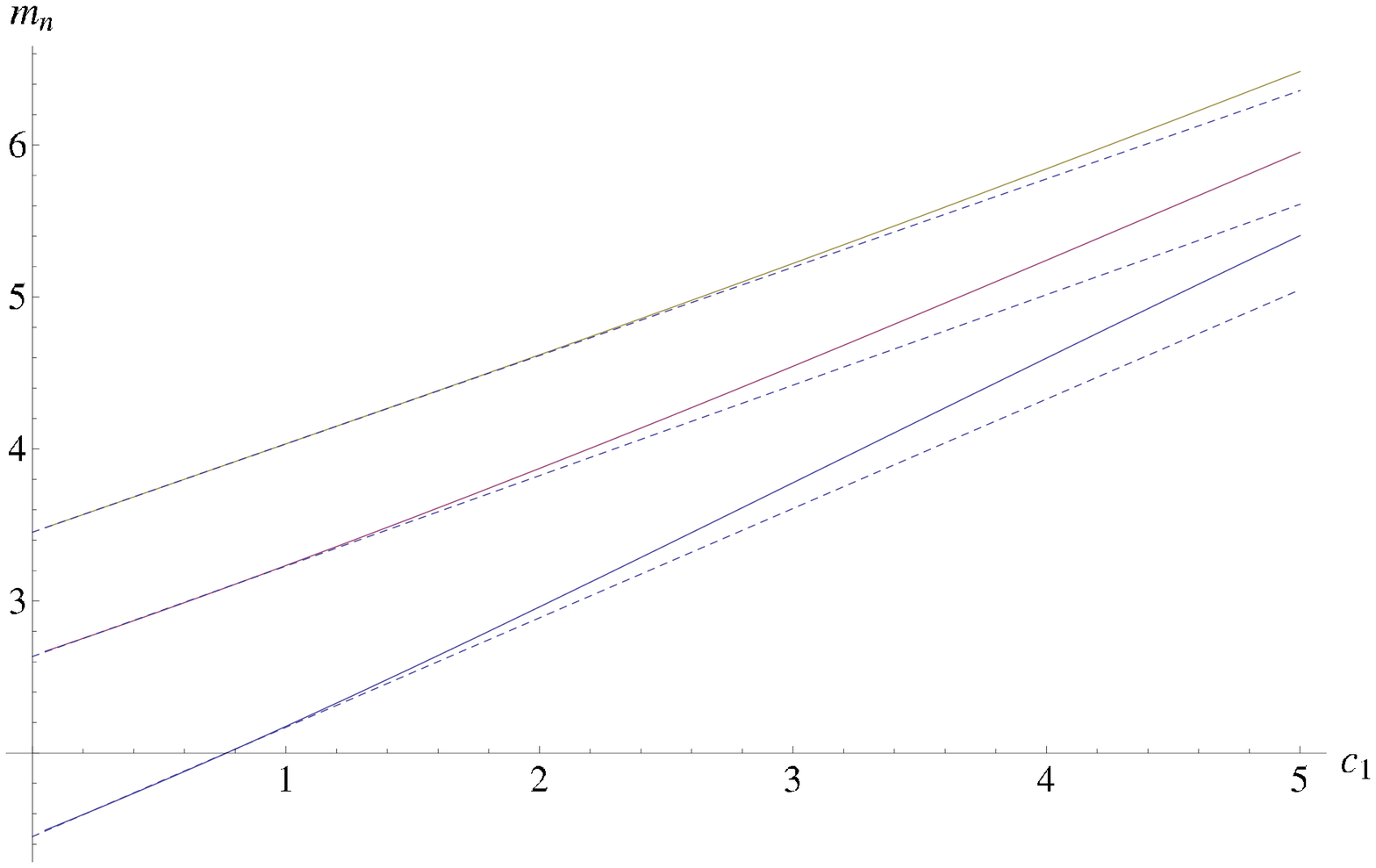}
	\caption{Vector meson masses as a function of $c_1$ (proportional to the quark mass).
	On the left, we plot the fitted straight lines  together with
	several points computed numerically, for the seven lowest-lying vector modes,
	in the range $c_1<1$. On the right, we go to larger values of $c_1$ and, for clarity, only plot
	the three ligthest states. The dashed lines correspond to the linear fits valid
	for small $c_1$ whereas the
	solid line are the actual values found numerically.}
		\label{fig:massvectors}
\end{figure}

By using standard numerical shooting techniques, we have computed the mass spectrum.
In particular, we have computed the first seven states, changing the quark mass parameter
in the range $0<c_1<5$. We plot some results in figure \ref{fig:massvectors}. It turns
out that for small $c_1$ the growth of meson mass on the quark mass is linear.
This is just what one expects from a Taylor expansion if we consider the meson masses
as function of the quark masses.
This was a result already found in \cite{Iatrakis:2010zf}. We have:
\bear
z_\Lambda\,m_V^{(1)} &\approx& 1.45 + 0.718 c_1 \sp
z_\Lambda\,m_V^{(2)} \approx 2.64 + 0.594 c_1 \sp
z_\Lambda\,m_V^{(3)}  \approx 3.45 + 0.581 c_1 \sp  (c_1 \leq 1)\rc
 z_\Lambda\,m_V^{(4)} &\approx& 4.13 + 0.578 c_1 \sp
 z_\Lambda\,m_V^{(5)} \approx 4.72 + 0.577 c_1 \sp
z_\Lambda\,m_V^{(6)} \approx 5.25 + 0.576 c_1 .
\label{vectormassesfit}
\eear

At around $c_1 \geq 1$, the graphs start departing from the straight line, as can
be seen on the second plot in figure \ref{fig:massvectors}.

\subsubsection{Current-current correlator and normalization of the action}
\label{veccorrel}

We have discussed the vector spectrum, but we are also interested in the decay constants of
 each state. In order to compute them, we have to fix the multiplicative
constant associated to the normalization
of the action associated to the vector modes. We will follow the reasoning of
\cite{Erlich:2005qh,Da Rold:2005zs} and match the correlator $\Pi_V$ to the
quark bubble perturbative computation at large Euclidean momentum.
In fact, all the discussion of this subsection is completely parallel to
\cite{Erlich:2005qh,Da Rold:2005zs}, since it only depends on the asymptotically
AdS structure. We however repeat the argument in the present notation for the
sake of clarity.

The current-current correlator is defined as:
\be
\int d^4x\, e^{iqx} \langle J_\mu (x) J_\nu (0) \rangle =
(\eta_{\mu\nu} q^2 - q^\mu q^\nu) \Pi_V (q^2)\,\,.
\ee
As usual, we compute it holographically from the on-shell action.
Integrating by parts in (\ref{vectoracti}) and adding the counterterm
from (\ref{allcounterterms}), we find:
\bear
S_{V}={(2\pi\alpha')^2 {\cal K}\over g_{V}^{4}}\int {d^{4}q\over (2\pi)^4} e^{-{1\over2}\mu^{2}\tau^2}
g_{xx}\tilde g_{zz}^{-{1\over2}}V_{\mu}(q,z)\partial_{z}V^{\mu}(-q,z)\Big|_{z=\epsilon}+\rc
-\frac{{\cal K}R^5 (2\pi\alpha'^2)}{g_V^4} \int {d^4q\over (2\pi)^4}\,\left(
 q^2 V_{\mu}(q,\epsilon) V^{\mu}(-q,\epsilon)(\log \epsilon + \frac12)
\right)
\label{sv}
\eear
where $V_\mu (q,z)=\psi^V(q,z)V_0^\mu (q)$, and $\psi^V(q,z)$ is the solution to
(\ref{vectoreom}) subject to $V(q,\epsilon)=1$ and with normalizable behaviour in the IR.
At small $z$, the solution for $\psi^V(q,z)$ can be expanded in terms of two integration
constants as:
\be
\psi^V = b_1(q) + \left( b_2(q) + \frac12 b_1(q) q^2 \log z\right)\,z^2 + \dots
\label{psiVUV}
\ee
Substituting  this expression into (\ref{sv}) and taking two derivatives with respect to
$V_0^\mu (q)$, we find that:
\be
\Pi_V (q^2) = - 4 \frac{{\cal K} R (2\pi\alpha')^2}{g_V^4} \frac{b_2}{q^2}
\label{correlatorofb2}
\ee
where we have set $b_1(q)=1$ consistent with the two-point function prescription and the non-trivial $q^2$-dependence comes through $b_2(q)$,
which has to be found by integrating numerically and demanding the physical
IR behaviour.

\begin{figure}[ht]
\centering
	\includegraphics[width=.55\textwidth]{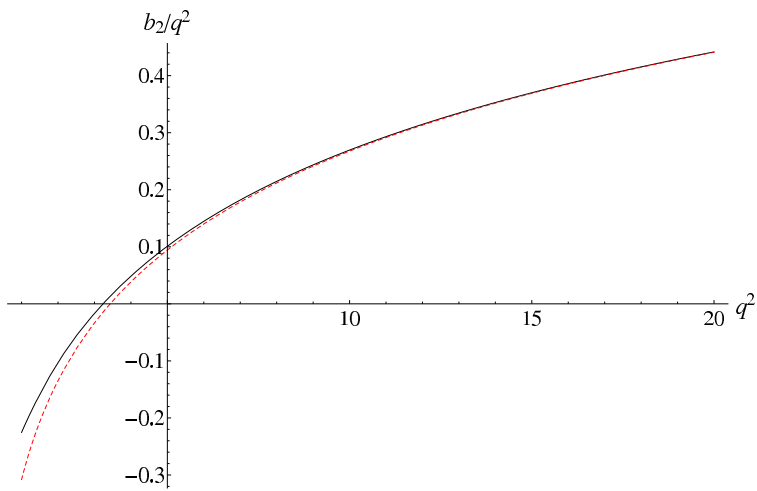}
	\caption{We plot the value of $b_2/q^2$, proportional to the vector
	current-current correlator. The plot
	is a  comparison of the approximation found in the text from a simplified
	Schr\"odinger problem, Eq.({\protect\ref{b2overq2}})
	 (red dashed line) to the actual numerical result
	(solid line).
	The numerical plot was made by taking $c_1=0.1$, $\mu^2=\pi$.
	}
	\label{fig:euclideancorrelator}
\end{figure}

Before entering into numerical integration, we are interested in computing
the limiting behaviour for $\Pi_V$ for large $q^2$.
In order to do this, we consider again the equation written in Schr\"odinger form and
notice that, for small $z$:
\be
u\simeq z \,\,,\qquad \alpha(u)\simeq u^{-{1\over2}}\psi^{V}(u)
\label{utoz}
\ee
The leading large $q$ behaviour is not affected by the
details of the Schr\"odinger potential, so we may just approximate it by an
expression that interpolates between its UV and IR behaviours, as discussed
in subsection \ref{subsecschrovec}.
Namely, we can just write:
\be
-\partial_{u}^{2}\alpha+\left({3\over 4 u^2}+c^{2} u^{2}\right)\alpha
+ q^{2} \alpha=0\,,
\label{largeq2eq}
\ee
where we should take $c^{2}={9\over 4 z_\Lambda^4}$. However, we will see that
the value of $c^2$ does not matter for the normalization we want to make.
(\ref{largeq2eq}) is nothing else than the soft wall model of \cite{Karch:2006pv}.
The general solution  of (\ref{largeq2eq}) is:
\be
\alpha(u)=k_{1}{e^{-c u^2 \over 2} \over \sqrt{u}} U({q^2 \over 4 c},{0},c u^2)
+k_{2}{e^{-c u^2 \over 2} \over \sqrt{u}} L_{-q^2 \over 4 c}^{-1}(c u^2)
\label{scsolvs}
\ee
where $U$ stands for the confluent hypergeometric function and $L$ for a generalized
Laguerre polynomial.
IR normalisability requires $k_2=0$. We now substitute in  (\ref{utoz}) and
fix $k_1$ by demanding that
$\lim_{z\to 0} \psi^V (q,z) =\lim_{u\to 0} u^{\frac12} \alpha (u)= 1$.
\be
\psi^{V}(q,u)={q^2 \over 4 c}\Gamma\left({q^2 \over 4 c}\right)e^{-c u^2 \over 2} U({q^2 \over 4 c},{0},c u^2)
\ee
We can now expand this expression for small $u\approx z$  and compare to (\ref{psiVUV}), in
order to read $b_2$ and, accordingly $\Pi_V$ from (\ref{correlatorofb2}).
The leading pieces at large $q^2$ for $b_2$ read:
\be
\lim_{q^2 \to \infty}\frac{b_2}{q^2}=\frac14 \log q^2 - \frac14
(1+\log 4 -2 \gamma)-{c^2 \over 3 q^4}+\dots\,\,,
\label{b2overq2}
\ee
where $\gamma$ is Euler's constant. Therefore,  the leading piece which we can compare to
the quark bubble via (\ref{correlatorofb2}) is:
\be
\Pi_V(q^2) = - \frac{{\cal K} R (2\pi\alpha')^2}{g_V^4}\log q^2
\label{pivlog}
\ee
By matching this expression to the perturbative result, we find\footnote{Notice that
we are dealing with abelian flavor symmetry. There is a factor of $\half$ difference with
respect to \cite{Erlich:2005qh} since in that paper they deal with a non-abelian case and
define $\tr (t^a t^b)= \frac12 \delta^{ab}$. This also makes different the definition of the
decay constants, for instance the $f_\pi$ defined in \cite{Erlich:2005qh} is the $f_\pi$ we will
use divided by $\sqrt2$.}:
\be
{(2\pi \a')^{2}{\cal K} R\over g_{V}^{4}}={N_{c}\over 12 \pi^{2}}
\label{gvqcd}
\ee
One may wonder how good  the results
obtained from the simple Schr\"odinger problem we have discussed
(\ref{largeq2eq}) are as an approximation  to the full problem (\ref{vectoreom}).
In figure \ref{fig:euclideancorrelator}, we compare (\ref{b2overq2}) to the
value of $b_2 (q^2)/q^2$ computed numerically.

\subsubsection{Decay constants}
\label{sec:decay}

\begin{figure}[ht]
\centering
	\includegraphics[width=.55\textwidth]{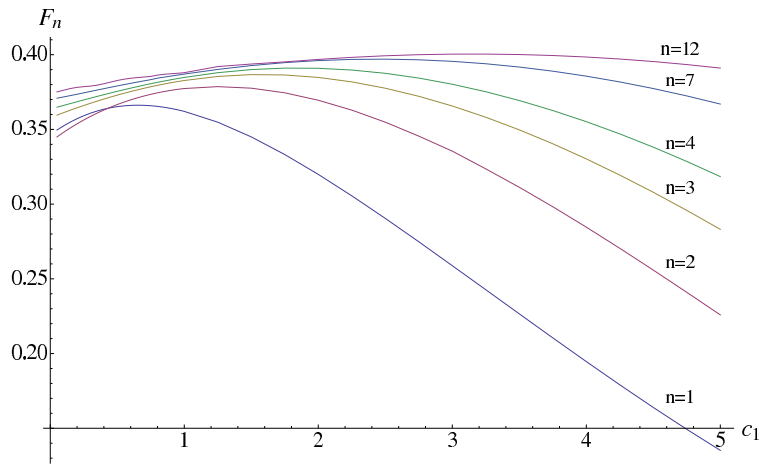}
	\caption{The decay constant, in units of $z_\Lambda^{-1}$ for
          the four lowest-lying, the seventh and the twelve-th vector
          mode (from bottom to top), as a function of $c_1$.
	The numerical plot was made by taking  $\mu^2=\pi$ and $N_c=3$.
	}
	\label{figdecay}
\end{figure}

We are now interested in determining the decay constants of our mesonic states.
We start by writing the current-current correlator as a sum rule.
\be
\Pi_V(q^2) =
 \sum_n \frac{F_n^2}{(q^2 + m_n^2 - i \epsilon)}
 \label{sumvectors}
\ee
The idea is to derive the form of the sum rule holographically. In appendix \ref{app:2point}
we give  a general description  on
how to write holographically a two-point correlator as an infinite sum.
Then we use properties of the normalizable modes in order to determine the
values of $F_n$. The argument  follows
\cite{Erlich:2005qh,Da Rold:2005zs} so we directly quote the result in the present
notation\footnote{Notice our definition of $F_n$ is different from \cite{Erlich:2005qh}.}:
\be
F_n^2= \frac{N_c}{6\pi^2} \frac{R}{m_n^2}\left(\frac{d^2 \psi_V^{(n)}}{dz^2}\Big|_{z=0}\right)^2
\label{computeFn}
\ee
where $\psi_V^{(n)}$, $n=1,2,\dots,\infty$ are the solutions of (\ref{vectoreom}) normalized as:
\be
\int B(z) (\psi_V^{(n)})^2 dz =1
\label{normcond}
\ee
with $B(z)$ given in (\ref{ABdefs}).
Again, we can compute numerically the values of the decay constants given by the model.
We have plotted them in figure \ref{figdecay}. One can see that the dependence on excitation
number is rather mild for small quark masses and for a large number of modes.

\subsubsection{Regge trajectories for vector mesons and linear confinement }

\begin{figure}[ht]
\centering
\includegraphics[width=.45\textwidth]{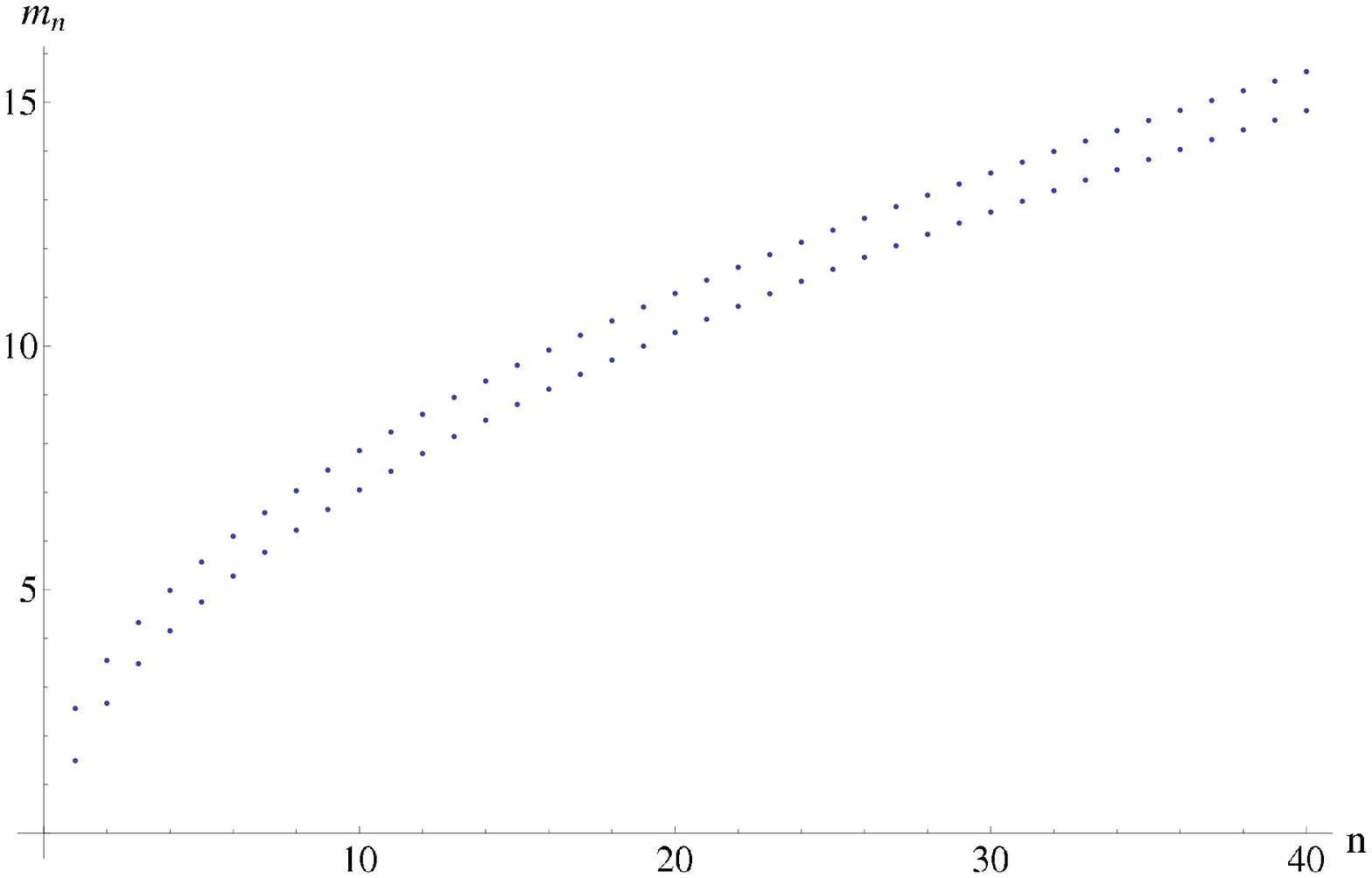}
	\includegraphics[width=.45\textwidth]{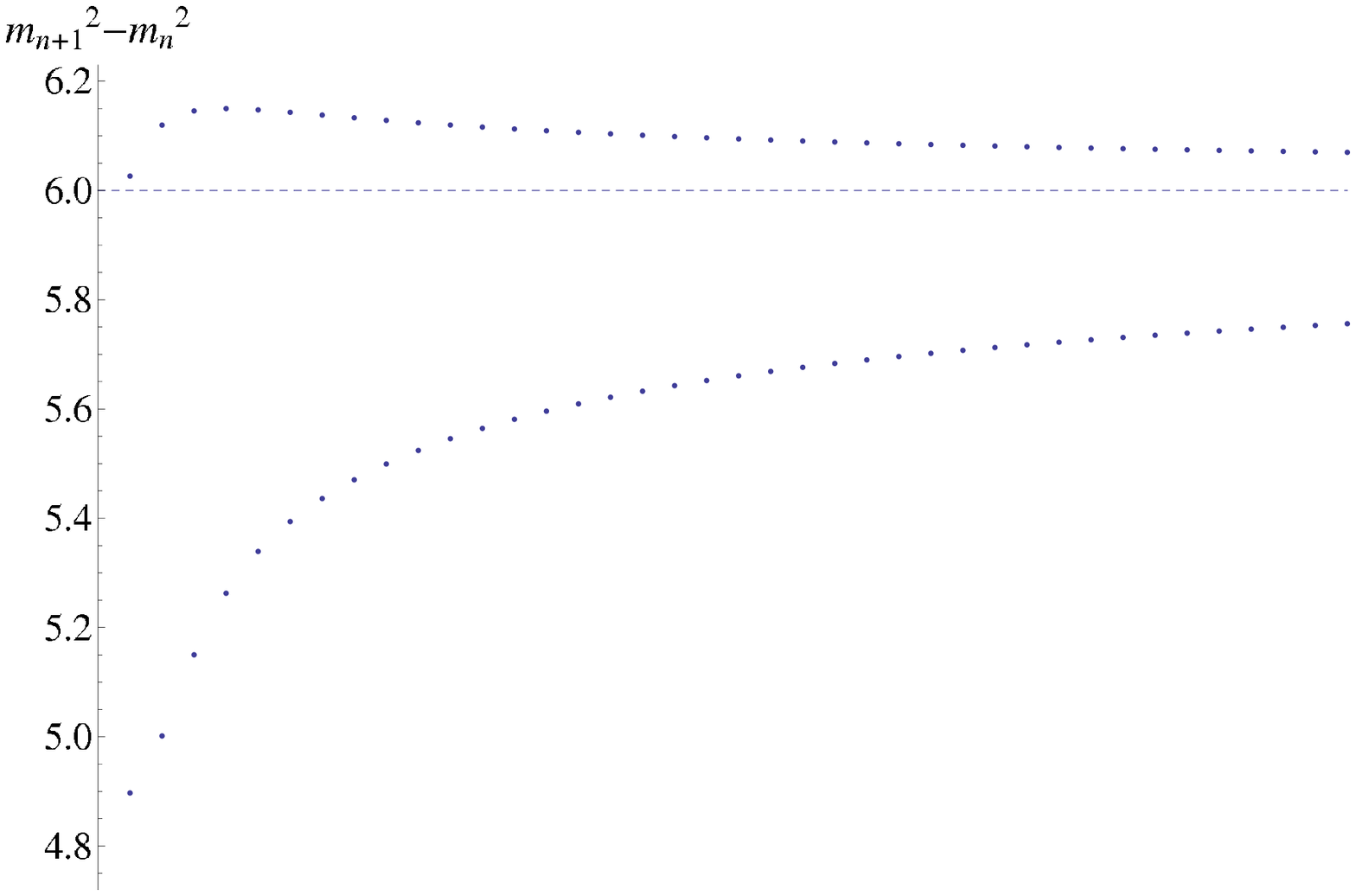}
	\caption{Results corresponding to the forty lightest vector states with
	$c_1=0.05$ and $c_1=1.5$. On the right, the horizontal line signals the asymptotic
	value 6 of the Regge trajectory, the lower line corresponds to $c_1=0.05$ and the upper line to
	$c_1=1.5$. Masses are given in units of $z_\Lambda^{-1}$.
	}
	\label{fig:Regge}
\end{figure}

Typical holographic models lead to a behaviour of the masses with the excitation
number as $m_n^2\, \propto \, n^2$, for large $n$ \cite{Schreiber:2004ie}.
However, experiment and semiclassical
quantization of a hadronic string (assuming linear confinement) suggest that
$m_n^2\, \propto \, n$ in QCD.
Circumventing this problem was the
motivation for developing the soft-wall model \cite{Karch:2006pv}.
As pointed out above and also in \cite{ckp}, a model including an open string tachyon
with action (\ref{generalact}) and gaussian tachyon potential, naturally implements this
behaviour. In figure \ref{fig:Regge}, we plot the results of some numerical computations
which display this feature. We remind the reader that, as we saw in section
\ref{subsecschrovec}, for vector mesons $\lim_{n\to \infty} m_{n+1}^2 - m_n^2 =
6/z_\Lambda^2$. This seems to be born out by the figure.

\subsection{Axial-vector mesons}

The quadratic action corresponding to the axial vector mesons that comes from expanding
(\ref{generalact}) picks an extra term with respect to (\ref{vectoracti}), coming
from the covariant derivative of the tachyon:
\be
S_A = - {(2\pi\alpha')^2 \over g_{V}^{4}}{\cal K}\int d^4x\, dz
e^{-\frac12 \mu^2 \tau^2}
\left[ \frac12 \tilde g_{zz}^\frac12 A_{\mu\nu}A^{\mu\nu} + g_{xx}
\tilde g_{zz}^{-\frac12} \partial_z A_\mu \partial_z A^\mu
+\frac{4 R^2 g_V^4}{3(2\pi \alpha')^2}\mu^2 \tau^2 g_{xx} \tilde g_{zz}^{\frac12}
A_\mu A^\mu
\right]
\label{axacti}
\ee
The equation of motion can be  derived to be:
\be
\frac{1}{ e^{-\frac12 \mu^2 \tau^2} \tilde g_{zz}^{\frac12}}
\partial_z \left( e^{-\frac12 \mu^2 \tau^2}
g_{xx}
\tilde g_{zz}^{-\frac12} \partial_z \psi^A (z)\right)
- k \frac{\mu^2\tau^2}{z^2} \psi^A (z)
- q^2  \psi^A (z) = 0
\label{axialeom}
\ee
where we have introduced a new constant $k$ as the combination:
\be
k=\frac{4R^4 g_V^4}{3 (2\pi\alpha')^2}
\label{kaval}
\ee
We observe that $\tau$ only enters through the combination $\mu\,\tau$ so $\mu$ is
immaterial since it can be rescaled away. On the other hand, the constant $k$, that did
not enter the parity even sector does affect the physics. In fact, by comparing
(\ref{vectoreom}) to (\ref{axialeom}), one can see that the difference between the equation
for the vectors and the one for the axials in controlled by $k$. Therefore, it is natural
to guess that $k$ somehow enhances or suppresses the effects of  chiral symmetry breaking on the P-odd spectra.
In the following, we will see how the physics depends on this parameter. The model of our
previous work \cite{Iatrakis:2010zf} was more constrained since, in terms of the present
notation, $k$ was fixed to $\frac{12}{\pi^2}$.

\subsubsection{Schr\"odinger formalism and the mass spectrum}
\label{subsecschroax}

The functions for converting to a Schr\"odinger problem $A(z)$, $B(z)$ are as before
(\ref{ABdefs}). On top of that, we have here a non-trivial $M(z)$ given by
$M(z)=B(z)\,k\,\mu^2\tau^2/z^2$. It is easy to check that the leading piece
in the UV of the Schr\"odinger
potential is $3/4u^2$ as for the vectors. However, the leading IR behaviour is modified
due to the term proportional to $k$ to $V_{IR}(u)= \frac{9}{4z_\Lambda^4}\left(1+\frac{4k}{3}\right)u^2$.
From this observation, one can immediately realize that the model gives different Regge slopes
for vectors and axials and that the leading behaviour of $\Pi_A(q^2)$ at large Euclidean momentum
coincides, consistently, with the vector one (\ref{pivlog}). We will later comment further on these
issues.

The qualitative appearance of the Schr\"odinger potentials for the axial excitation is similar to the
ones for the vectors. But the value of the potentials in the axial case is always higher due to
the terms coming from $M(z)$. Thus, for equal excitation number and quark mass, the axial mode
is always heavier than the vector mode (with the difference controlled by $k$).
For small values of $c_1$, the dependence of the meson masses on the quark masses is
linear

\bear
z_\Lambda\,m_A^{(1)} &\approx& 2.05 + 1.46 c_1 \,\,,\ \qquad
z_\Lambda\,m_A^{(2)} \approx 3.47 + 1.24 c_1 \,\,,\ \qquad
z_\Lambda\,m_A^{(3)} \approx 4.54 + 1.17 c_1\,\,,  \qquad (c_1 \leq 1)\rc
z_\Lambda\,m_A^{(4)} &\approx& 5.44 + 1.13 c_1\,,\qquad\
z_\Lambda\,m_A^{(5)} \approx 6.23 + 1.11 c_1\,,\qquad\
z_\Lambda\,m_A^{(6)} \approx 6.95 + 1.10 c_1
 \,\,.
\label{axialmassesfit}
\eear
For this calculation, we used $k={18\over \pi^2}$ as it is found by
the fit of the parameters in section (\ref{sec: pheno}), whereas in
\cite{Iatrakis:2010zf}, we used $k={12\over \pi^2}$.
For larger $c_1$, the plots of the meson mass dependence on the quark mass for the axial excitation
look similar to the vector case, see the plot on the right of figure \ref{fig:massvectors}.

\subsubsection{Current-current correlator and the pion decay constant}

By explicit computation it is easy to check that the UV expansion of
the solution to (\ref{axialeom}) is given in terms of the two integration constants as:
\be
\psi^A = b_1 + \left( b_2 + \frac12 b_1  (q^2 + \pi \,k\,c_1^2)  \log z\right)\,z^2 + \dots
\label{psiaxUV}
\ee

For the case of the axial vector, the corresponding sum rule
generalising (\ref{sumvectors}) reads:
\be
\Pi_A (q^2)=  \frac{f_\pi^2}{q^2} +
\sum_n \frac{F_n^2}{(q^2 + m_n^2 - i \epsilon)}
\ee
where of course now the $n$ run over the axial resonances.
The $F_n$ here are computed in essentially
the same way as for the vector case, namely using
(\ref{computeFn}) with a normalization condition (\ref{normcond}).

Now, we are also be able to compute the value of $f_\pi$.
We do this by directly computing
the 2-point function at zero-momentum, namely:
\be
f_\pi^2 = - \frac{N_c}{6\pi^2}
b_2|_{q=0}
\label{fpi}
\ee
where we have used the expansion (\ref{psiaxUV}), which up to that order, is also valid
for the axial case.
The value of $b_2$ to be inserted in (\ref{fpi}) is found numerically
by solving (\ref{axialeom}) with $q^2=0$, with initial condition $\psi^A|_{z=\epsilon}=1$ and
demanding IR normalizability.
\begin{figure}[ht]
\centering
\includegraphics[width=.40\textwidth]{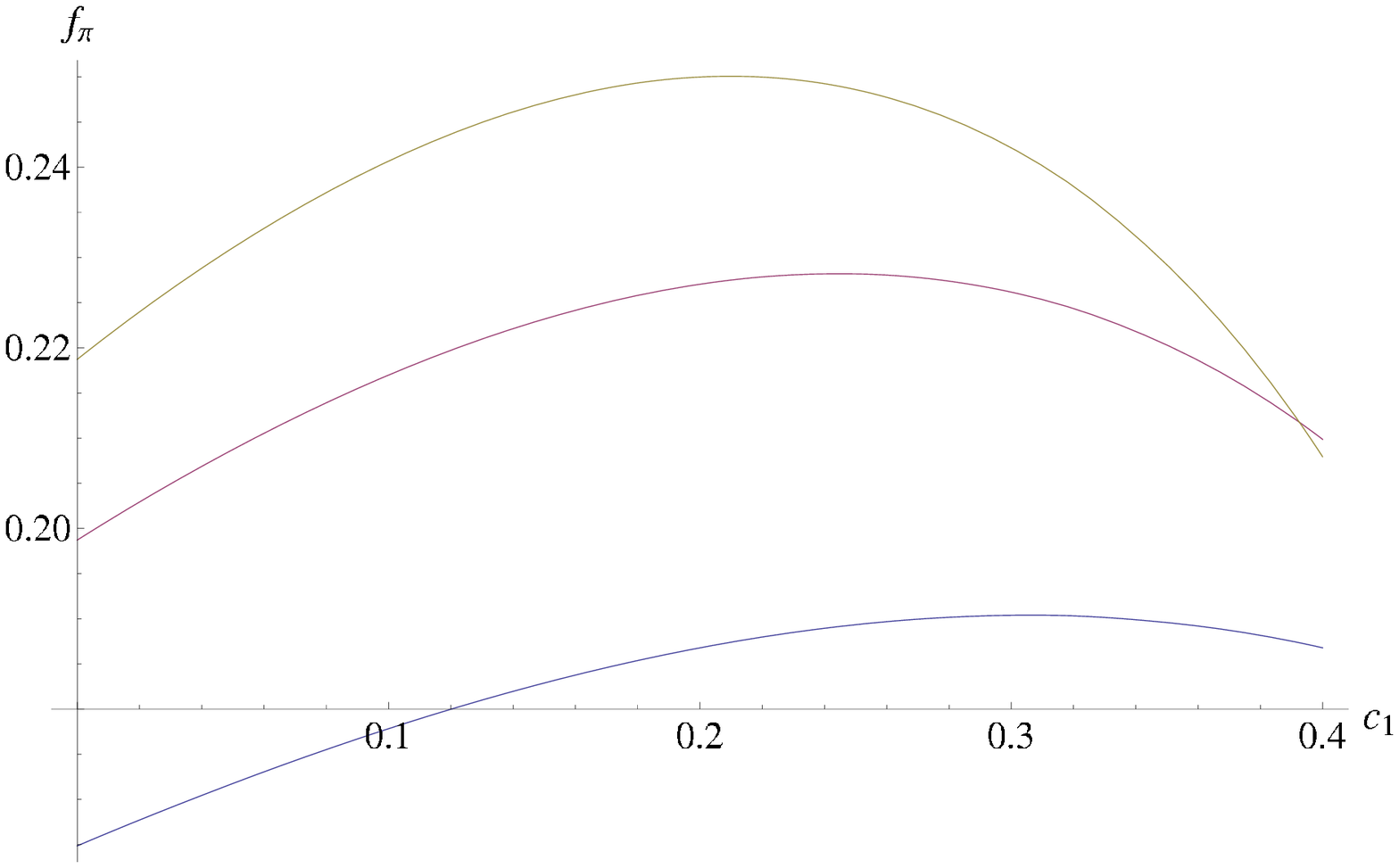}\ \ \
\includegraphics[width=.40\textwidth]{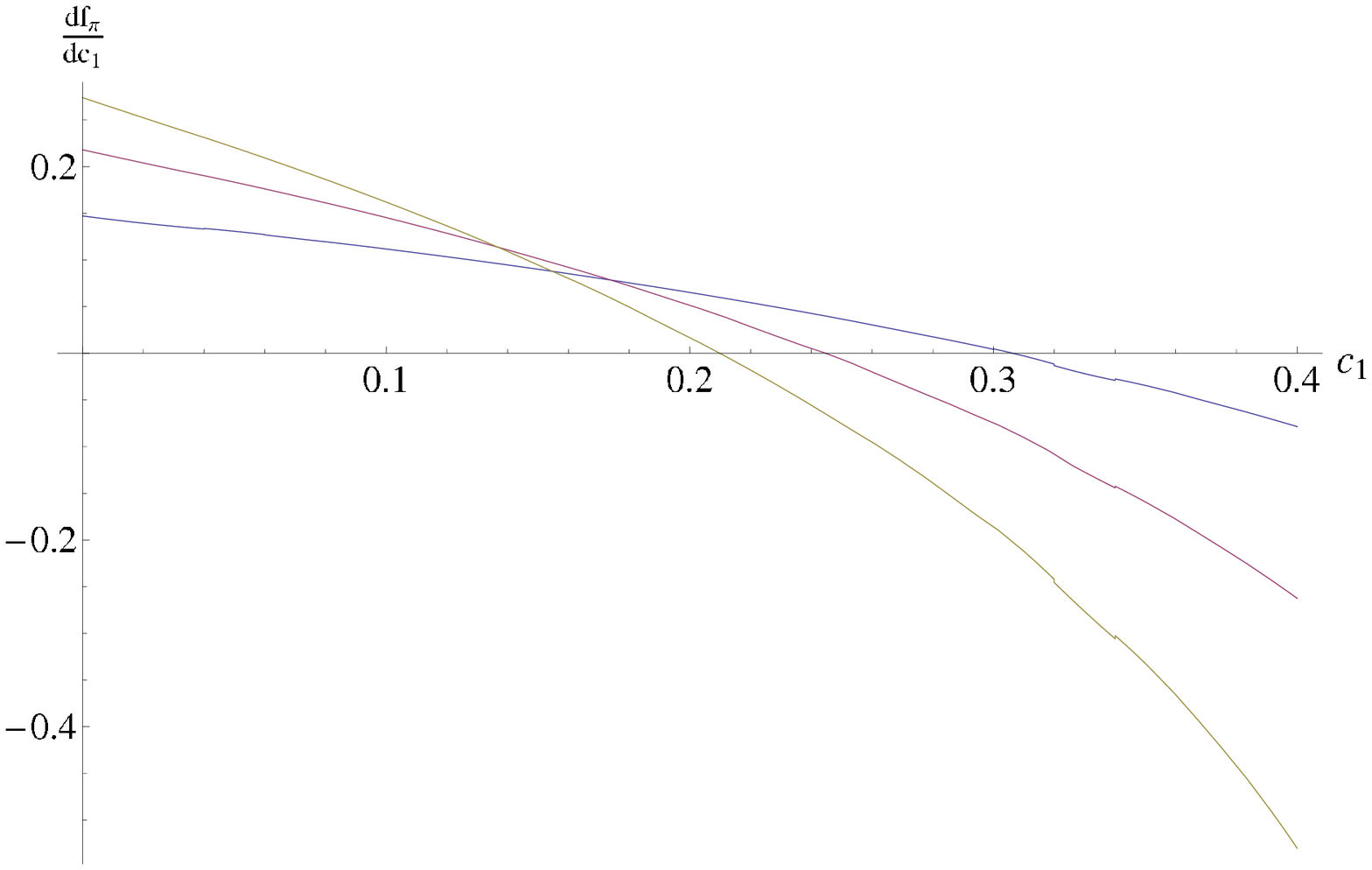}
	\caption{The pion decay constant and its derivative
	as a function of $c_1$ - the quark mass. The different lines
	correspond to different values of $k$. From bottom to top
	(on the right plot, from bottom to top in the vertical axis)
	 $k=\frac{12}{\pi^2},\frac{24}{\pi^2},
	\frac{36}{\pi^2}$. The pion decay constant comes in units of $z_\Lambda^{-1}$.
}
	\label{fig:fpi}
\end{figure}

From the figure, we observe that the decay constant grows with the quark mass for small quark
masses and then starts decreasing.
We can also observe that increasing the parameter $k$, increases the value of the
pion decay constant. This is in agreement with the intuitive notion given above that $k$ somehow
controls the amount of chiral symmetry breaking.

\subsection{Scalar mesons}
\label{sec:scalarmes}

We now deal with the scalar excitation.
The quadratic action reads:
\begin{eqnarray}
S&=&-2 \pi \alpha' {\cal K}\, \lambda\int d^4 x dz e^{-\frac12 \mu^2 \tau^2}
\Big[  g_{xx}^2 g_{zz} \tilde g_{zz}^{-{3\over2}} (\partial_{z} s(x,z))^{2}
-2 \mu^{2}  g_{xx}^{2}\tilde g_{zz}^{-{1\over 2}}\tau(z) \tau(z)' s(x,z)\partial_{z} s(x,z) \nonumber \\
&+&{\mu^2 \over 2\pi\alpha'\,\lambda} (\mu^{2}\tau(z)^{2}-1)g_{xx}^{2}\tilde g_{zz}^{1\over 2}s(x,z)^{2}
+  g_{xx}g_{zz}\tilde g_{zz}^{-{1\over 2}} (\partial_{\mu} s(x,z))^{2} \Big]
\end{eqnarray}
From (\ref{generalact}), it can be seen that there is also a linear term in the bulk action, but can
be easily shown to be a total derivative.

\begin{figure}[ht]
\centering
\includegraphics[width=.45\textwidth]{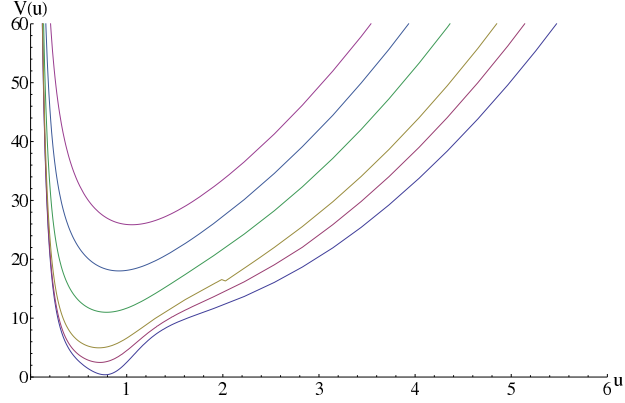}
	\caption{The Schr\"odinger potentials associated to the scalar excitation for
	$c_1=0,0.5,1,2,3,4$.
	}
	\label{fig:schroscalars}
\end{figure}

We can read the functions that are used to rewrite this problem is Schr\"odinger form, as defined in appendix
\ref{app: schro}:
\bear
A(z)=e^{-\frac12 \mu^2 \tau^2}g_{xx}^2 \frac{g_{zz}}{\tilde g_{zz}^{3/2}} \,,\qquad \qquad \qquad \qquad
B(z)=e^{-\frac12 \mu^2 \tau^2}g_{xx} \frac{g_{zz}}{\tilde g_{zz}^{1/2}} \,,\qquad \qquad \qquad \rc
C(z)=-2\mu^2 e^{-\frac12 \mu^2 \tau^2} \frac{g_{xx}^2}{\tilde g_{zz}^{1/2}}
\tau(z) \partial_z \tau(z)\,,\qquad
M(z)=\frac{\mu^2 }{2\pi\alpha' \,\lambda} e^{-\frac12 \mu^2 \tau^2}(\mu^2 \tau^2-1)g_{xx}^2
\tilde g_{zz}^{\frac12}\,.
\eear
Notice that $B(z)/A(z)$ takes the same value as for the vector and axial excitations, which means that
the definition of the $u$-radial coordinate is the same as in those cases. The expression built from
$B(z)$, $C(z)$ and $M(z)$ which enters the Schr\"odinger potential takes a remarkably simple form:
\be
\frac{1}{B(z)} \left( M(z) - \frac12 \partial_z C(z)
\right) = - \frac{3}{z^2}
\ee
We will find the UV and IR limiting behaviour of the associated Schr\"odinger potential.
At small $z\approx u$, we find $\Xi \approx R^{\frac32} /u^{\frac32}$ and one can immediately
compute from (\ref{schrlike}) the UV leading term to be $V(u)=\frac{3}{4u^2}$. Similarly, the
term that dominates for large $u$ is quadratic $\frac{9u^2}{4z_\Lambda^2}$. Thus, we have found that
the UV and IR asymptotics are the same as for the vector case and, as a first approximation, we can
use the soft wall equation (\ref{largeq2eq}). Thus, we can again match the asymptotic behaviour of the
current-current correlator to the perturbative result. For that, we make an argument similar
to \cite{DaRold:2005vr} assume that the $q\bar q$ operator is dual to the tachyon rescaled by some
constant $\beta$, such that the boundary coupling is, schematically, $\int \beta (\tau/z) q\bar q$.
This is what we anticipated in the relation between $c_1$ and the quark mass
(\ref{mqc1}). Then, matching the large $q^2$ result
$\Pi_S(q^2)=\frac{N_c}{8\pi^2} q^2 \log q^2$ and reasoning  as in the the vector case, we find:
\be
{(2\pi \a'){\cal K} R^3 \lambda \over \beta^2}={N_{c}\over 8 \pi^{2}}
\label{lamqcd}
\ee
In figure \ref{fig:schroscalars}, we depict the associated Schr\"odinger potential for different
values of $c_1$. Comparing figure \ref{fig:schroscalars} to figure \ref{fig:schrovectors}, one can
check that the potentials for the scalars are above those of the vectors. Thus, for equal excitation
number, scalar mesons are typically heavier than vectors in the present model.
\\
The relation of the lowest scalar meson masses to $c_1$ follows
\bear
z_\Lambda\,m_S^{(1)} &\approx& 2.47 + 0.683 c_1 \,\,,\ \qquad
z_\Lambda\,m_S^{(2)} \approx 3.73 + 0.488 c_1 \,\,,\ \qquad
z_\Lambda\,m_S^{(3)} \approx 4.41 + 0.507 c_1\, \,, \qquad (c_1\leq 1)
\rc
z_\Lambda\,m_S^{(4)} &\approx& 4.99 + 0.519 c_1\,,\qquad\
z_\Lambda\,m_S^{(5)} \approx 5.50 + 0.536 c_1\,,\qquad\
z_\Lambda\,m_S^{(6)} \approx 5.98 + 0.543 c_1
 \,\,.
\label{scalarmassesfit}
\eear
We point out that the scalar meson masses do not depend on the
parameter $k$.

\subsection{Pseudoscalar mesons}

We now focus on the pseudoscalar mesons. With respect to the previous modes, there
is an extra complication because the physical modes are a combination of two bulk
fields $\theta$ and $A_\mu^\lVert$. However, we will see that it is possible to find
a combination of the fields for which one obtains a standard Sturm-Liouville problem.

The quadratic action reads:
\begin{eqnarray}
S&=&-(2 \pi \alpha')^2 {\cal K}\int d^4 x dz e^{-\frac12 \mu^2 \tau^2}
\Big[ {1\over g_{V}^{4}}g_{xx}\tilde g_{zz}^{-{1\over 2}}(\partial_{z} A^{\lVert}_\m)^{2}
\nonumber \\
&+&
{\lambda \over 2\pi \alpha'} \tau^2 g_{xx}^{2} \tilde g_{zz}^{-{1\over 2}} (\partial_{z}\theta)^{2}
+ {\lambda \over 2 \pi \alpha'}\tau^2 g_{xx}
\tilde g_{zz}^{1 \over 2} (\partial_{\mu}\theta+2 A^{\lVert}_\m)^2 \Big]
\label{pseudoac}
\end{eqnarray}
Defining $\varphi (z)$ and $\vartheta (z)$ as in (\ref{somedefsnorm}) and Fourier transforming
${\cal P}(x^\mu)$,
we can write the equations of motion for $A^{\lVert}_\m$ and $\theta$ as:
\begin{eqnarray}
{1 \over  e^{-\frac12 \mu^2 \tau^2} \tilde g_{zz}^{{1\over 2}}}
\partial_z (e^{-\frac12 \mu^2 \tau^2} g_{xx} \tilde g_{zz}^{-{1\over 2}} \partial_z \varphi(z))
-k \frac{\mu^2 \tau^2}{z^2} ( \varphi(z)-\vartheta(z) )=0
\label{pseom1}
\\
k \frac{\mu^2 \tau^2}{z^2} \partial_z \vartheta(z) + q^2 \partial_z \varphi(z)=0
\label{pseom2}
\end{eqnarray}
These two equations can be combined into one by solving (\ref{pseom1}) for $\vartheta$ and inserting this into
(\ref{pseom2}).
\begin{eqnarray}
e^{-\frac12 \mu^2 \tau^2} \tau^{2} g_{xx}^{2} \tilde g_{zz}^{-{1\over 2}}
\partial_{z} \left[{ 1 \over e^{-\frac12 \mu^2\tau^2} \tau^2 g_{xx}\tilde g_{zz}^{1\over 2}}\partial_{z}  \psi^{P}(z)\right]
-k\frac{\mu^2 \tau^2}{z^2}\psi^{P}(z)-q^2 \psi^{P}(z)=0
\label{pseom3}
\end{eqnarray}
where we have defined:
\begin{equation}
\psi^{P}(z)=-e^{-\frac12 \mu^2 \tau^2} g_{xx} \tilde g_{zz}^{-{1\over 2}} \partial_{z} \varphi(z)
\end{equation}
and we have used the definition of $k$ in (\ref{kaval}).
We can transform equation (\ref{pseom3}) to a Schr\"odinger form following appendix \ref{app: schro}.
Comparing (\ref{pseom3}) to (\ref{lalala}) (and inserting $C(z)=0$) , we find:
\be
A(z)=e^{\frac12 \mu^2 \tau^2} \tau^{-2} g_{xx}^{-1} \tilde g_{zz}^{-\frac12}
\,,\qquad
B(z)=e^{\frac12 \mu^2 \tau^2} \tau^{-2} g_{xx}^{-2} \tilde g_{zz}^{\frac12}\,,\qquad
\frac{M(z)}{B(z)}= k\frac{\mu^2 \tau^2}{z^2}\,.
\ee
Notice that the value of $B/A$ coincides with those for the rest of modes and, therefore,
the Schr\"odinger coordinate $u$ is the same for all the possible excitations.
Let us compute the IR (large $u$) leading behaviour of such a Schr\"odinger potential.
It turns out to be $V_{IR}(u)\approx \frac{9}{4z_\Lambda^4}(1+\frac{4k}{3})u^2$, as for the
axials. The coefficient of this quadratic term is what controls the slope of the Regge
trajectories for highly excited mesons. Thus, the outcome of the present model in this respect
is that vectors and scalars have the same Regge slope, whereas the slopes for axials and pseudoscalars
coincide and are larger than the vector one.

An important observation is that the natural normalization condition is not
(\ref{normpsi}) but, looking for the kinetic term of the pseudoscalar field in
(\ref{pseudoac}), we obtain:
\be
(2 \pi \alpha')^2 {\cal K}\int_0^{z_\Lambda} dz e^{-\frac12 \mu^2 \tau^2}
\Big[ {1\over g_{V}^{4}}g_{xx}\tilde g_{zz}^{-{1\over 2}}(\partial_{z} \varphi_n(z))^{2}
+ {4\lambda \over 2 \pi \alpha'}\tau^2 g_{xx}
\tilde g_{zz}^{1 \over 2} (\vartheta_n(z)-\varphi_n(z))^2 \Big]=\frac12\,\,.
\ee
Rewriting this expression in terms of $\psi^{P}$, we find:
\bear
\frac12=\frac{(2 \pi \alpha')^2 {\cal K}}{g_V^4}\int_0^{z_\Lambda} dz \, e^{\frac12 \mu^2 \tau^2}
g_{xx}^{-1}\left(\tilde g_{zz}^{\frac12} \psi^{P}_n(z)^2 +
\frac{2\pi\alpha'}{4\lambda g_V^4 \tau^2}\tilde g_{zz}^{-\frac12}(\partial_z \psi^{P}_n(z)
)^2\right)=\rc
=\frac{(2 \pi \alpha')^2 {\cal K}}{g_V^4}
\int_0^\infty du \left(
g_{xx} \tau^2 \alpha(u)^2 +
\frac{2\pi\alpha'}{4\lambda\,g_V^4}
e^{\frac12 \mu^2 \tau^2} g_{xx}^{-\frac32} \tau^{-2}
\left[ \partial_u \left(
e^{-\frac14 \mu^2 \tau} g_{xx}^{\frac34} \tau \alpha(u)
\right)\right]^2
\right)
\label{normpseudo}
\eear
where in the last line we have changed to the Schr\"odinger variables following the conventions
of appendix \ref{app: schro}.
There
are some subtleties related to the UV behaviour of
(\ref{normpseudo}) which are worth explaining. Since the leading UV behaviour of our model is the same as in
 the hard wall \cite{Erlich:2005qh}, \cite{Da Rold:2005zs} or soft wall
\cite{Karch:2006pv} models, the following arguments are analogous in all these cases. However, we
are unaware of any reference where the discussion below is explicitly shown.
It turns out that this UV behaviour is qualitatively
different for massless ($c_1=m_q=0$) or massive ($c_1\sim m_q >0$) quarks. We will
 study both cases separately below.

\subsubsection{The $m_q=0$ case}

We will now study the qualitative properties of the physical spectrum for $m_q=0$.
In this case, the $\tau \sim u^3$ near the UV and therefore $\Xi=(AB)^{\frac14}
\sim u^{-\frac32}$, which implies that
\be
V_{UV}(u)=\frac{15}{4u^2}+\dots
\ee
The first correction represented by the dots comes at order $u^3$.
One can then find the UV expansion for $\alpha(u)$ that solves (\ref{schroeqap})
in terms of the two integration constants which we denote $k_1,k_2$ as
$\alpha(u)=k_1 u^{-\frac32} + \frac14 k_1 m_n^2 u^{\frac12}
- \frac{1}{16} k_1 m_n^4 u^{\frac52} \log u
+ k_2 u^{\frac52} + \dots $ We now want to insert this in the last line of
(\ref{normpseudo}) and check whether the integral converges near $u=0$. The first term
is always convergent so we focus on the second term which behaves as
$\int_0 du u^{-3} [\partial_u (u^{\frac32} \alpha(u) )]^2$. Therefore, for $m_n=0$, this
mode is UV-normalizable irrespective of the values of $k_1$ and $k_2$. Thus, one can always
tune $k_2/k_1$ in order to find a solution that is well-behaved in the IR. This means that
for $m_q=0$ there is always a normalizable solution with $m_n=0$, which corresponds to the
expected massless Goldstone boson. On the other hand, if $m_n \neq 0$, one has to impose
$k_1=0$ in order to have UV-normalizability. Then, as in a standard Sturm-Liouville problem,
there will be a discrete set of massive modes, where $m_n$ is selected by matching the
normalizable UV and IR behaviours.

In summary, for $m_q=0$ the UV structure of the Schr\"odinger potential and normalizability
condition ensures the existence of
a massless Goldstone boson
together with a discrete tower of massive excitations, as expected.

\subsubsection{The $m_q \neq 0$ case}

Near the UV, we now have $\tau \sim u$  and therefore $\Xi=(AB)^{\frac14}
\sim u^{\frac12}$, and
\be
V_{UV}(u)=-\frac{1}{4u^2}+\dots
\ee
where the first correction in the dots is ${\cal O}(u^0)$.
We can find again the UV solution in terms of two integration constants
$\alpha(u)=k_1 u^{\frac12} \log u+ k_2 u^{\frac12} + {\cal O}(u^{\frac52})$.
Now, requiring that the last term of (\ref{normpseudo}) is UV-finite requires setting
$k_1=0$. Again, one has a Sturm-Liouville problem with a discrete
spectrum.

\begin{figure}[ht]
\centering
\includegraphics[width=.45\textwidth]{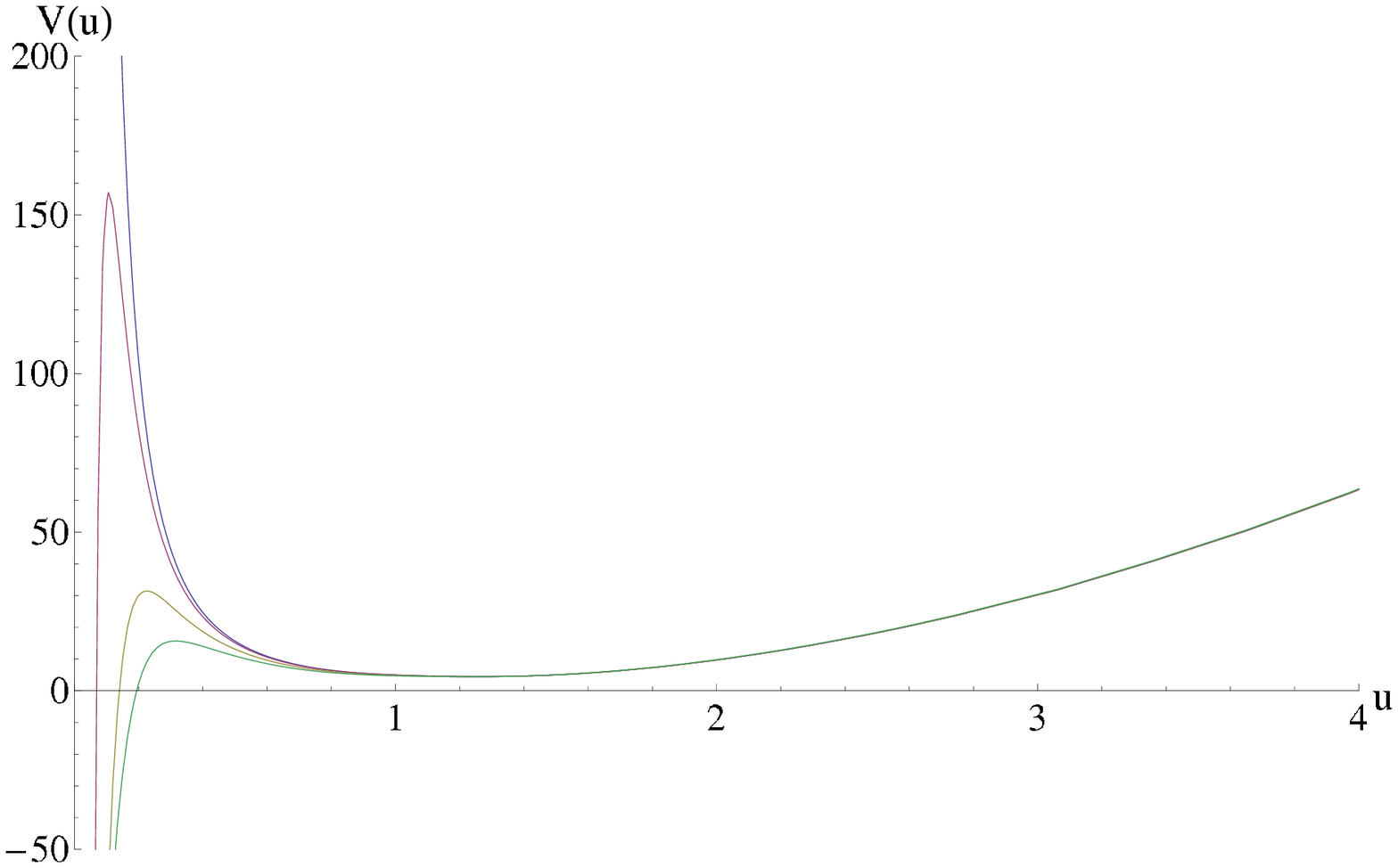}
\includegraphics[width=.45\textwidth]{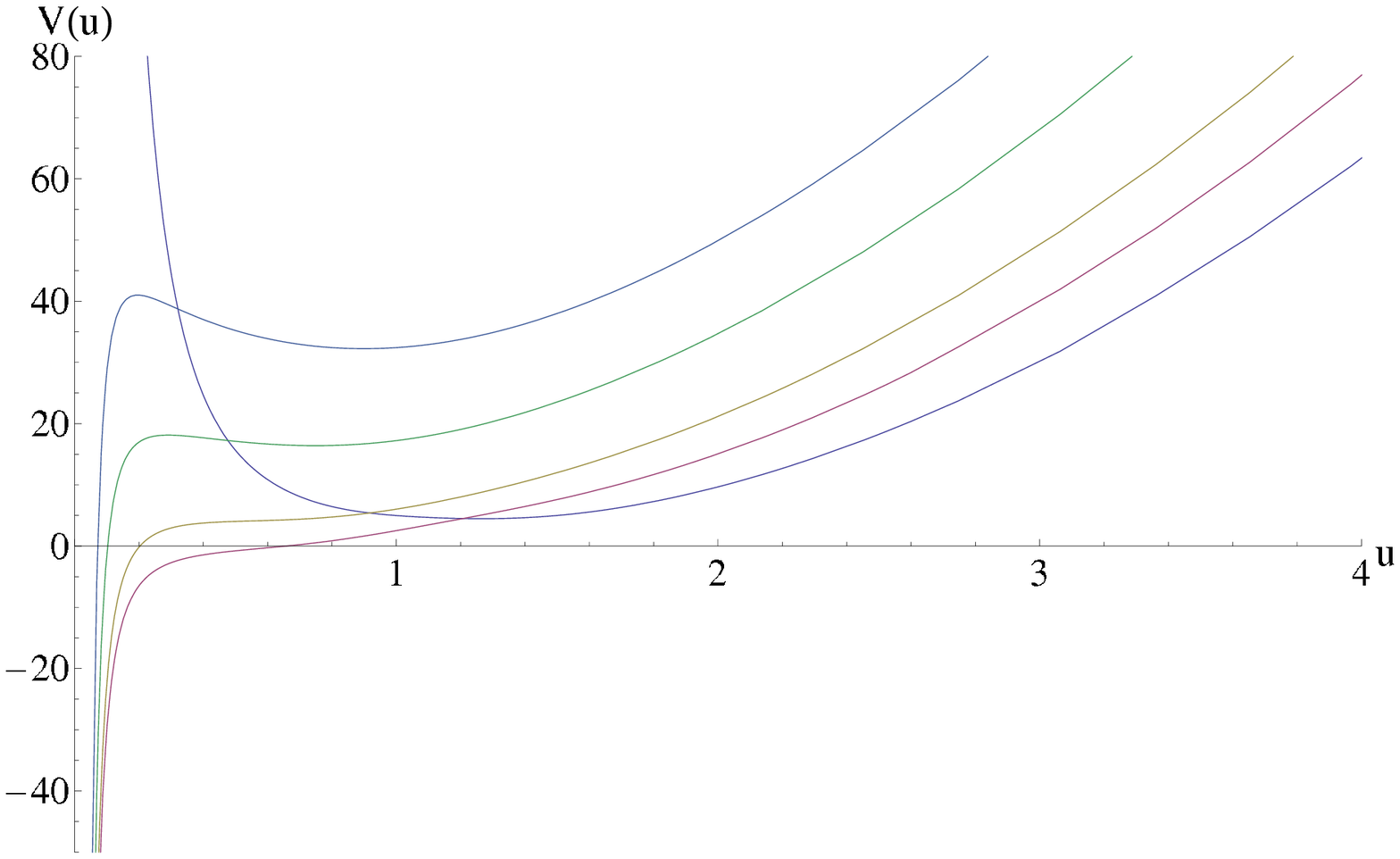}
	\caption{On the left, we plot the associated Schr\"odinger potential for
	small quark mass, concretely $c_1=0,0.001,0.005,0.01$. Even if the UV behaviour
	 is completely different for $c_1=0$, we see that for small $c_1$, the potential
	looks very similar to the massless case except precisely around $u=0$.
	On the right, we plot the same but with larger values of $c_1$, in particular
	$c_1=0,0.5,1,2,3$. All plots have been done taking $k=\frac{12}{\pi^2}$. As for
	the axial case, increasing $k$ amounts to pushing up the IR part of the potential.
	}
	\label{fig:schrops}
\end{figure}

Figure \ref{fig:schrops} depicts a few Schr\"odinger potentials for the pseudoscalar mode, for
different values of $c_1$.

\begin{figure}[ht]
\centering
\includegraphics[width=.45\textwidth]{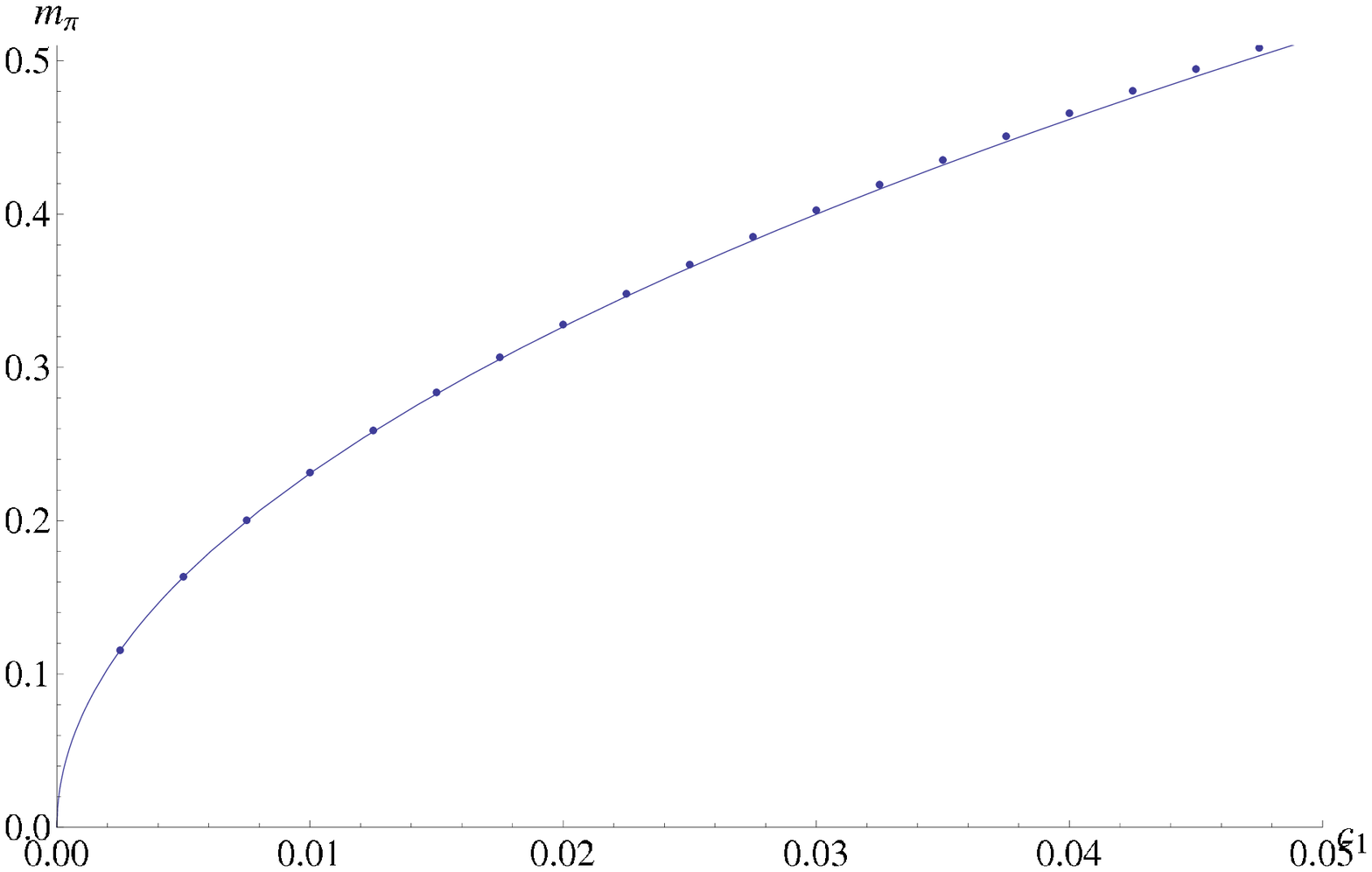}
\includegraphics[width=.45\textwidth]{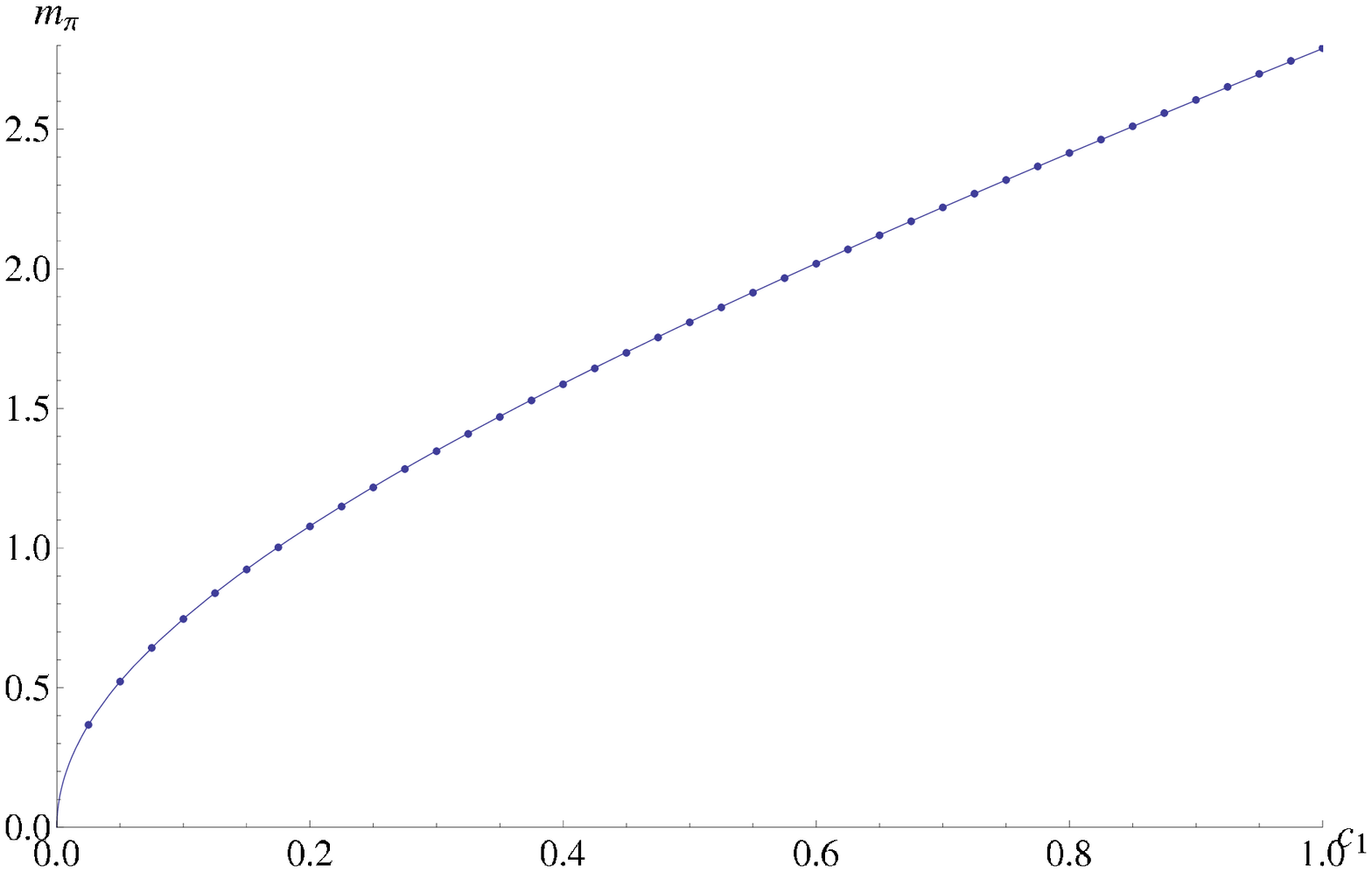}
	\caption{We plot the mass of the lowest lying pseudoscalar as a function of $c_1$
	(namely, the quark mass). On the
	left, we crosscheck the GOR relation (where for $\langle q\bar q \rangle$,
	$f_\pi$ we have introduced the result found numerically in the chiral limit) to some points computed numerically.
	The approximation is very good for small $c_1$, and deviations start seeing visible around $c_1=0.03$.
	On the right, we have fitted $b$ in the expression $m_\pi = \sqrt{\frac{- 4 m_q \langle q\bar q \rangle}{f_\pi^2}
	+ b \,m_q^2}$ and checked that the fit is rather good up to $c_1=1$.
	the parameter $k$ has been taken to be $\frac{12}{\pi^2}$  in these plots.
	}
	\label{fig:GMOR}
\end{figure}

Let us now look at the lowest-lying excitation when $m_q$ is small. This should be
a pseudo-Goldstone boson with its mass given by the Gell-Mann-Oakes-Renner relation.
One indeed can find this following the argument of \cite{Erlich:2005qh}:
for zero quark mass, the $q^2=0$ solution of (\ref{pseom1}), (\ref{pseom2}) is
given by $\vartheta(z)=-1$, $\varphi(z)=\psi^A_{q^2=0}(z)-1$, where we have defined
$\psi^A_{q^2=0}(z)$ as the solution at zero momentum of (\ref{axialeom}) with boundary
condition $\psi^A_{q^2=0}=1$. Consequently, regarding (\ref{fpi}),
$z^{-1} \partial_z \psi^A_{q^2=0}|_{z=0}=-\frac{6\pi^2}{N_c}f_\pi^2$.
We now find perturbatively the small $m_\pi^2$ solution just by integrating
(\ref{pseom2}) and we obtain:
\be
-1= m_\pi^2 \int \frac{z^2}{\mu^2 k \tau^2}\partial_z \psi^A_{q^2=0} dz
\ee
Using that this integral is dominated by the small $z$ region and taking into account
$\int \frac{z^3}{\tau^2} \approx \int \frac{z^3}{(c_1 z + c_3 z^3)^2} \approx \frac{1}{2c_1c_3}$,
we can substitute the relations between $c_1,c_3$ and $m_q,\langle \bar qq\rangle$
(\ref{mqc1}), (\ref{quarkcon}) together with (\ref{dimcon}), (\ref{gvqcd}) and (\ref{kaval})
to find the GOR relation \cite{GellMann:1968rz}:
\be
- 4 m_q \langle q\bar q \rangle = m_\pi^2 f_\pi^2
\label{GOR}
\ee
We have obtained this expression by making a series of approximations. However, we can crosscheck
it with the values for the mass obtained by the standard numerical computation, see figure
\ref{fig:GMOR}.
(\ref{GOR}) is very accurate for small masses.
When going to larger masses (up to $c_1\approx 1$), we can fit the mass of the
lowest lying pseudoscalar to $\sqrt{a \,m_q + b \, m_q^2}$. We include here the masses of the first six pseudoscalar modes in terms of $c_1$
\bear
z_\Lambda\,m_P^{(1)} &\approx& \sqrt{3.53 c_1^2 + 6.33 c_1} \,\,,\qquad
z_\Lambda\,m_P^{(2)} \approx 2.91 + 1.40 c_1 \,\,,
\qquad
z_\Lambda\,m_P^{(3)} \approx 4.07+ 1.27 c_1 \,\,, \qquad (c_1\leq 1) \rc
 z_\Lambda\,m_P^{(4)} &\approx& 5.04 + 1.21 c_1 \,\,,\qquad\
 z_\Lambda\,m_P^{(5)} \approx 5.87 + 1.17 c_1 \,,\,\,\qquad
 z_\Lambda\,m_P^{(6)} \approx 6.62 + 1.15 c_1\
.
\label{psmassesfit}
\eear
where we have also set $k={18\over \pi^2}$ for this calculation.

\subsection{Mesonic excitations: a brief phenomenological analysis}
\label{sec: pheno}

We now make a phenomenological analysis of our model by comparing our
results for the spectrum and the decay constants of {\it light unflavored
mesons} to their experimental values. An extensive study of meson
spectrum appeared in \cite{Iatrakis:2010zf}, without
including the decay constants. We will fit the three parameters
of the model,  $z_{\Lambda}$, $c_1$ and $k$, using mesons
with isospin 1 and $J^{PC}=1^{--}, 1^{++}, 0^{-+}, 0^{++}$.
Since $z_\Lambda \sim \Lambda_{QCD}^{-1}$ and $c_1 \sim m_q$, it turns out that there is
 a single phenomenological parameter $k$, apart from those inherent of QCD physics.

The experimental values of the meson masses
which are used are quoted by \cite{Amsler:2008zzb}.

We fit the three parameters of our model to the masses of the light
mesons which appear in table [\ref{tablemass}] and the decay constants
appearing in [\ref{tabledeccon}]. To make the fit we minimize the rms
error

\be
\epsilon_{rms}= \left( {1\over n}\sum_{i}\left( \delta O_i \over O_i\right)^2
\right)^{1\over 2}
\ee
where $n$ is the number of the observables minus the number of the
fitted parameters, $n=9-3$. The values of the parameters minimizing $\epsilon_{rms}$ read

\be
z_{\Lambda}^{-1}=549~{\mbox MeV} \sp c_{1 l}z_\Lambda =0.0094 \sp k={18 \over \pi^2}
\label{1fit}
\ee
The rms error then is $\epsilon_{rms}=14.5 \%$ and the comparison
between the experimental and model values appears in table (\ref{tablemass}),
for the masses and in table ({\ref{tabledeccon}}), for the decay constants.

\begin{table}[h]
\begin{center}
\begin{tabular}{|c|c|c|c|c|}
\hline
$J^{CP}$ & Meson & Measured (MeV) & Model (MeV) & $100 |\delta O| / O$ \\
\hline
$1^{--}$ & $\rho(770)$  & 775 &  800 & 3.2\% \\
\cline{2-5}
 & $\rho(1450)$  & 1465 &  1449 & 1.1\% \\
\cline{1-5}
$1^{++}$ & $a_1 (1260)$  & 1230 &  1135 & 7.8\% \\
\cline{1-5}
$0^{-+}$ & $\pi_0$  & 135.0 &  134.2 & 0.5\% \\
\cline{2-5}
 & $\pi(1300)$  & 1300 &  1603 &23.2\% \\
\cline{1-5}
$0^{++}$ & $a_0(1450)$  & 1474 &  1360 & 7.7\% \\
\hline
\end{tabular}
\end{center}
\caption{The results of the model and the experimental
  values for light unflavored meson masses.}
\label{tablemass}
\end{table}

\begin{table}[h]
\begin{center}
\begin{tabular}{|c|c|c|c|c|}
\hline
$J^{CP}$ & Meson & Measured (MeV) & Model (MeV) & $100 |\delta O| / O$ \\
\hline
$1^{--}$ & $\rho(770)$  & 216 &  190 & 12\% \\
\cline{1-5}
$1^{++}$ & $a_1 (1260)$  & 216 &  228.5 & 5.8\% \\
\cline{1-5}
$0^{-+}$ & $\pi_0$  & 127 &  101.3 & 20.2\% \\
\hline
\end{tabular}
\end{center}
\caption{A comparison of the results to the experimental values for
  the decay constants of light unflavored mesons.}
\label{tabledeccon}
\end{table}

\section{Meson melting in the deconfined phase}
\label{sec:melting}

We briefly discuss in this section the fate of the mesonic modes when the gauge theory
undergoes a deconfining phase transition \cite{Hoyos:2006gb}, namely when we use the background of
equation (\ref{adsbh6b}). The first observation is that, as we saw in section
\ref{secdeconf}, the tachyon cannot diverge at any point in this case and, therefore,
the brane reaches the horizon, and we only have ``black hole embeddings", in analogy
with the terminology introduced in \cite{Mateos:2006nu}. This means that there is no
discrete spectrum above the deconfining phase transition.
When we are considering small quark masses, this is perfectly realistic.

However, in the real world,
charmonium and bottomonium
do survive the QCD phase transition.
We want to study this problem in the present model, and therefore we will compute the
spectral functions at different values of $m_q/T$. In particular, we will focus in the
vector excitation.

\begin{figure}[ht]
\centering
\includegraphics[width=.29\textwidth]{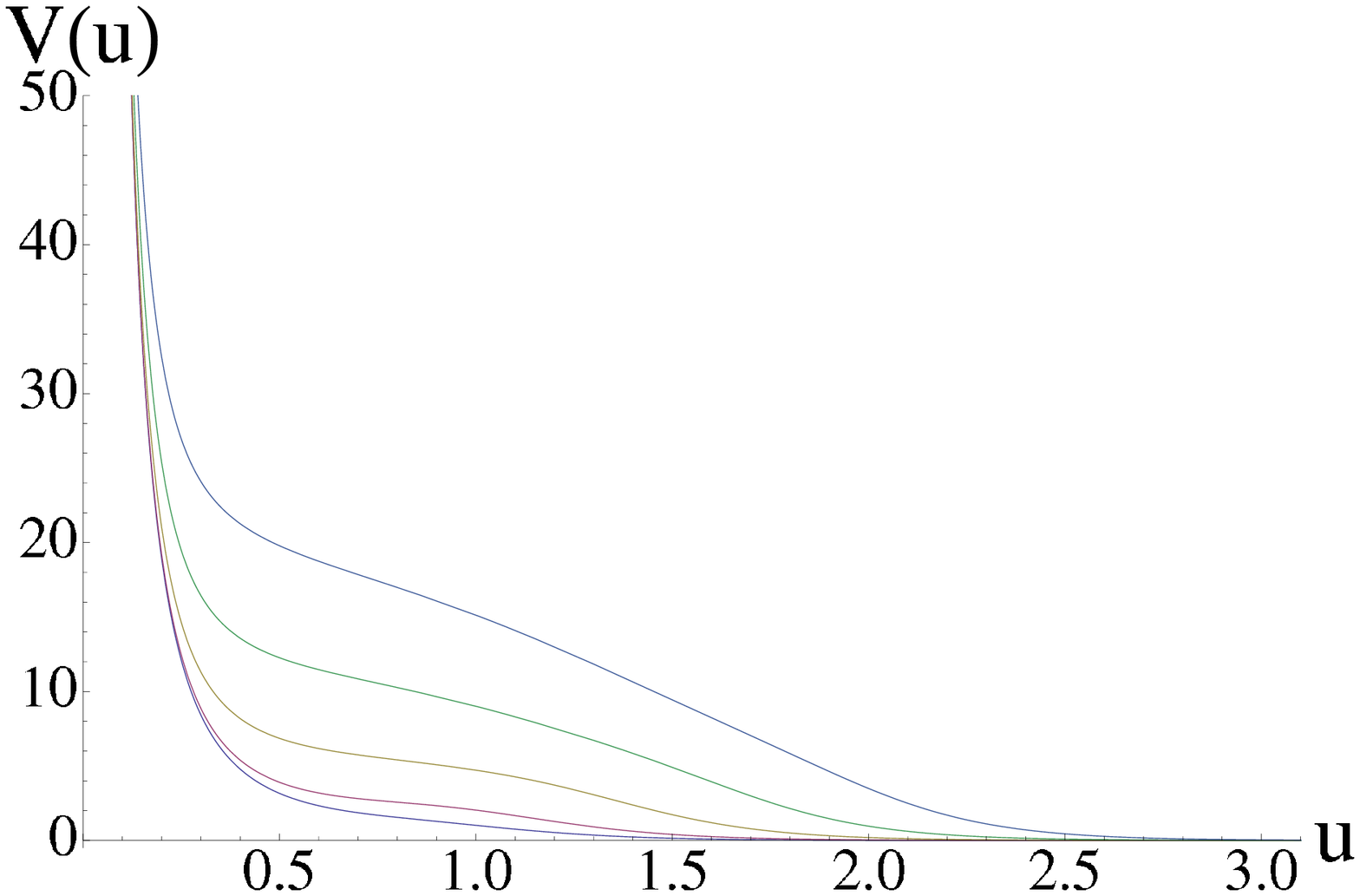}\quad
\includegraphics[width=.29\textwidth]{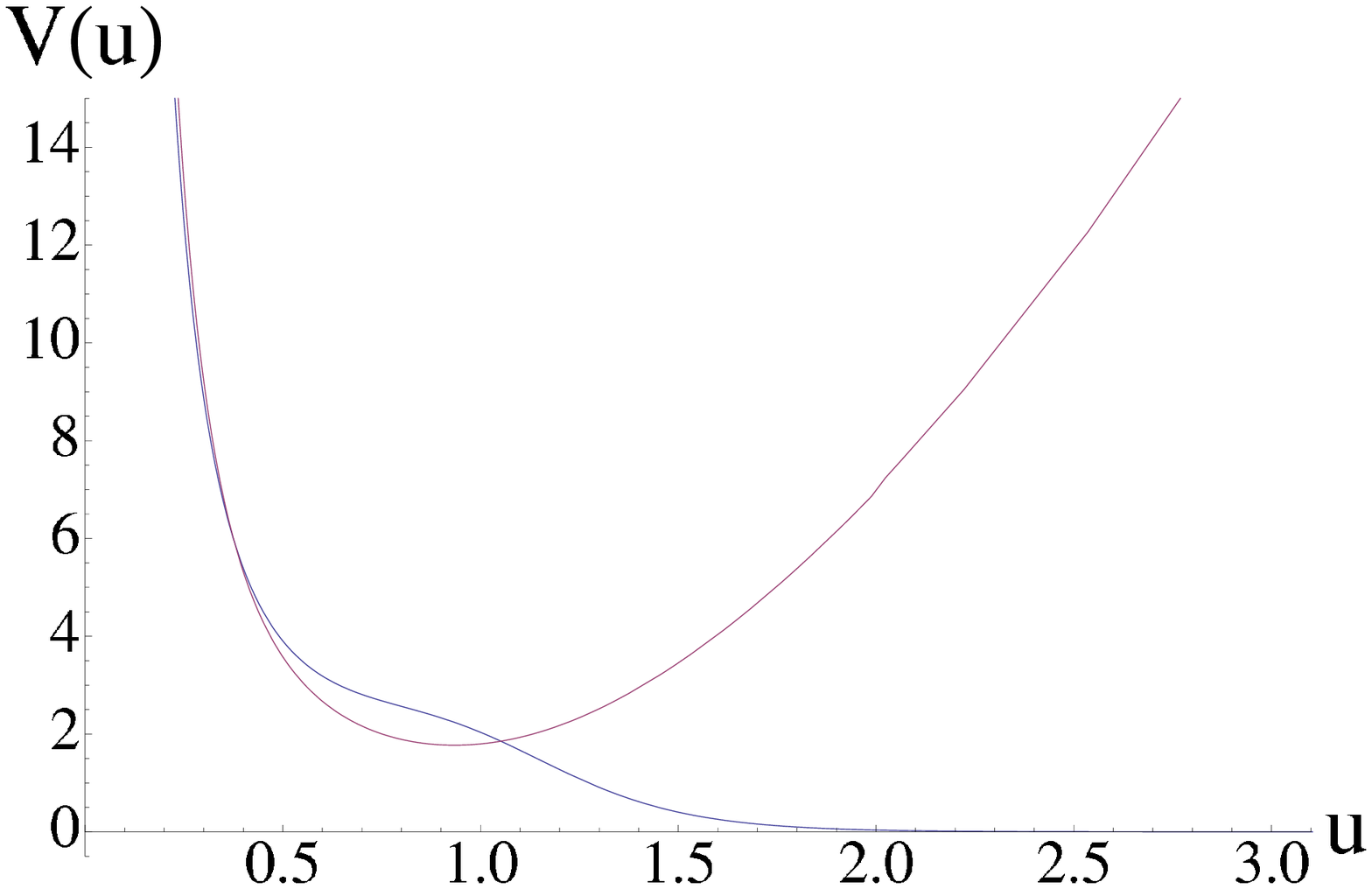}\quad
\includegraphics[width=.29\textwidth]{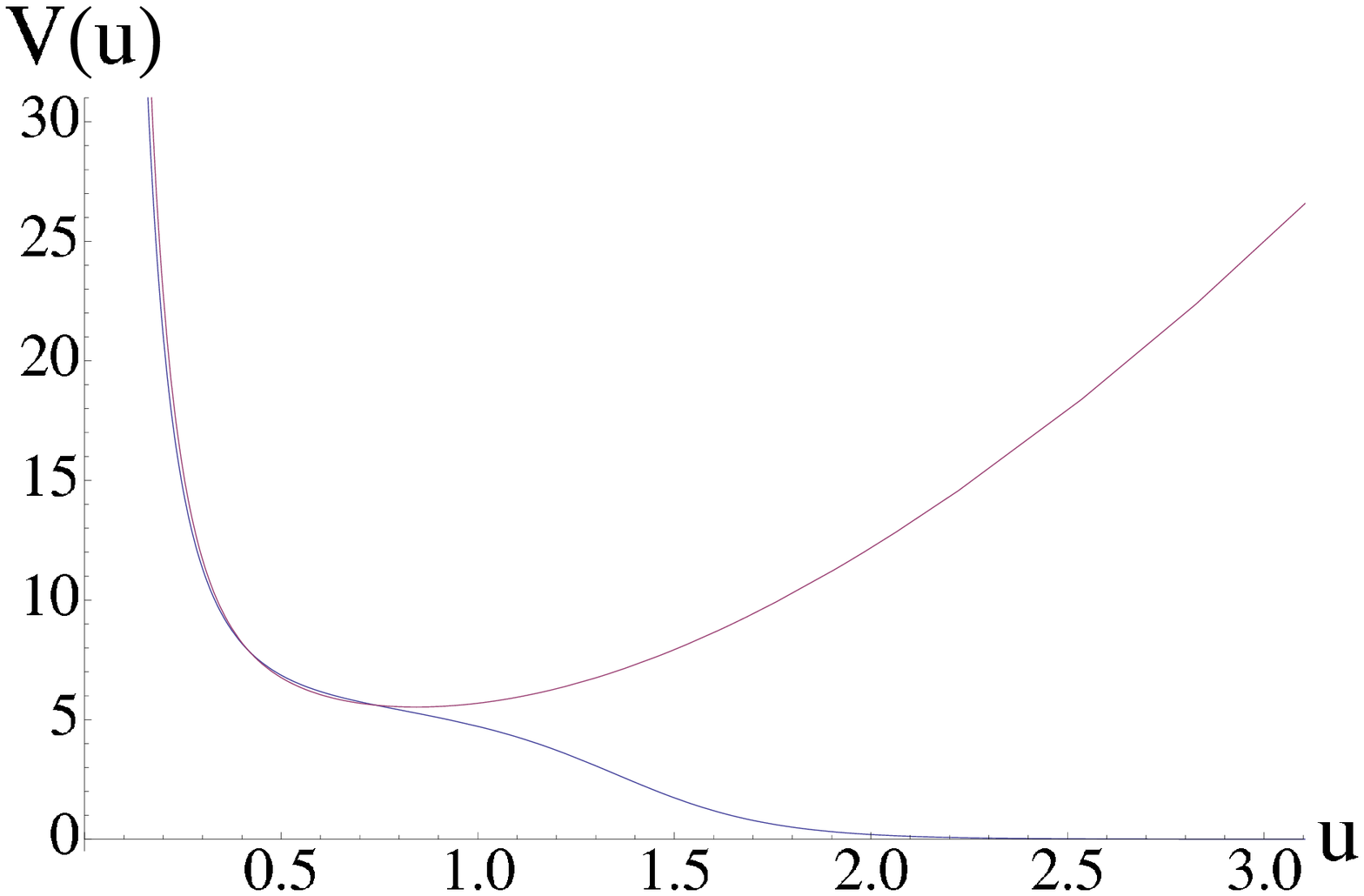}
	\caption{The Schr\"odinger potentials associated to the vector excitation in the deconfined
	phase, at zero momentum, for different values of $c_1 \sim m_q/T$. In the first plot $c_1=0.01,1,2,3,4$.
	The second and third plot (respectively $c_1= 1,2$) make a comparison with the potentials
	in the confined phase for the same values of $c_1$.
	}
	\label{fig:schrodec}
\end{figure}

We start by discussing the associated Schr\"odinger potential for the vector excitation
in the deconfined background,
at zero momentum. The expressions in (\ref{ABdefs}) are modified to:
\begin{equation}
A(z)=e^{-\frac12 \mu^2 \tau^2}
g_{xx}^{\frac12} g_{tt}^{\frac12}
\tilde g_{zz}^{-\frac12}\,\,,\qquad
B(z)=e^{-\frac12 \mu^2 \tau^2} g_{xx}^{\frac12} g_{tt}^{-\frac12} \tilde g_{zz}^{\frac12} \,\,,
\label{ABdefsdeconf}
\end{equation}
where one should remember that now $g_{\mu\nu}$ refers to the metric (\ref{adsbh6b}).
Notice that $\sqrt{B/A}$ diverges at $z=z_T$ as a single pole, such that $\int \sqrt{B/A} dz$
diverges and the horizon $z=z_T$ corresponds to $u=\infty$ in the Schr\"odinger coordinate.
$V(u)$ is exponentially decreasing for large $u$.
For completeness, we write in appendix \ref{app:deconf} the functions determining the
potential for the rest of modes.
In figure \ref{fig:schrodec}, we show several examples of potentials computed numerically,
for different values of  $c_1\sim m_q z_T \sim m_q/T$.
In the second and third plots, we also compare it to the potentials
in the confined phase for the same value of $c_1$. Namely, we show how the potentials for the
vector excitations are modified at the phase transition. They share the same UV behaviour, but
are drastically modified in the IR due to the different behaviour of the tachyon and the metric.

In \cite{Paredes:2008nf}, it was shown that a step-like potential gives quasi-particle behaviour.
Moreover, if the position of the step coincides with that of a barrier in the confined phase, the
quasi-particle mass is related to the mass of the meson before the phase transition. The third plot
seems to point along that line. However, in the present model the potentials
for the deconfined phase
are never
step-like enough, nor present bumps, and thus do not
create sharp peaks in the spectral function, as we will see below.

\begin{figure}[ht]
\centering
\includegraphics[width=.45\textwidth]{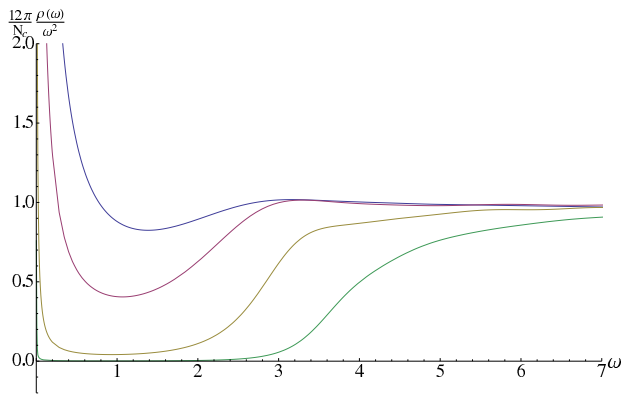}
	\caption{The spectral function (divided by $\omega^2$) in
          units of  ${N_c \over 12 \pi}$ for $c_1=0.01,1,2,3,4$, from top to bottom.
	}
	\label{fig:spectral}
\end{figure}

Once we have the potentials, it is straightforward to compute the spectral function from the
retarded correlator which is computed following the
prescription of \cite{Son:2002sd}.
In practice, what one has to do is to impose ingoing boundary conditions at the horizon
($\alpha(u) \sim e^{i\,\omega\, u}$ near $u=\infty$)\footnote{Even if we use the notation with
$u$ since it is better for illustrative purposes, we have found easier to perform the numerical
computations in the $z$-coordinate.} and find the behaviour of the wavefunction near the boundary,
namely compute $b_1,b_2$ matching the numerical result to
 the UV-expansion (\ref{psiVUV}). Then, the spectral function is given by:
\be
\rho(\omega)=- {\rm Im}~G_{R}(\omega)={N_{c}\over 3 \pi^2}{\rm Im} \left(
  {b_2\over b_1} \right)
\ee
where we have replaced the parameters of our model by $N_c \over 3 \pi^2$, using
(\ref{gvqcd}).
We show in figure \ref{fig:spectral} some examples of this
computation. We have plotted ${12 \pi \over N_c}{\rho(\omega) \over
  \omega^2}$ in terms of $\omega$ for various values of $c_1$.

In \cite{Grigoryan:2010pj}, a bottom-up
holographic model was built in order to discuss the physics of the $J/\Psi$ above the
phase transition (see also \cite{Fujita:2009ca} for related work).
In the model of \cite{Grigoryan:2010pj}, there is also no discrete spectrum but, for low enough
temperature, by engineering the potential, the $J/\Psi$
shows up as a peak in the spectral function (or, alternatively, as a small negative
imaginary part of the associated quasinormal frequency). Namely, it has quasi-particle
behaviour.

We now discuss the results comparing to \cite{Grigoryan:2010pj}. The authors of  \cite{Grigoryan:2010pj} showed how the Schr\"odinger potential should
look like to find some qualitative physical properties of the $J/\Psi$ meson. In our model,
the potentials are found dynamically, and look similar to \cite{Grigoryan:2010pj}
except for the important fact that they fail to have the dip of \cite{Grigoryan:2010pj}.
As a consequence, the $J/\Psi$ decay constant is too low
(see figure \ref{figdecay}) and a clear peak in the spectral function
is absent. One may compare our figure \ref{fig:spectral} to figure  3  in \cite{Grigoryan:2010pj}.

It does not seem possible to introduce the dip feature in our model without inserting at least a new scale in
the problem. Notice that the $\Lambda_{QCD}$ does not play any role in the deconfined case
for the present model, apart from setting the value of $T$ at which the transition takes place.
In other words, the $z_\Lambda$ does not appear at all in the deconfined background.
We speculate that one should modify the background by including some dependence on $\Lambda_{QCD}$
in the deconfined phase
 in order to improve the  model in these respects.
It would
be interesting to investigate whether analyzing the tachyon action in
the less simple but well motivated backgrounds of
``improved holographic QCD" models \cite{review}
 might ameliorate the issue along this path.

\section{Discussion}
\label{sec:discussion}

\subsection{Summary and general comments on the model}
\label{sec:summary}

In this paper, we have analyzed in detail several issues of the holographic model presented in
\cite{Iatrakis:2010zf}. The main ingredient is to describe the open string
(meson) physics by using tachyon condensation, captured by Sen's action \cite{Sen:2003tm}, as first advocated in \cite{ckp}.
The ``tachyon", namely the lowest lying bifundamental scalar of the brane-antibrane system,
is dual to the quark mass scalar bilinear operator and its condensation corresponds to chiral symmetry
breaking. In order to make explicit computations, it is necessary to give an explicit form
of the tachyon potential and to choose a curved gravitational background. We have considered
extremely simple possibilities for both, see (\ref{tachyonpot}) and (\ref{adsbh6}).

As we have already remarked, the model is inspired by string theory but is phenomenological.
It does not provide a well
controlled approximation to string theory in any limit. In this sense, it should be considered
a bottom-up model.
Nevertheless,
the main point we want to make in this work is that it can be
very useful to incorporate
top-down derived ingredients as (\ref{generalact}) into bottom-up models.
We will make a comparison between qualitative properties of our model and the soft wall model \cite{Karch:2006pv} and modifications thereof.
 Successes of \cite{Karch:2006pv} include
a reasonable qualitative and quantitative matching of properties of
the measured mesonic states, including the Gell-Mann-Oakes-Renner relation and
the
Regge trajectories. These are also found in our model, together with the following extra appealing
properties, which are automatic in our general set-up:

\begin{itemize}

\item The model incorporates  confinement in the sense that the quark-antiquark potential
computed with the usual AdS/CFT prescription \cite{Sonnenschein:1999if} confines.
Moreover, magnetic quarks are screened.
The background solution stems from a gravitational action, that allows, for instance,
to compute thermodynamical quantities. All of this are properties associated to the
background geometry and were already discussed in  \cite{ks2}.

\item The string theory nature of the bulk fields dual to the quark bilinear currents is readily identified:
they are low-lying modes living in a brane-antibrane pair.

\item Chiral symmetry breaking is realized dynamically and consistently, because of the
tachyon dynamics.
See \cite{Sui:2009xe} for discussion and possible solutions in the soft-wall model
context.

\item In the present model, the mass of the $\rho$-meson grows with increasing quark mass, or,
more physically, with increasing pion mass. This welcome physical feature is absent in the soft wall model, \cite{Karch:2006pv}.
It occurs here  because the tachyon potential multiplies the full action and in particular
the kinetic terms for the gauge fields, which therefore couple to the chiral symmetry breaking vev.
In our previous work \cite{Iatrakis:2010zf}, we exploited this fact in order to fit the strange-strange mesons
together with the light-light mesons, with rather successful results.
In \cite{Shock:2006qy}, the authors added the strange quark mass to the hard
wall model and computed the dependence of vector masses on the quark mass. In that case however,  this dependence
of the vector masses originated  only from the non-abelian structure and therefore misses at
least part of the physics\footnote{On the other hand, the quark mass dependence of the $\rho$-meson 
can be seen in different top-down models, see \cite{cdk} for a recent work in the context of the
Klebanov-Strassler model.}.

\item The soft wall requires assuming a quadratic dilaton in the closed string theory background.
It has been shown that such a quadratic dilaton behaviour can never be derived from a gravitational action while keeping
the geometry to be that of AdS.\footnote{This was shown in the second reference of  \cite{ihqcd2}. In
\cite{ihqcd} such behavior can be implemented for glue, but the metric changes appropriately, an important ingredient for implementing confinement in the glue sector.}.
That the background is not found as a solution is a shortcoming if for instance one wants to study the
thermodynamics of the underlying glue theory. The thermodynamics of the soft wall model is therefore ill-defined.
 In the present  model, we found the background as a solution
of a two-derivative approximation to non-critical string theory, see section \ref{sec: backgr}.
In order to obtain Regge behaviour, we also needed a further  assumption: that the
tachyon potential is asymptotically gaussian. However, this is rather natural since this potential has appeared
in the literature, for instance \cite{Kutasov:2000aq,Takayanagi:2000rz}. Still, we warn the reader that the
formalism of \cite{Kutasov:2000aq,Takayanagi:2000rz} cannot be directly and controllably connected to the present setup.

\item Considering that the dynamics is controlled by a tachyon world-volume action automatically
provides the model with a WZ term of the form given in
\cite{Kennedy:1999nn,Kraus:2000nj,Takayanagi:2000rz}. We have not discussed this term at all in this
work, but in \cite{ckp} it was shown that properties like discrete symmetries (parity and charge
conjugation)
and anomalies are,
in general,
correctly described by analyzing this term.

\end{itemize}

In summary, we regard our model as being in the general framework of \cite{Karch:2006pv},
but with several qualitative improvements due to the dynamics built in the action (\ref{generalact}).
Moreover, our starting assumptions are rather simple and well motivated from a top-down
perspective, such that the {\it ad hoc} input is scarce.
It is also encouraging to find that
quantitative fits to experimental data are reasonably good, see \cite{Iatrakis:2010zf} and
section \ref{sec: pheno}, but those are not the main aim of the present work.

We have also found several aspects in which the model does not capture features which are known
from perturbative QCD.

The main issue is the leading large $q^2$ contribution to $\Pi_V(q^2)-\Pi_A(q^2)$ as will be discussed in
section \ref{sec: OPEs}. We have also seen that our model does not work that  well for large quark
masses\footnote{Perhaps this is not surprising since for heavy quarkonium perturbative methods and
in particular non-relativistic QCD (see \cite{Brambilla:2004jw} for reviews)
are accurate and it may
be naive to expect that a dual theory can be a good approximation to the physics.}
as it would grossly underestimate the decay constants for charmonium and no clear peaks are
observed in the corresponding spectral functions in the deconfined phase (see the discussion of
section \ref{sec:melting}). It would be interesting to know whether mild corrections of the model
could ameliorate these issues or whether these are unsurmountable differences of models of this
class to actual QCD physics.

The behaviour of $\frac{d\langle \bar qq\rangle}{dm_q}$
for small $m_q$
(figure \ref{div_tac_fig_2}) has been investigated in  earlier classic works \cite{svz}.
However, the leading IR divergence-free correction, is of order $1/N^2$ at large N, originating from a pion loop, \cite{pich}.
The leading-N corrections come from the four-derivative terms and are dominated by the scalar meson contribution.
The behavior is qualitatively similar to what we find.
For large $m_q$ we have not been able to calculate the asymptotic behavior but generically speaking we do 
not expect it to necessarily match that of QCD, as discussed in \cite{svz}.
The reason is that the UV asymptotics of the bulk gravity solution are not necessarily the same as in QCD.

\subsection{Comments on effective actions for the open string tachyon}
\label{othertachyons}

The notion that a scalar bifundamental in a brane-antibrane system should be
the holographic dual of QCD-like chiral symmetry breaking is rather simple
and robust, see for instance
\cite{Sugimoto:2004mh,BCCKP,Cotrone:2007gs,Aharony:2008an,k}.

What is not obvious, however, is which effective action is best in order to describe
a brane-antibrane system if curved spacetime. We have used the simple proposal
of \cite{Sen:2003tm}, but one should keep in mind that other alternatives might also
be useful. We provide here a short guide to the literature on the issue.

Garousi and collaborators, starting from
the early work \cite{Garousi:2000tr}, have tried to use explicit string theory computations
in order to constrain the tachyon generalization of the DBI and WZ actions \cite{Garousi:2007fk},
see also \cite{Gutierrez:2008ya}.
In \cite{Garousi:2007fn}, an action for a D$p$-$\bar{{\rm D}p}$ system based on a particular
symmetriced trace prescription was proposed\footnote{In \cite{Garousi:2007fn}, a trace is needed
even for a single brane-antibrane pair since the degrees of freedom are 2$\times$2 matrices.
In order to non-abelianize the flavor group in our case, one should also deal with the problem
of how to implement traces on a non-abelian generalization of Sen's action (\ref{generalact}).
Investigating the physical consequences on the dual theory of this non-abelianization and
of different proposals for the effective action \cite{Garousi:2007fk} would be very interesting, but is
beyond the scope of the present work.}. There are subtle differences between the proposal of
\cite{Garousi:2007fn} and the one by Sen \cite{Sen:2003tm}, which may have dramatic consequences.
In fact, we have checked that using the symmetrized trace action of \cite{Garousi:2007fn} for our
model, one still finds Regge trajectories for vector and axial mesons but the slope for the axials
changes, see appendix \ref{garousiaction}. The study of other physical properties inferred from
the action of
\cite{Garousi:2000tr} is beyond the scope of the present work.

In \cite{garousi1}, it was discussed how to take into account the brane-antibrane distance in
the string action. This is important for holographic duals, specially if one wants to insist in
top-down approaches. A reason is that, in the weakly coupled picture, if one has brane and antibrane
on top of each other in flat space, the tachyon would create a real instability (it cannot be compensated
by the AdS curvature). Therefore one should think of separated branes as in the Sakai-Sugimoto model
\cite{Sakai:2004cn}. With this in mind, generalizations of \cite{Sakai:2004cn}, based on the action
proposed in \cite{garousi1} where constructed in \cite{Bergman:2007pm}.

In a beautiful recent paper \cite{Niarchos:2010ki}, building on the work
\cite{Erkal:2009xq}, Niarchos proposed a different way of building the effective action which
should better capture the physics of
a separated brane-antibrane system. This may be useful in holographic modeling and in particular
to improve the Sakai-Sugimoto model.
We also refer the reader to \cite{Niarchos:2010ki} for a more exhaustive overview of the
literature of tachyon effective actions.

\subsection{On the OPEs and the slope of the Regge trajectories}
\label{sec: OPEs}

There has been some debate in the literature on whether the large Euclidean
momentum
behaviour of two-point correlators can be used to constrain the behaviour of the QCD
spectrum of excited mesons. The main point is to compare infinite sums like
(\ref{sumvectors}) (or, more precisely, differences of such sums: vector minus axial or
scalar minus pseudoscalar)
to the large $q^2$ behaviour expected from the operator product expansion (OPE).
In particular, there is the question of whether different Regge slopes in the
vector and axial (or scalar and pseudoscalar) channels
\be
(m^{V,A}_n)^2 \sim \Lambda_{V,A}^2 n + const\qquad \textrm{ for large $n$}
\label{mVAlVA}
\ee
are consistent with the OPEs.
Notice that this is a theoretical question, irrespective of the experimental observation
of the spectra. Let us give a brief and incomplete overview about the debate regarding this
issue.
 For instance, in \cite{Golterman:2001nk}, a model with $\Lambda_V\neq \Lambda_A$ was
put forward. Later, in \cite{Beane:2001uj}, it was claimed that this
model was inconsistent with the OPEs, but the arguments of \cite{Beane:2001uj}
were called into question in \cite{Golterman:2002mi}, due to subtleties in the regularization
of infinite sums like (\ref{sumvectors}).
More recently, works like \cite{Shifman:2005zn,Shifman:2007xn,Andrianov:2008ra} claim that
the Regge slopes should be equal whereas the opposite conclusion was reached in
\cite{Cata:2006fu,Mondejar:2007dz,Mondejar:2008dt}.

We have found above that in our model, there are different asymptotic Regge slopes
$\Lambda_A > \Lambda_V$. However, the coefficients of the leading
logarithms in the large $q^2$ correlator for vectors and axials coincide, consistently.
In order to illustrate this fact, let us remember that, asymptotically in the UV, our model resembles
the soft wall of \cite{Karch:2006pv}, see equation (\ref{largeq2eq}). In the soft wall
model, the Regge slope is controlled by the constant $c$ of (\ref{largeq2eq}), but the
quotient $F^2/\Lambda^2$ is independent of $c$ \cite{Cata:2006ak}. This quotient is indeed
what controls the coefficient of the leading logarithm \cite{Golterman:2002mi}.
We thus have $\lim_{q^2\to \infty} \left( \Pi_V (q^2) - \Pi_A(q^2) \right)=0$ together
with different Regge slopes. However, this is not enough to comply with the OPEs.
The leading contribution to $\left( \Pi_V (q^2) - \Pi_A(q^2) \right)$ at large $q^2$
should be of order $q^{-4}$ because QCD does not have dimensionful quantities that allow
to rewrite for instance a $q^{-2}$ term.
We can resort  to numerics to compute
$ \left( \Pi_V (q^2) - \Pi_A(q^2) \right)$ in our model and the result does not comply with
the $q^{-4}$ expectation. This fails in the axial channel as shown in appendix \ref{axcorrel}.

 The obvious guess is that, since our
holographic model is clearly not
exactly QCD, it includes operators or condensates which are absent in QCD,
modifying the subleading pieces of the correlators.

The same kind of problem is present for any holographic model we are aware of, see
\cite{Cata:2006ak} for a discussion concerning the soft wall\footnote{A.P.
thanks O. Cata for a discussion on this subject.}.
However, this seems to us more of a technical problem that may be resolved by finding the
appropriate potentials than a general obstruction to this class of models. Settling these issues
requires further work.

\subsection{Outlook}
\label{sec:outlook}

As we have discussed, our simple model is quite successful in describing many features of QCD.
A lot of effort has been devoted in bottom-up models to estimate other QCD related properties
as for example form factors, see for instance \cite{formfactors}. To reproduce such computations
in the present setting and compare the results is an interesting problem that we leave open
for the future. We have not studied non-trivial baryon number or chemical potentials, which
would clearly be worthy extensions of the model.

There are some aspects of the present that would be worth improving, like the physics
associated to heavy quarks (compared to the QCD scale). We have just explored the
result of working with Sen's action in the simplest confining holographic background available in the
literature \cite{ks1,ks2}. Therefore, it is still left to understand the consequences of implementing the tachyon action
in different backgrounds, as for instance those which go under the name of improved holographic QCD
\cite{ihqcd,review} or
modifications thereof.

It could also be interesting to try to introduce the quarks beyond the quenched approximation and therefore
compute the backreaction of the tachyon action for the fundamental fields on the gravity background.
For a review of unquenched flavor in critical (ten-dimensional) string theory backgrounds,
see \cite{Nunez:2010sf}.

Of course, it would be worth to provide a non-abelian generalization of this model,
for which one should provide a technical prescription on how to take traces in the action.
Another line of obvious interest would be to use the model for baryons. Since quark masses play
a more dynamical role than in other bottom-up approaches, this could be interesting for the physics
of the sigma-term.

Finally, and most importantly, we would like to point out that using an effective open
string action like (\ref{generalact}) in the
framework of holography can well have interesting applications beyond the realm of strong interactions.
For instance, many bottom-up technicolor models have appeared in the literature, see
\cite{Piai:2010ma} for a review. It is a very
interesting question to understand whether chiral symmetry breaking controlled by an action like
(\ref{generalact}) may offer new dynamical possibilities for the modeling of electroweak symmetry breaking.
On the other hand, in the last years, many phenomenological models have been constructed in order to address
some issues of superconductors and other condensed matter systems, for
a review see \cite{Hartnoll:2009sz}. Again, we would like to remark that (\ref{generalact}) could hopefully lead to interesting new dynamics in different set-ups.

\section{Acknowledgements}\label{ACKNOWL}

We would like to thank for useful conversations and correspondence.
R. Casero, O. Cata, A. Cotrone, S. Eydelman,
U. Gursoy, D. Mateos, V. Niarchos, F. Nitti, M. Panero, T. Pich, A. Pomarol, M. Shifman.

E. Kiritsis would like to thank  CERN,  ENS, ESI and Perimeter Institute for hospitality during the present work.
The research of A.P is
supported by grants FPA2007-66665C02-02 and DURSI
2009 SGR 168, and by the CPAN CSD2007-00042 project of the
Consolider-Ingenio
2010 program.
This work was  partially supported by  a European Union grant FP7-REGPOT-2008-1-CreteHEPCosmo-228644. I. Iatrakis was supported by an Onassis
graduate fellowship.

\newpage
\appendix
\section*{APPENDIX}
\setcounter{equation}{0}
\renewcommand{\theequation}{\Alph{section}.\arabic{equation}}
\addcontentsline{toc}{section}{APPENDIX\label{app}}

\section{Book-keeping summary of the parameters of the model}
\label{app:bookkeeing}

 We summarize here the parameters of the our model. There are two
 parameters coming from the background, the AdS radius $R$
 and the position of the cigar tip $z_{\Lambda}$.

  The action of the flavor brane-antibrane pair also includes
$\alpha'$ and two more parameters $g_V$ and $\lambda$ which are related to the
 normalization of the vector and tachyon fields respectively.
 The tachyon potential also includes two constants, ${\cal
   K}$ which is an overall factor in front of the action and $\mu^2$.  It should also noticed that
 $\mu$ can be absorbed in $\tau(z)$ by redefining $\tilde \tau \rightarrow \mu \tau$. Then this parameter disappears from all equations. We used $\mu^2=\pi$,
for the numerics through our analysis, but this does not
affect any physical results of our model.

Another parameter which exists in the model is
 $c_1$ which appears in the UV asymptotic of the tachyon expectation
 value (\ref{tauUVexpan}). $c_1$ is proportional to the quark mass, with
 a proportionality constant $\beta$ which was introduced in (\ref{mqc1}),
 however $\beta$ does not appear in the equations for the spectrum or
 the decay constants, so its value is not relevant for the model predictions.

  In
 total we have the following parameters $R$, $z_{\Lambda}$, $\alpha'$,
 $g_V$,  $\lambda$, ${\cal K}$ and $c_{1}$. $R$,
 $\alpha'$ and $\lambda$ are related by equation (\ref{dimcon}), which relates the
 tachyon mass to the  dimension of its dual operator. Then, we relate
 $g_V$ and $\lambda$ to the number of colors $N_c$ in QCD by matching
 the results of the vector and scalar two point functions as
 calculated in bulk on the one hand and in QCD on the other hand. The
 results are given in (\ref{gvqcd}), (\ref{lamqcd}), which relate
 $g_V$ to $\lambda$. Hence, finally the spectrum and the decay
 constants depend on  $z_{\Lambda}$, $c_1$ and a combination of $R$, $g_V$ and $\alpha'$
 which was named $k$ and is given by, (\ref{kaval}),
 \be
k=\frac{4R^4 g_V^4}{3 (2\pi\alpha')^2}
\label{kaval1}
\ee

\section{Analysis of singularities in the tachyon differential equation}
\label{app:singul}
\subsection{Confining Background}
In this appendix, we will investigate the existence of singular solutions of the tachyon equation of motion (\ref{taueq}). It was argued already that $\tau$ can diverge only at the tip of the cigar.
 Therefore what is left to investigate is  solutions where $\tau'(z)$ diverges at a point $z_0 \in [0,z_{\Lambda}]$, but $\tau(z)$ remains finite at the same $z_0$. We call these solutions ``spurious".
Taking into account $\tau'(z)\gg \tau(z)$ in the neighborhood of $z_0$, the leading terms of Eq.(\ref{taueq}) are the first and the second one. Hence, the leading order equation is (we set $\mu=1$ in the sequel as it can be absorbed in the normalization of $\tau$ and $z_{\Lambda}=1$)

\be
\tau''(z)-{4\over 3}z f(z)\tau(z)=0
\label{taueqsing1}
\ee
with solution in the vicinity of the divergence
\be
\tau'(z)={1\over \sqrt{g(z)}}={1\over \sqrt{C-{4\over 3}z^2\left(1-{2\over 7}{z^5}\right)}}\sp \tau(z)=\int_0^{z} {dz\over \sqrt{g(z)}}+\tau_0
\label{sinsollea}
\ee
where $g(z)=C-{4\over 3}z^2\left(1-{2\over 7}{z^5}\right)$. The function $g(z)$ has either one or three real roots. In particular, there are the following three cases

\begin{enumerate}
\item $\mathbf{C<0 :}$ There are no roots of $g(z)$ in the interval $[0,1]$, since the only one root is at $z_0>1$. It should also pointed out that for $z\in[0,1]$, $g(z)$ is negative so the solution does not exist.
\item $\mathbf{0<C<{20\over 21} :}$ There is a single real zero at $z_0\in [0,1]$. While, the other two real zeros lie outside that interval.

    \item When $\mathbf{C={20\over 21}}$ the divergence happens exactly at the tip of the cigar.

\item $\mathbf{C>{20\over 21} :}$  Again there is no real zero in $[0,z_0]$.
\end{enumerate}

If $g(z)$ has a real root $z_0\in [0,1]$, then it follows from (\ref{sinsollea})
that $\tau'(z)$ diverges at $z=z_0$, but $\tau(z)$ is regular there.
Only in case that $z_0$ is a double root of $g(z)$, both $\tau(z)$ and $\tau'(z)$ diverge at the same point. We are particularly interested in the above case where the acceptable solution diverges at some point $z_0$ in order to obtain the effect of chiral symmetry breaking in the dual quantum field theory. This is only managed if we tune the initial conditions ($C$ here).
There are two possibilities which lead to a double root of $g(z)$, in the context of the above approximation

\begin{enumerate}
\item $\mathbf{C=0 :}$ In that case the double root is at $z_0=0$. Then, Eq.(\ref{taueqsing1}) has not real solutions, so it is not considered here.
\item $\mathbf{C={20\over 21} :}$ The double root now is at $z=1$.
\end{enumerate}

The only rigorous result of the aforementioned analysis is that ``spurious" singularities are generic if  $0<C<{20\over 21}$.

If the tachyon diverges as a power law $\tau\sim (z-z_0)^{-a}$ with $a>0$, ${\tau'\over \tau}\sim {1\over z-z_0}$ and the approximation described above is still valid, provided $z_0\not =1$. But then, this is not a valid solution since the solution in the above approximation is of the form (\ref{sinsollea}). This excludes such a divergence if $z_0\not =1$.

The only other option of divergence of $\tau(z)$, and/or $\tau'(z)$ at a point $z_0 \in [0,1]$  is the case where $\tau'^2\tau$ term is dominating. The relevant equation then is
\be
\tau''(z)+\tau'(z)^2  \tau =0
\label{taudiv2}
\ee
which leads to

\be
\tau'(z)=C~e^{-{1\over 2}\tau(z)^2}
\ee
For $\tau(z)\gg 1$, an approximate solution to Eq.(\ref{taudiv2}) is
\be
{1\over \tau(z)}e^{{1\over 2}\tau(z)^2}\simeq Cz+\ldots
\ee
Therefore, $\tau(z)$ diverges only if $z\to \infty$ which is not allowed in the present geometry as $z\in [0,1]$. Hence this case is excluded.

From the above mentioned, we conclude that the only place where  $\tau(z)$ diverges is at $z=1$. In order to find the solution of diverging $\tau(z)$ we must tune the initial conditions in the UV, see
section \ref{section3}.
We also showed that ``spurious" singularities in the interior of the interval $[0,1]$ are generic for a range of initial conditions.

\begin{figure}[htbp]
\centerline{\includegraphics[width=.48\textwidth]{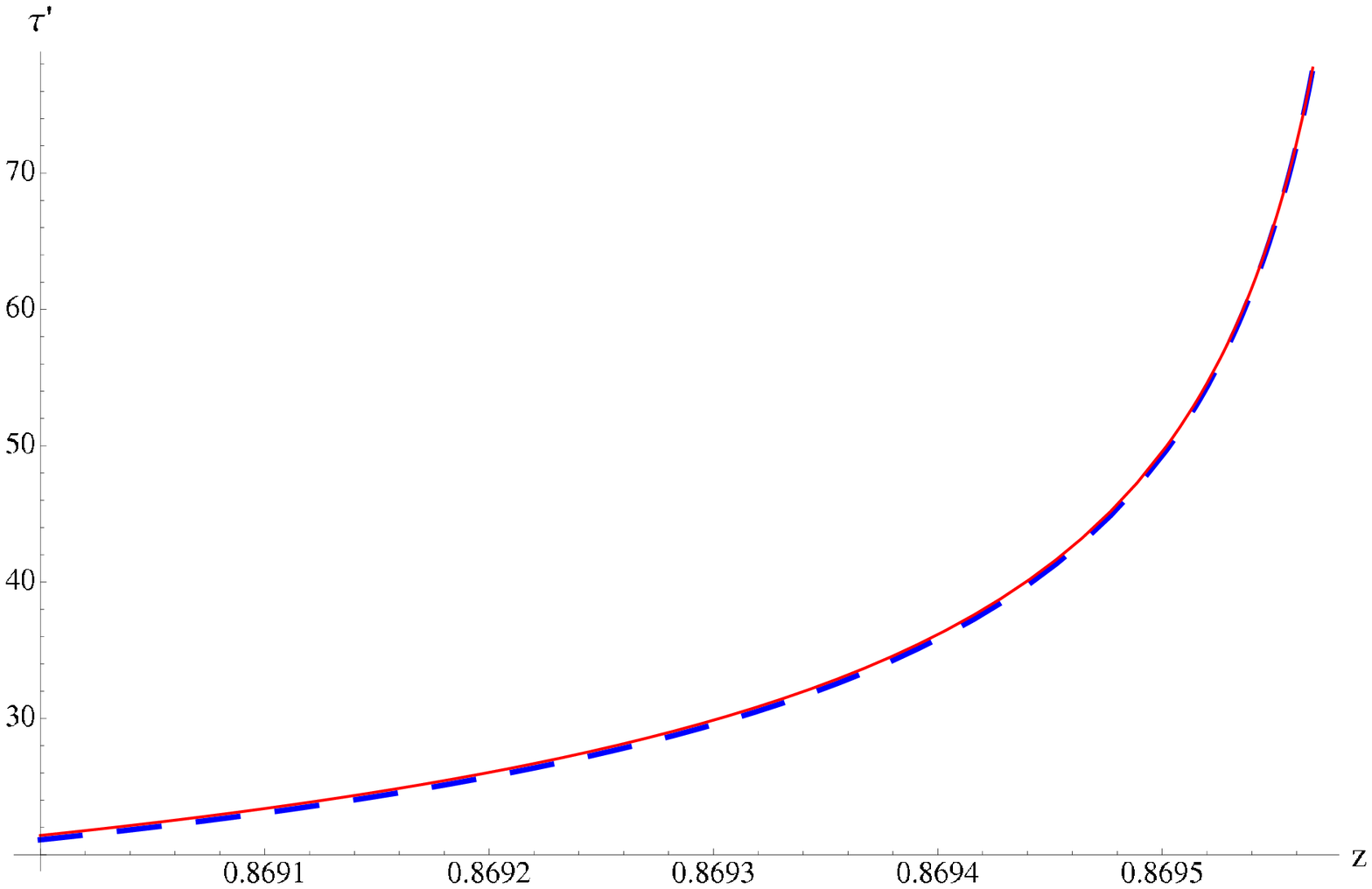}
\includegraphics[width=.48\textwidth]{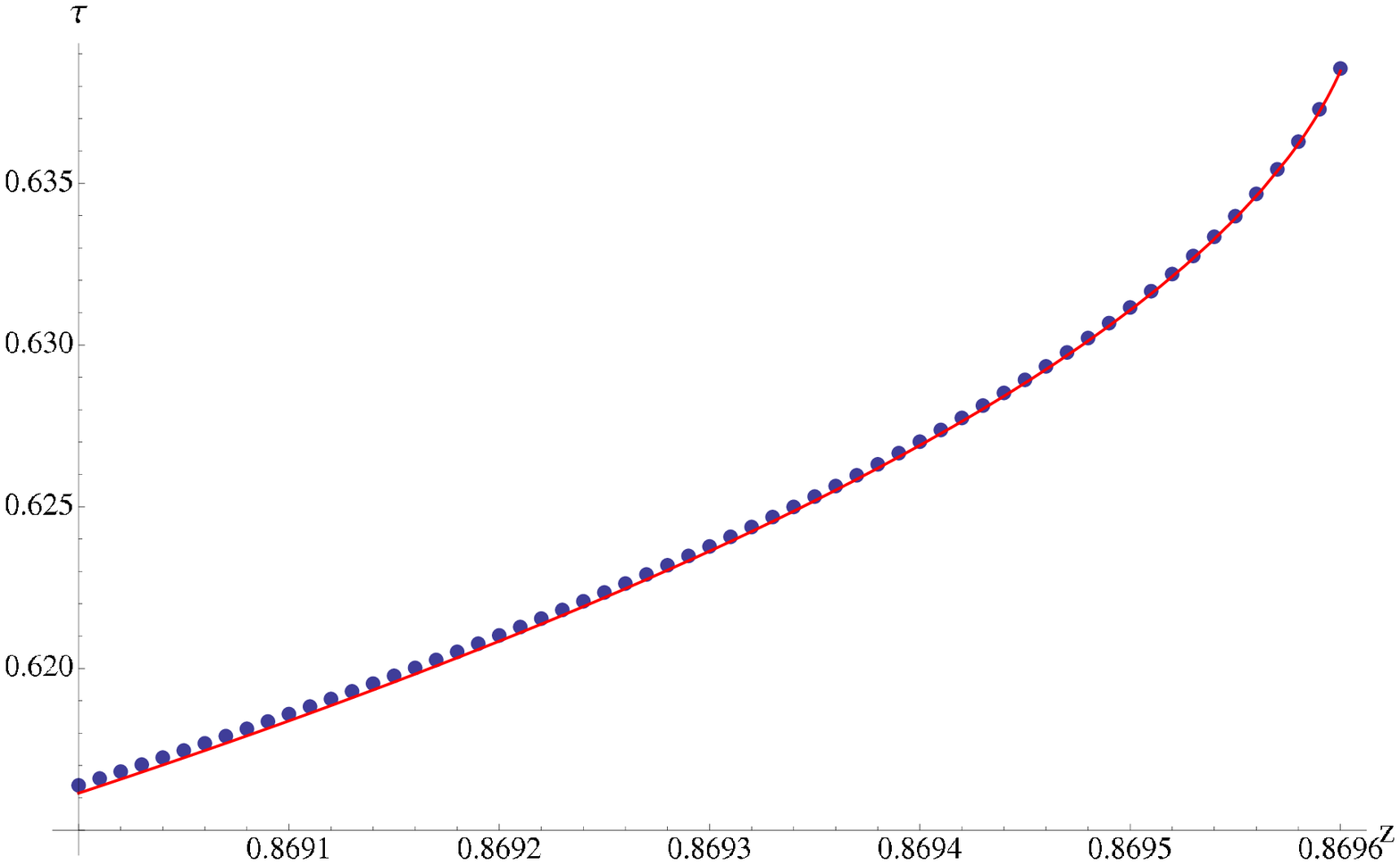}}
\caption{An example of a solution of type
which diverges at $z_{0}=0.8696 < 1$ was found numerically. The numerical solution and its derivative are compared to the asymptotic solution
}
\label{fig:spursing}
\end{figure}

 The existence of ``spurious" singularities has been verified numerically, and it fits the asymptotics (\ref{sinsollea}). An example of this behavior is shown in Fig. \ref{fig:spursing}. We solve numerically  eq.
(\ref{taueq}) for arbitrary initial conditions, meaning that the mass and the vev are not tuned according to the plot in
Fig. \ref{div_tac_fig_2}.
 In particular, we have chosen $c_1=0.1$, $c_3=0.439$. In this case we notice that $\tau'(z_0) \gg \tau(z_0)$ at $z_0=0.8696<1$. The right part of Fig.(\ref{fig:spursing}) includes the plot of the derivative of the numerical solution (red line) and
 the expression for $\tau'(z)$ given in Eq.(\ref{sinsollea}) (dashed blue line), near $z_0$. On the left part we have plotted $\tau(z)$ and the asymptotic solution (\ref{sinsollea}) (blue dots). The parameters of the expressions in (\ref{sinsollea}) are $C=2.71758<{20\over 21}\pi$ and $\tau_0=-0.28$. For those values the asymptotic solution (\ref{sinsollea}) fits the numerical solution of the full equation near $z_0$.

\subsection{Deconfined Background}
\label{singulT}
We now look for singular solutions of the tachyon equation of motion in the deconfined background, Eq.(\ref{taueqbh}). Considering again that $\tau'(z)\gg \tau(z)$ at the vicinity of $z_0$, with $z_0$ being the point where $\tau'(z_0)\rightarrow \infty$, the leading terms of Eq.(\ref{taueqbh}) near $z_{0}$ are

\begin{equation}
\tau''(z) + {z^2 \over 3} f(z)\left( -{4\over z}+{f'(z)\over 2 f(z)} \right) \tau'(z)^{3}=0\,\,,
\label{taueqbhsing}
\end{equation}
where $f(z)=1-{z^5\over z_T^5}$. The solution reads

\begin{equation}
\tau(z)=\int_{0}^{z}{dz \over \sqrt{C-{4  \over 3}z^{2}\left( 1-{3 \over 28} {z^5\over z_T^5}\right) }}+\tau_{0}
\end{equation}

The function $g(z)=C-{4 \over 3}z^{2}\left( 1-{3 \over 28} {z^5\over z_T^5 }\right)$ has a maximum in $z=0$ and no other extrema in the interval $[0,z_T]$. If this function is zero at some point $z_0$, then $\tau'(z_0)$ is infinite and $\tau(z_0)$ is finite except if $g(z_0)=0$ has a double root.
Similarly to the confining background case, there are three choices

\begin{enumerate}
\item $\mathbf{C<0 :}$ $g(z)$ is not zero in the interval $[0,z_T]$.
\item $\mathbf{0<C < {25 \over 21}z_T^2 :}$ $g(z)$ has one real root in the interval $[0,z_T]$ which is not a double root. When, $C={25 \over 21}$ the root of $g(z)$ is at $z_0 =1$.
\item $\mathbf{C>{25 \over 21}z_T^2:}$ There is no root of $g(z)$ in $[0,z_{T}]$.
\end{enumerate}
So, for a suitable range of the initial conditions ($C$) we may have a solution with diverging $\tau'(z)$ and finite $\tau(z)$ at some point $z_0 \in [0,1]$.
A double root is possible to be found in case of $C=0$, and it is at $z=0$.

The discussion about the case where the term $\tau'^2\tau$ of Eq.(\ref{taueqbh}) is leading, remains the same as the one that follows Eq.(\ref{taudiv2}) in the previous appendix. Therefore, in case of the deconfined background tachyon cannot diverge in the interval $[0,1]$ but ``spurious" singularities of the form $\tau'(z) \gg \tau(z)$ exist in general.

\section{Scheme dependence of the condensate and the constant $\alpha$}
\label{app: alpha}

We have introduced an arbitrary constant $\alpha$ in (\ref{Sct}) associated to a counterterm
that gives a finite contribution and therefore cannot be fixed by demanding the cancellation
of divergences. This is a common situation in holographic renormalization and is related to
scheme dependence of renormalization in the field theory side, as we now discuss.

The gauge invariant composite operator $q_L^{\dagger}q_R$ must be defined by a subtraction in QFT.
To asses what enters in such subtractions we must study the OPE $q_L^{\dagger}(x)q_R(y)$ as $x\to y$. The operator can then be defined
by point splitting, subtracting divergent contributions, and then taking the limit $x=y$.

Apart from the identity operator, all other operators that can appear in the OPE $q_L^{\dagger}(x)q_R(y)$ do not provide divergences.
These include the operator itself $:q_L^{\dagger}(x)q_R(x):$ as well higher dimension operators. Therefore this composite
operator can be defined by normal ordering.
To make this precise we use a Dirac basis, so that we have the real part $\bar q(x)q(y)$ and the imaginary part $\bar q(x)\gamma^5 q(y)$.
The imaginary part is finite and no subtraction is needed.
For the real part we introduce a momentum cutoff $\Lambda$ as we will be working in momentum space.
We will therefore define
\be
:\bar q q:=\lim_{\Lambda\to \infty} \left[\bar q q-a_1\Lambda^3-a_2m\Lambda^2-a_3m^2\Lambda-a_4m^3\log{\Lambda^2\over m^2}-a_5m^3\right]
\label{39}\ee
where on the left-hand side we subtracted the most general expression of scaling dimension 3.
This should be enough in a conformally invariant theory. If the theory is asymptotically free, as in QCD, more subtractions are necessary,
as there is one more relevant scale entering the problem, namely $\Lambda_{QCD}$.
However in the model we consider in this paper, the physics in the UV is conformal so these
subtractions are enough.

To establish the coefficients in (\ref{39}) we must require that the (perturbative, short distance) part of the vev is finite when we remove the cutoff.
We have
\be
\langle:\bar q q:\rangle=\lim_{\Lambda\to \infty} \left[\langle\bar q q\rangle-
a_1\Lambda^3-a_2m\Lambda^2-a_3m^2\Lambda-a_4m^3\log{\Lambda^2\over m^2}-a_5m^3\right]
\label{40}\ee
We calculate (in Euclidean space)

\be
\langle\bar q q\rangle=-N_cTr\int {d^4p\over (2\pi)^4}{-i \qslash+m\over q^2+m^2}=-4N_cm \int {d^4p\over (2\pi)^4}{1\over q^2+m^2}
\label{41}\ee
$$
=-2N_c\Omega_3 m\int_0^{\Lambda^2}{p^2dp^2\over p^2+m^2}=-2N_c\Omega_3 m\left[\Lambda^2-m^2\log{\Lambda^2+m^2\over m^2}\right]
$$
$$
=-2N_c\Omega_3 m\left[\Lambda^2-m^2\log{\Lambda^2\over m^2}+{\cal O}(\Lambda^{-2})\right]
$$
where $\Omega_3$ is the volume of th unit 3-sphere.

To renormalize we must choose
\be
a_1=0\sp a_2=-2N_c\Omega_3=-a_4\sp a_3=0
\label{42}\ee
while $a_5$ can be arbitrary.
It is this arbitrariness that is reflected in the coefficient $\alpha$ in the holographic renormalization setup
in (\ref{quarkcon}).\footnote{In a theory like QCD, the arbitrariness involves the addition of a finite function $m^3f\left({m\over \Lambda_{QCD}}\right)$
that reflects the presence of an extra mass scale.}

In some cases, this scheme dependence can be fixed on physical grounds. For instance, in supersymmetric
cases, one can demand that the renormalized on-shell action is zero, see \cite{Karch:2005ms} for an example.
We could not find any convincing prescription to fix $\alpha$ in the present case. A possibility would be
to demand that the condensate vanishes for asymptotically large quark masses, but we have checked numerically
that it is not possible. This is not surprising, since we have seen that our model works much better for small
masses than for large ones.

\section{Appendix: converting to Schr\"odinger form}
\label{app: schro}

In many cases, it is useful to write down the Sturm-Liouville problem
which determines the spectrum of any given mode as a Schr\"odinger-like equation.
Let us start by writing a generic quadratic five-dimensional action for a field
$\Psi(x^\mu,z)$:
\be
S = -\frac12 {\cal K}_\Psi \int d^4 x dz \left(
A(z) (\partial_z \Psi)^2 + B(z) \eta^{\mu\nu} \partial_\mu \Psi \partial_\nu \Psi +
C(z) \Psi \partial_z \Psi + M(z) \Psi^2
\right)\,\,,
\label{quadracti}
\ee
where we have allowed an arbitrary constant multiplying the action.
Let us consider $\Psi = e^{iqx} \psi(z)$ and define as $ m_n^2$ the discrete
set of values of $-q^2$ which satisfy the appropriate normalizability conditions of the
Sturm-Liouville problem.  The discrete set of solutions satisfy the equation of motion
extracted from (\ref{quadracti}):
\be
-\frac{1}{B(z)} \partial_z \left( A(z) \partial_z \psi_n (z)\right)
+ h(z) \psi_n(z) = m_n^2 \psi_n(z)
\label{lalala}
\ee
where we have introduced:
\be
h(z)\equiv \frac{1}{B(z)} \left( M(z) -\frac12 \partial_z C(z)\right)\,\,.
\label{defhofz}
\ee
We can define the orthonormality condition:
\be
 \int dz B(z) \psi_n(z) \psi_m (z) = \delta_{mn}
\label{normpsi}
\ee
We now define a new radial variable $u$, and a rescaled field $\alpha$
in terms of a function $\Xi$ as:
\be
du = \sqrt{\frac{B(z)}{A(z)}} dz\,\,,\qquad
\alpha = \Xi\, \psi \,\,,\qquad
\Xi = (A(z)B(z))^\frac14\,\,,
\label{defsschrei}
\ee
The Sturm-Liouville problem now takes the Schr\"odinger form:
\be
- \frac{d^2 \alpha_n(u)}{du^2} + V(u) \alpha_n(u) = m_n^2 \alpha_n (u)
\label{schroeqap}
\ee
where the Schr\"odinger potential is:
\be
V(u) = \frac{1}{\Xi}\frac{d^2 \Xi}{du^2} + h(u)
\label{schrlike}
\ee
Substituting (\ref{normpsi}) in (\ref{defsschrei}), we find that in the new variables,
the normalization condition is the canonical one:
\be
 \int du \alpha_n(u) \alpha_m(u) = \delta_{mn}
\label{normalpha}
\ee
In order to estimate the mass of the modes with large eigenvalues, it is
sometimes useful
to employ a WKB formula:
\be
\frac{d(m_n^2)}{dn}=2\pi \left[ \int_{u_1}^{u_2} \frac{du}{\sqrt
{m_n^2-V(u)}}
\right]^{-1}
\label{WKB}
\ee
where $u_1$ and $u_2$ are the classical turning points.

\section{Two-point function and sum rules}
\label{app:2point}

We describe here how, typically, the bulk solution needed to compute a two-point correlator from
the gravity side can be written in terms of an infinite sum. Physically, two-point functions at
arbitrary momenta are expressed as a sum over the discrete set of physical states. This  is
not a new result but, however, we believe that explicitly making the discussion as shown below can be
illustrative. We remark that the argument of this appendix is not general in the sense that we have not included in
the reasoning the possibility of having counterterms or other subtleties which may to be dealt with
in a case by case basis.

Let us start with the equation:
\be
-\frac{1}{B(z)} \partial_z \left( A(z) \partial_z \psi (z)\right)
+h(z) \psi(z) = -q^2 \psi(z)
\label{lalala2}
\ee
where we have used the notation of appendix \ref{app: schro}. We want to
find  a solution $\psi_q(z)$ such that in the boundary
$\psi_q(0)=1$, and which is IR-normalizable.
Our goal is to write $\psi_q(z)$ in terms of the discrete infinite set of solutions
$\psi_n(z)$
of the Sturm-Liouville problem (\ref{lalala}), normalized as (\ref{normpsi}).
Let us momentarily change to the notation with $\alpha(u)$, in which the problem
is converted to:
\be
- \frac{d^2 \alpha(u)}{du^2} + V(u) \alpha(u) = -q^2 \alpha (u)
\label{schroeqap2}
\ee
and the discrete spectrum is $\alpha_n(u)$
with (\ref{normalpha}) as normalization and the completeness relation:
\be
\sum_n \alpha_n(u) \alpha_n(u') = \delta(u-u')
\ee
We introduce:
\be
G(u,u')=-\sum_n\frac{\alpha_n(u) \alpha_n(u')}{q^2 + m_n^2}
\ee
such that it is a Green function, $\left[\frac{d^2}{du^2} -V(u) -q^2\right]
G(u,u') = \delta(u-u')$.
Let us assume now that the UV boundary is at $u=0$ and that UV-normalizability implies
that $\alpha_n(0)=0$, such that $G(0,u')=0$.
Regarding (\ref{defsschrei}), $\alpha_q= \Xi \,\psi_q$, such that
for generic momentum
the UV condition is $\alpha_q(0)=\Xi(0)$. The solution we are looking for reads:
\be
\alpha_q(u) = \Xi(u)  + \int_0^\infty G(u,u') (h(u') +q^2 \Xi(u'))du'
\ee
We can translate this back to the original variables. After some manipulations, we get
our final result:
\be
\psi_q(z)=1-\sum_n \frac{\psi_n(z)}{m_n^2} \int_0^\infty \frac{h(z')B(z')}{\Xi(z')} \psi_n(z') dz'
-q^2 A(0)\,\sum_n\frac{\psi_n(z)\psi_n(0)}{m_n^2 (q^2 + m_n^2)}
\label{psiqsol}
\ee
Two-point correlators are built from the on-shell action associated to this solution, which can be
typically found by computing the derivatives fo $\psi_q(z)$ at the boundary.
From the last term in (\ref{psiqsol}) one can find the decay constants of the states in the spectrum,
as in section \ref{sec:decay}.
The second term is $q$-independent and in fact it can be thought of as the $q^2=0$ contribution.
For the axial excitation, this is related to the pion decay constant whereas for the vector excitation,
this term is absent since $h(z)=0$ in that case.

\section{Axial current-current correlator}
\label{axcorrel}

The axial current-current correlator is now derived following the same
procedure as in section (\ref{veccorrel}). We are interested in the two
point function in the limit of large Euclidean momenta. We expect that
the leading term will be the same as in the vector current correlator,
but the subleading term is different.

We define the correlator as

\bea
\int_x e^{iqx} <J_\mu(x)J_\nu(0)> =
(q^2 \eta_{\mu\nu}-q_\mu q_\nu) \Pi_A(q^2)
\label{ax2ptdef}
\eea
This is calculated by differentiating twice the on-shell bulk action.
Integrating by parts (\ref{axacti}) we find

\be
S_{A,reg} =  {(2\pi\alpha')^2 \over g_{V}^{4}}{\cal K}
\int {d^4q\over (2 \pi)^4} \left(
e^{-\frac12 \mu^2 \tau^2}
 g_{xx}
\tilde g_{zz}^{-\frac12}  A_\mu(q,z) \partial_z A^\mu(-q,z)
\right)_{z=\epsilon}
\label{onshellact1}
\ee
where $A_{\mu}(q,z)=\psi^A(q,z)A_{0}(q)$. Then, we insert the
asymptotic solution (\ref{psiaxUV}) into the action

\bea
S_{A} =
{(2\pi\alpha')^2 \over g_{V}^{4}}{\cal K} R
\int {d^4q \over (2 \pi)^4}A^0_\mu(q) A_0^\mu(-q)
\left( 2 b_2(q) +(q^2+ k \mu^2 c_1^2 )({1\over 2} + \log \epsilon)
 \right)
\label{onshellact2}
\eea
where we have set $b_1=1$. The last term is cancelled by the
corresponding counterterm from (\ref{allcounterterms}), so after differentiating twice with respect to
$A_0(q)$ we find the final answer

\be
\Pi_A(q^2)=-4{{\cal K} R (2\pi\alpha')^2 \over g_{V}^{4}}{b_2(q) \over q^2}
\ee
We now compute $b_2(q)$ for large $q^2$, similarly with
section (\ref{veccorrel}). We convert (\ref{axialeom}) to Schr\"odinger
form. Then, the new variable $u$ reads

\be
u \simeq z  \;,\;\; u\simeq {z_{\Lambda} \over \sqrt{3}}\mu\tau
\ee
in the UV and IR respectively. We calculate the asymptotic behavior of the Schr\"odinger potential
in the UV and IR

\be
V_{UV}\simeq {3 \over 4 u^2}+k \mu^2 c_1^2  \;,\;\; V_{IR}\simeq c^2 u^2
\ee
where $c^2={1 \over z_{\Lambda}^4}\left( {9 \over 4}+3 k \right)$.
Adding these two contributions we finally find the equation of motion the
axial modes

\be
-\partial_{u}^2 \alpha+\left( {3 \over 4 u^2}+k \mu^2 c_1^2 + c^2
  u^2\right)\alpha+q^2 \alpha=0
\label{axeomsch}
\ee
where $\alpha(u)\simeq u^{-{1\over 2}}\psi^A (u)$ near $z=0$. Its general solution reads

\be
\alpha(u)=k_{1}{e^{-c u^2 \over 2} \over \sqrt{u}} U\left({q^2+k \mu^2 c_1^2 \over 4 c},{0},c u^2\right)
+k_{2}{e^{-c u^2 \over 2} \over \sqrt{u}} L^{-1}\left({q^2 + k \mu^2 c_1^2\over 4 c},0,c u^2\right),
\label{scsolvsbis}
\ee
we set $k_2=0$, since the generalized Laguerre polynomial is going to
infinity in the IR. $k_1$ is  found such that $\lim_{z\rightarrow 0} \psi^{A}(q,z)=1$

\be
\psi^A={q^2 +k \mu^2 c_1^2 \over 4 c} \Gamma\left({q^2 + k \mu^2 c_1^2
  \over 4 c}\right)e^{-c u^2 \over 2} U\left({q^2+k \mu^2 c_1^2 \over 4 c},0,c u^2\right)
\ee
By expanding the solution for large momenta we find

\be
\lim_{q^2 \rightarrow \infty}{b_2 \over q^2}={1 \over 4}\log\, q^2-{1\over 4}(1+ \log\, 4 -2
\gamma)+ {k \mu^2 c_1^2(2 \gamma -\log\,4+\log\, q^2) \over q^2}+{3 k^2
  \mu^4 c_1^4- 8 c^2 \over 24 q^4}+\ldots
\ee
and eventually

\be
\Pi_A(q^2)=-{{\cal K} R (2\pi\alpha')^2 \over g_{V}^{4}}\left(\log\, q^2-(1+ \log\, 4 -2
\gamma)+ 4 {k \mu^2 c_1^2(2 \gamma -\log\,4+\log\, q^2) \over q^2}+{3 k^2
  \mu^4 c_1^4- 8 c^2 \over 6 q^4}+\ldots \right)
\ee
We notice that for $c_1=0$ the result coincides with the vector
current two point function. The $1/q^2$ term, which is absent in
the QCD result, comes from the mass term of the axial field $(\sim
\tau^2 A_{\mu}A^{\mu})$, see (\ref{axacti}).

\section{Excitation equations in the deconfined phase}
\label{app:deconf}
We assemble here the the Shr\"odinger functions, as defined in Appendix B,
for the equations of motion of field excitation modes  in the deconfined
background (\ref{adsbh6b}), at vanishing spatial momentum.
These modes satisfy an equation of the form (\ref{lalala})
but there is no discrete spectrum.

In case of the vector excitations, we have already mentioned the functions
giving rise to the Schr\"odinger potential approach in (\ref{ABdefsdeconf}).
Then, the  variable $u$ reads
\be
u=\int_0^z \sqrt{\frac{\tilde g_{zz}(\tilde z)}{g_{tt}(\tilde z)}}
d\tilde z\,\, ,
\label{uofzdeconf}
\ee
which remains the same for all different excitations.

For  axial-vector mesons $A(z)$ and $B(z)$ are the same as for the vectors
(\ref{ABdefsdeconf}),
but now we also have:
\be
h(z)=\frac{M(z)}{B(z)}
=  k \mu^2  {\tau^2\over z^2} f_T(z) \,.
\ee
The Schr\"odinger functions appearing in the equations of motion of
scalar excitation modes are
\bear
A(z)=e^{-\frac12 \mu^2 \tau^2}g_{xx}^{3 \over 2}g_{tt}^{\frac12}
{g_{zz}\over  \tilde g_{zz}^{-{3\over2}}}
\,,\qquad \qquad \qquad \qquad
B(z)=e^{-\frac12 \mu^2 \tau^2}g_{xx}^{3\over 2} g_{tt}^{-{1\over2}} \frac{g_{zz}}{\tilde g_{zz}^{1/2}} \,,\qquad \qquad \qquad \rc
C(z)=-2\mu^2 e^{-\frac12 \mu^2 \tau^2} \frac{g_{xx}^{3\over 2}
  g_{tt}^{1\over 2}}{\tilde g_{zz}^{1/2}}
\tau(z) \partial_z \tau(z)\,,\qquad
M(z)=\frac{\mu^2 }{2\pi\alpha' \,\lambda} e^{-\frac12 \mu^2
  \tau^2}(\mu^2 \tau^2-1)g_{xx}^{3\over 2}g_{tt}^{1\over 2}
\tilde g_{zz}^{\frac12}\,.
\eear
The function $M(z)$, $B(z)$ and $C(z)$ combine to give a quite simple expression for the
$h(z)$ defined in (\ref{defhofz}):
\be
h(z)=-\frac{3}{z^2} f_T(z)
\ee

Finally, for the  pseudoscalars we  have
\be
A(z)=e^{\frac12 \mu^2 \tau^2} \tau^{-2} g_{xx}^{-{3\over 2}}g_{tt}^{\frac12} \tilde g_{zz}^{-\frac12}
\,,\qquad
B(z)=e^{\frac12 \mu^2 \tau^2} \tau^{-2} g_{xx}^{-2} \tilde g_{zz}^{\frac12}\,,\qquad
h(z)=\frac{M(z)}{B(z)}= k\frac{\mu^2 \tau^2}{z^2}f_T(z)\,.
\ee

\setcounter{equation}{0}
\section{The action with the symmetric trace and Regge slopes}
\label{garousiaction}

In \cite{Garousi:2007fn}, Garousi proposed an effective action for
the brane-antibrane system which has subtle difference with respect
to Sen's one \cite{Sen:2003tm}, which we have used in this work.
One may wonder what would be the physical consequences of using such
an action in our model. We focus in this appendix on the behaviour of
the spectra of highly excited vectors and axial vectors. The equations
for the vectors are not modified with respect to the main text.
The equations for the axials are different. It turns out that they still
obey a Regge law $m_n^2 \propto n$ for large excitation number $n$ but with
different slope compared to the main text. This slope is still larger than
the one for vectors. We will not deal in the present work with other
possible phenomenological implication of this different tachyon action.

Garousi's action reads\footnote{We adapt it to our present framework,
for instance defining the covariant derivative with a different sign and
disregarding the $B_{\mu\nu}$ field. With respect to the main text, we will
fix the value of some of the constants that we have defined, namely
$g_V^2 = 2\pi\alpha'$, $\lambda=1/(2\pi\alpha')$, ${\cal K}=1$, $\mu^2=2$.
Regarding (\ref{dimcon}), this implies $R^2=3/2$.
This is
inessential (the constants can be easily restored)
and we have done it for the sake of clarity of the equations.
Our convention will be
that indices $M,N$ running over the five space-time coordinates are contracted with
the metric $g_{MN}$ whereas indices $\mu,\nu$ running over the Minkowski directions
are contracted with the flat metric $\eta_{\mu\nu}$.}:
\be
S = - \STr \int d^4x dz \ e^{-\hat T^2}
\sqrt{-\det(g_{MN}+\hat F_{MN} +  D_M \hat T
D_N \hat T)}
\label{straction}
\ee
where hatted symbols are 2x2 matrices:
\be
\hat T = \begin{pmatrix} 0 & T \\ T^* & 0 \end{pmatrix}\,,\quad
\hat F_{MN} = \begin{pmatrix} F_{MN}^{(L)} & 0 \\ 0 &
 F_{MN}^{(R)} \end{pmatrix}\,,\quad
 D_M \hat T = \begin{pmatrix} 0 & D_M T \\ (D_M T)^* &
 0 \end{pmatrix}\,.\quad
\ee
with $F_{MN}^{(i)} = \partial_M A_N^{(i)} -  \partial_N A_M^{(i)}$
and $D_M T = \partial_M T + i (A_M^{(L)}-A_M^{(R)})T=
\partial_M T + 2i A_M T$ the usual
field strength and covariant derivative, where
the definition (\ref{VAdefs}) has been substituted.
 The STr means that one has
to symmetrice in $\hat F_{MN}$, $D_M \hat T$,
$\hat T$ after expanding the square root, and then take the trace.

The expression (\ref{straction}) is quite involved but
we will see that in the particular case
we are interested, one can deal with it: we will consider
quadratic excitations of the gauge fields, while the tachyon
phase is set to a trivial constant and the tachyon modulus
is a non-trivial $z$-dependent function (so we have to keep
all orders in $\tau$, $\partial_z \tau$). We again take a gauge
with $A^{(i)}_z=0$. This is enough to compute the vector and
axial spectrum in the tachyon background.

So let us compute the quadratic expansion in gauge fields.
There are terms in $A_\m^2$ coming from the covariant derivatives
and terms with $F^{(i)2}$. In principle, there could be terms with,
schematically, $i \tau \partial_z \tau A F$ coming from a $DT DT\, F$
product, but these terms would make the action complex and are
 removed by the symmetric trace prescription.
 In the following, we make the computation in two steps: we first compute
the $A_\m^2$ terms and  then compute the $F^2$ terms.

In order to compute the $A_\m^2$ terms, we can consider the
action:
\bear
S_{A^2} = - \STr \int d^4x dz\ e^{-\hat T^2}
\sqrt{-\det(g_{MN}+  D_M \hat T
D_N \hat T)} =\rc
= - \int d^4x dz\
\sqrt{-\det g}\ \STr \left[e^{-\hat T^2}
\sqrt{ \det (\delta^M_N +  D^M \hat T
D_N \hat T)}\right]
\label{stractionnoF}
\eear
We now compute the determinant. Being
inside a $\STr$, the $\hat T$ matrices can be considered as
commuting objects, so $\sqrt{\det (\delta^M_N +
D^M \hat T D_N \hat T )}=
\sqrt{
1 + D^M \hat T
D_M \hat T}$.
We have to expand the square root.
In order to simplify notation, let
us define $s_j$, the coefficients of such expansion:
\be
\sqrt{1+\xi} = \sum_{j=0}^\infty (-1)^{j+1} \frac{(2j -3)!!}{j! 2^j} \xi^j \equiv
\sum_{j=0}^\infty s_j \xi^j
\ee
Thus, also expanding the exponential:
\be
{\cal L}_{A^2} = -
\sqrt{-\det g} \sum_{k=0}^\infty \frac{(-1)^k}{k!}
\sum_{j=0}^\infty s_j
\STr \left[ \hat T^{2k} (D^M \hat T D_M \hat T)^j \right]
\ee
The next step is to perform the symmetriced trace, and a major simplification
 comes out because of the particular computation we are doing.
Define:
\be
\hat J_1 = \begin{pmatrix} 0 & 1 \\ 1 & 0 \end{pmatrix}\,,\quad
\hat J_2 = \begin{pmatrix} 0 & -1 \\ 1 & 0 \end{pmatrix}
\ee
such that (use $T= T^* = \tau$):
\be
\hat T = \tau \hat J_1 \,\qquad D_M \hat T = \partial_M \tau \hat J_1
- 2 i A_M \tau \hat J_2
\ee
The order
zero term in $A_M$, {\it i.e.} the action for the tachyon modulus
is just
\be
{\cal L}_\tau =- 2
\sqrt{-\det g}\, e^{-\tau^2} \sqrt {1 +
\partial^M \tau \partial_M \tau},
\ee
 the same used in the main
text. This means that the discussion of section \ref{section3} still
holds.

We now isolate the quadratic term in $A_M$.
One out of the $j$ factors of
$(D^M \hat T D_M \hat T)^j$ has to be $g_{xx}^{-1}(- 2 i \tau \hat J_2)^2
A_\m A^\m$
while the other $j-1$ factors are $\hat J_1^2 g_{zz}^{-1} (\partial_z \tau)^2$
each. Notice there cannot be crossed terms
because $\partial_M \tau A^M =0$ in the case we are conisdering.
Thus:
\be
{\cal L}_{A^2} = - g_{xx}^{-1}
\sqrt{-\det g} \sum_{k=0}^\infty \frac{(-1)^k}{k!}
\sum_{j=0}^\infty s_j
 j\ \tau^{2k}
(g_{zz}^{-1} (\partial_z \tau)^2)^{j-1}
(- 4 A_\m A^\m \tau^2)
\STr \left[ \hat J_1^{2k+2j-2} \hat J_2^2 \right]
\ee
The $j$ at the beginning of the second line of course comes
because the $A_\m A^\m$ term can be chosen from any of the
$j$ factors in $(D^M \hat T D_M \hat T)^j$.
In order to perform the symmetriced trace, notice that, in general:
\be
\STr[\hat J_1^{2n} \hat J_2^2]= \frac{1}{2n+1}
\left((n+1) Tr [\hat J_2^2] + n \Tr[\hat J_1 \hat J_2
\hat J_1 \hat J_2]\right)= -\frac{2}{2n+1}
\ee
where we have used that $\hat J_1^2$ is the 2x2 identity matrix
and $\Tr[\hat J_1 \hat J_2
\hat J_1 \hat J_2] = - Tr [\hat J_2^2] = 2$.
Substituting:
\be
{\cal L}_{A^2} = - 8 g_{xx}^{-1}
\sqrt{-\det g} \ \sum_{k=0}^\infty \frac{(-1)^k}{k!}
\tau^{2k+2}
\sum_{j=0}^\infty s_j
 j\
(g_{zz}^{-1} (\partial_z \tau)^2)^{j-1}
A_\n A^\n
\frac{1}{2k + 2j -1}
\label{la2int}
\ee
We should now resum the series. Let us use the identity\footnote{In order to prove this, notice that
$\sum_{j=0}^\infty j s_j y^j = \frac{y}{2\sqrt{1+y}}$ and
consider an auxiliary function $g(a)=\sum_{i=0}^{\infty} \frac{1}{i!} x^i \sum_{j=0}^\infty
\frac{j \ s_j}{2i + 2j -1} y^j a^{2i+2j-1}$ such that
$\partial_a g(a)= a^{-2} e^{x a^2}\frac{a^2\ y}{2\sqrt{1+a^2\ y}}$.
Since $g(0)=0$ and $g(1)$ is what we want to compute, we arrive at
(\ref{sumid1}).}:
\be
\sum_{i=0}^{\infty} \frac{1}{i!}x^i \sum_{j=0}^\infty
\frac{j \ s_j}{2i + 2j -1} y^j = \frac{y}{2}
\int_0^1 \frac{e^{xa^2}}{\sqrt{1+y\ a^2}} da
\label{sumid1}
\ee
Using $x=-\tau^2$ and $y=g_{zz}^{-1} (\partial_z \tau)^2$, we finally find:
\be
{\cal L}_{A^2} = - 4 g_{xx}^{-1}
\sqrt{-\det g} \
\tau^{2}A_\n A^\n
\int_0^1 \frac{e^{-\tau^2 a^2}}{\sqrt{1+g_{zz}^{-1} (\partial_z \tau)^2
 a^2}} da
\ee
Let us now compute the $F^2$ terms.
We want to expand the determinant of
(\ref{straction}) to second order in $\hat F$
but to all orders in $D_z \hat T D_z \hat T$.
The determinant reads:
\be
-\det(g_{MN}+\hat F_{MN}+ D_M \hat T D_N \hat T)=
g_{xx}^4 (g_{zz} + D_z \hat T D_z \hat T)
+ \half \hat F_{\m\n} \hat F^{\m\n} g_{xx}^2
(g_{zz} +  D_z \hat T D_z \hat T)
+g_{xx}^3 \hat F_{\m z}^2
\ee
and thus the $F^2$ contribution to the square root is:
\be
\frac14 g_{zz}^\frac12 \sqrt{(1 +g_{zz}^{-1}
 D_z \hat T D_z \hat T)}
\hat F_{\m\n} \hat F^{\m\n}
+ \frac12 g_{xx} g_{zz}^{-\frac12}
\frac{1}{\sqrt{1 + g_{zz}^{-1} D_z \hat T D_z \hat T}}
\hat F_{\m z}^2
\ee
Let us start by computing the term with $\hat F_{\m\n} \hat F^{\m\n}$.
We have to expand before taking the symmetrized trace. Notice
that now $D_z \hat T = \partial_z \tau \hat J_1$ up to subleading
terms. To shorten notation, we define:
\be
x \equiv - \tau^2\,,\qquad y =g_{zz}^{-1} (\partial_z \tau)^2
\label{xydefs}
\ee
The $\hat F_{\m\n} \hat F^{\m\n}$ term in the lagrangian density
(\ref{straction})
then reads:
\be
{\cal L}_{F_{\m\n}^2} = -\frac14
g_{zz}^\frac12
 \sum_{k=0}^\infty \frac{x^k}{k!}
\sum_{j=0}^\infty s_j y^j
\STr[\hat J_1^{2k + 2j} \hat F_{\m\n} \hat F^{\m\n}]
\label{fmnint}
\ee
It is now easy to compute the symmetriced trace:
\be
\STr [\hat J_1^{2k + 2j} \hat F_{\m\n} \hat F^{\m\n}]=
\frac{1}{2k+2j+1}\left(
(k+j+1) \Tr [\hat F_{\m\n} \hat F^{\m\n}] +
(k+j) \Tr [\hat J_1 \hat F_{\m\n} \hat J_1 \hat F^{\m\n}]\right)
\ee
Now, $\Tr [\hat F_{\m\n} \hat F^{\m\n}] = F_{\m\n}^{(L)}
F^{\m\n (L)} + F_{\m\n}^{(R)} F^{\m\n (R)}$ which, splitting
in vector and axial part and using notation of section
\ref{sec:mesons} gives $\Tr [\hat F_{\m\n} \hat F^{\m\n}]=
2 V_{\m\n} V^{\m\n} + 2 A_{\m\n} A^{\m\n}$.
Similarly, $\Tr [\hat J_1 \hat F_{\m\n} \hat J_1 \hat F^{\m\n}]=
2 F_{\m\n}^{(L)} F^{\m\n (R)} = 2 V_{\m\n} V^{\m\n} - 2 A_{\m\n} A^{\m\n}$
and we find $\STr [\hat J_1^{2k + 2j} \hat F_{\m\n} \hat F^{\m\n}]=
2 V_{\m\n} V^{\m\n} + \frac{2}{2i+2j+1} A_{\m\n} A^{\m\n}$.
Inserting this in (\ref{fmnint}):
\bear
{\cal L}_{F_{\m\n}^2} = -\frac14
g_{zz}^\frac12
 \sum_{k=0}^\infty \frac{x^k}{k!}
\sum_{j=0}^\infty s_j y^j
\left( 2 V_{\m\n} V^{\m\n} + \frac{2}{2i+2j+1} A_{\m\n} A^{\m\n}\right)=\rc
=-\frac12
g_{zz}^\frac12 \left[ e^x \sqrt{1+y} V_{\m\n} V^{\m\n}
+ \left(\int_0^1 e^{a^2x} \sqrt{1+a^2 y} da\right) A_{\m\n} A^{\m\n} \right]
\label{appH1}
\eear
The fact that for non-trivial tachyon the symmetric trace produces a coupling
between the left and right gauge fields was already pointed out in
\cite{Garousi:2007fn}. It results in different kinetic terms for vectors and axials.
We skip the details of the similar computation leading to $\hat F_{\mu z}^2$:
\bear
{\cal L}_{F_{\m z}^2} = -   g_{xx}
g_{zz}^{-\frac12}
\left[ e^x \frac{1}{\sqrt{1+y}} (\partial_z V_\m)^2
+ \left(\int_0^1 e^{a^2x} \frac{1}{\sqrt{1+a^2 y}}da \right) (\partial_z A_\m)^2 \right]
\label{appH2}
\eear
By comparing  (\ref{appH1}), (\ref{appH2})  to (\ref{vectoracti}),
we find that the quadratic action for the vector excitation is identical
regardless the choice between Sen's and Garousi's actions.
Nevertheless,
 the axial part changes. From (\ref{appH1}), (\ref{appH2}) it can be read
 that, introducing notation of appendix \ref{app: schro}:
\be
{\cal L}_{axial}=
- \left[
\frac12  B(z)
A_{\m\n}A^{\m\n}+ A(z)
(\partial_z A_\m)^2+
M(z) A_\m^2 \right]
\label{lagrAgarousi}
\ee
with:
\bear
A(z)&=&g_{xx}g_{zz}^{-\frac12}\int_0^1 e^{-\tau^2 a^2}
\left( \sqrt{1+a^2 g_{zz}^{-1} (\partial_z \tau)^2}\right)^{-1} da \,,
 \rc
B(z)&=& g_{zz}^{\frac12}\int_0^1 e^{-\tau^2 a^2} \sqrt{1+a^2 g_{zz}^{-1} (\partial_z \tau)^2} da \,,\rc
M(z)&=& 4 g_{xx} g_{zz}^{\frac12} \tau^2
\int_0^1 e^{-\tau^2 a^2} \left(\sqrt{1+a^2g_{zz}^{-1} (\partial_z \tau)^2} \right)^{-1} da
\label{apphABM}
\eear
In order to proceed further, we need estimate the integrals near the IR, where
$z\to z_\Lambda$ and
the
tachyon diverges, see section \ref{sec:confvev}.
We will use that in the limit where $-x \gg 1 $ and $y \gg 1$ with $-\frac{y}{x}\gg 1$,
it happens that:
\be
\int_0^1 e^{x\,a^2}\sqrt{1+y\,a^2}da \approx -\frac{\sqrt{y}}{2x}\,\,,\qquad
\int_0^1 e^{x\,a^2}(\sqrt{1+y\,a^2})^{-1}da \approx \frac{\log\left(\frac{y}{-x}\right)
-\gamma + 2 \log 2}{2\sqrt{y}}
\label{valueintegrals}
\ee
where $\gamma$ is Euler's constant. The first equality is just found by neglecting the 1
inside the square root. For the second computation, it is not possible to directly neglect the
1 since the result would be divergent, but one can express the integral as
$\int_0^1 (\sqrt{1+y\,a^2})^{-1}da +
\int_0^1 (e^{x\,a^2}-1)(\sqrt{1+y\,a^2})^{-1}da$, such that the first integral can be done
explicitly and in the second one  the 1 inside the square root can be neglected.

We are now ready to compute the leading IR behaviour of the Schr\"odinger potential which
will determine the behaviour of the highly excited states.
We will use that near the IR ($z_\Lambda -z \ll 1$), we have:
\bear
&&g_{xx}\approx \frac{R^2}{z_\Lambda^2}\,\,,\qquad
g_{zz} \approx \frac{R^2}{z_\Lambda^2} \frac{z_\Lambda}{5(z_\Lambda-z)}
\,\,,\qquad
\tau = \sqrt{-x} \approx C\,(z_\Lambda - z)^{-\frac{3}{20}}\rc
&&g_{zz}^{-1} (\partial_z \tau)^2=y \approx \frac{9z_\Lambda  C^2}{
80 R^2}(z_\Lambda -z)^{-\frac{13}{10}}
\label{apphthings}
\eear
From (\ref{apphABM})-(\ref{apphthings}) one can readily check that
$\lim_{z\to z_\Lambda} M(z)/B(z)=0$ and therefore the term $h(u)$ does not
contribute in the IR to the Schr\"odinger potential (\ref{schrlike}).
On the other hand, we can obtain the relation of the $z$-coordinate to the
$u$-coordinate of the Schr\"odinger problem (\ref{defsschrei}):
\be
\sqrt{\frac{B(z)}{A(z)}} \approx \frac{3z_\Lambda}{20}\ \frac{(z_\Lambda-z)^{-1}}{\sqrt{-\log(b(z_\Lambda-z))}}
\,\,,\qquad
u\approx \frac{3}{10}z_\Lambda \sqrt{-\log(b(z_\Lambda-z))}\,\,.
\ee
where $b$ is  a positive constant that will not be important in the following.
We also compute:
\be
\Xi = (AB)^{\frac14} \sim e^{-\frac{5u^2}{6 z_\Lambda^2}}
\ee
where we have not written multiplicative constants and powers of $u$ which do not
affect the leading IR behaviour of the Schr\"odinger potential. Finally, from
(\ref{schrlike}) we find
\be
V(u)\approx \frac{25}{9z_\Lambda^4}u^2
\ee
Since we have a quadratic potential in the IR, the behaviour for asymptotically highly
excited axials is still Regge-like. Unlike in the main text - section \ref{subsecschroax} -, the slope found using Garousi's
action does not depend on the constant $k$.
Comparing to the vector modes - section \ref{subsecschrovec} -,
we see that the Regge slope for the axials is slightly larger, in particular
$\Lambda_A^2/\Lambda_V^2 = 10/9$, where $\Lambda_{V,A}$ are defined as in (\ref{mVAlVA}).

\end{document}